\documentclass[12pt,preprint]{aastex}
\usepackage{amssymb,amsmath}
\usepackage{color,hyperref}
\definecolor{linkcolor}{rgb}{0,0,0.25}
\hypersetup{
  colorlinks=true,        % false: boxed links; true: colored links
  linkcolor=linkcolor,    % color of internal links
  citecolor=linkcolor,    % color of links to bibliography
  filecolor=linkcolor,    % color of file links
  urlcolor=linkcolor      % color of external links
}
\newcounter{address}
\newcounter{tabletwo}
\newcommand{\ie}{i.e.}
\newcommand{\etal}{et al.}
\newcommand{\eg}{e.g.}
\newcommand{\viz}{viz.}
\newcommand{\eqnname}{equation}

\newcommand{\sectionname}{Section}
\newcommand{\documentname}{Article}
\newcommand{\versiontag}[1]{\texttt{#1}}
\newcommand{\flag}[1]{\texttt{\lowercase{#1}}}

\newcommand{\dd}{\mathrm{d}}
\newcommand{\degree}{\ensuremath{^\circ}}

\newcommand{\project}[1]{\emph{#1}}
\newcommand{\sdss}{\project{SDSS}}
\newcommand{\sdssiii}{\project{SDSS-III}}
\newcommand{\galex}{\project{GALEX}}
\newcommand{\ukidss}{\project{UKIDSS}}
\newcommand{\panstarrs}{\project{Pan-STARRS}}
\newcommand{\lsst}{\project{LSST}}
\newcommand{\skymapper}{\project{Skymapper}}
\newcommand{\wfirst}{\project{WFIRST}}
\newcommand{\xdqso}{\project{XDQSO}}
\newcommand{\sdssxdqso}{\sdss-\xdqso}
\newcommand{\exd}{XD}
\newcommand{\Likelihood}{\emph{Likelihood}}
\newcommand{\boss}{\project{BOSS}}
\newcommand{\first}{\project{FIRST}}

\newcommand{\uv}{UV}
\newcommand{\ir}{IR}
\newcommand{\nir}{N\ir}
\newcommand{\core}{CORE}

\newcommand{\setofaj}{\ensuremath{\{a_j\}}}
\newcommand{\ugriz}{\ensuremath{ugriz}}
\newcommand{\classA}{\ensuremath{A}}
\newcommand{\classB}{\ensuremath{B}}
\newcommand{\setofajtrue}{\ensuremath{\{\tilde{a}_j\}}}
\newcommand{\iband}{$i$-band}
\newcommand{\starc}{`star'}
\newcommand{\qsoc}{`quasar'}
\newcommand{\galaxyc}{`galaxy'}
\newcommand{\setoffj}{\ensuremath{\{f_j\}}}
\newcommand{\fluxi}{\ensuremath{f_i}}
\newcommand{\setofrfj}{\ensuremath{\{f_j/\fluxi\}}}
\newcommand{\pqso}{\ensuremath{P(\mathrm{\qsoc})}}
\newcommand{\pstar}{\ensuremath{P(\mathrm{\starc})}}
\newcommand{\ncatentries}{160,904,060}

\newcommand{\ngauss}{20}
\newcommand{\nbins}{47}
\newcommand{\binwidth}{0.2}
\newcommand{\binspacing}{0.1}
\newcommand{\ilow}{17.7}
\newcommand{\ihigh}{22.5}
\newcommand{\zmin}{2.2}
\newcommand{\zmax}{3.5}

\begin{document}

\title{Think outside the color box: probabilistic target selection and
  the \sdssxdqso\ quasar targeting catalog}
\author{Jo~Bovy\altaffilmark{\ref{NYU},\ref{email}},
  Joseph~F.~Hennawi\altaffilmark{\ref{MPIA}},
  David~W.~Hogg\altaffilmark{\ref{NYU},\ref{MPIA}},
  Adam~D.~Myers\altaffilmark{\ref{MPIA},\ref{UIUC}},
  Jessica~A.~Kirkpatrick\altaffilmark{\ref{LBNL}},
  David~J.~Schlegel\altaffilmark{\ref{LBNL}},
  Nicholas~P.~Ross\altaffilmark{\ref{LBNL}},
  Erin~S.~Sheldon\altaffilmark{\ref{Brookhaven}},
  Ian~D.~McGreer\altaffilmark{\ref{Arizona}},
  Donald~P.~Schneider\altaffilmark{\ref{PennState}},
  and Benjamin A. Weaver\altaffilmark{\ref{NYU}}}
\altaffiltext{\theaddress}{\label{NYU}\stepcounter{address} Center for
  Cosmology and Particle Physics, Department of Physics, New York
  University, 4 Washington Place, New York, NY 10003, USA}
\altaffiltext{\theaddress}{\label{email}\stepcounter{address}
  Correspondence should be addressed to jo.bovy@nyu.edu~.}
\altaffiltext{\theaddress}{\label{MPIA}\stepcounter{address}
  Max-Planck-Institut f\"ur Astronomie, K\"onigstuhl 17, D-69117
  Heidelberg, Germany}
\altaffiltext{\theaddress}{\label{UIUC}\stepcounter{address}
  Department of Astronomy, University of Illinois at Urbana-Champaign,
  Urbana, IL 61801, USA}
\altaffiltext{\theaddress}{\label{LBNL}\stepcounter{address} Lawrence
  Berkeley National Laboratory, 1 Cyclotron Road, Berkeley, CA 92420, USA}
\altaffiltext{\theaddress}{\label{Brookhaven}\stepcounter{address}
  Brookhaven National Laboratory, Upton, NY 11973, USA}
\altaffiltext{\theaddress}{\label{Arizona}\stepcounter{address}
  Steward Observatory, University of Arizona, 933 North Cherry Avenue, Tucson, AZ 85721, USA}
\altaffiltext{\theaddress}{\label{PennState}\stepcounter{address}
  Department of Astronomy and Astrophysics, The Pennsylvania State
  University, 525 Davey Laboratory, University Park, PA 16802, USA}

\begin{abstract}
We present the \sdssxdqso\ quasar targeting catalog for efficient
flux-based quasar target selection down to the faint limit of the
Sloan Digital Sky Survey (\sdss) catalog, even at medium redshifts
($2.5 \lesssim z \lesssim 3$) where the stellar contamination is
significant. We build models of the distributions of stars and quasars
in flux space down to the flux limit by applying the
extreme-deconvolution method to estimate the underlying density. We
convolve this density with the flux uncertainties when evaluating the
probability that an object is a quasar. This approach results in a
targeting algorithm that is more principled, more efficient, and
faster than other similar methods. We apply the algorithm to derive
low- ($z < 2.2$), medium- ($2.2 \leq z \leq 3.5$), and high-redshift
($z > 3.5$) quasar probabilities for all \ncatentries\ point-sources
with dereddened \iband\ magnitude between 17.75 and 22.45 mag in the
14,555 deg$^2$ of imaging from \sdss\ Data Release 8. The catalog can
be used to define a uniformly selected and efficient low- or
medium-redshift quasar survey, such as that needed for the \sdssiii's
\emph{Baryon Oscillation Spectroscopic Survey} project. We show that
the \xdqso\ technique performs as well as the current best photometric
quasar selection technique at low redshift, and outperforms all other
flux-based methods for selecting the medium-redshift quasars of our
primary interest. We make code to reproduce the \xdqso\ quasar target
selection publicly available.
\end{abstract}

\keywords{
  catalogs
  ---
  galaxies: distances and redshifts
  ---
  methods: statistical
  ---
  quasars: general
  ---
  stars: general
  ---
  stars: statistics
}

\section{Introduction}\label{sec:intro}

Large samples of quasars are important tools in astrophysics and
cosmology for several reasons. They can be used to study the
properties of quasars and their co-evolution with galaxies, \eg,
through measurements of quasar clustering
\citep[\eg,][]{Arp70a,Hawkins75a,Osmer81a,Shen07a,Ross09a}, or they
can be used to study the intervening intergalactic medium
\citep[\eg,][]{Chelouche07a,Menard10a}. Recently, it has been shown
that the Lyman-$\alpha$ forest at $z \approx 2.5$ as revealed by
background quasars can be used to detect the baryon acoustic feature
at this redshift and thus provide an important measurement of the
angular diameter distance at a distance that cannot currently be
studied with galaxies \citep{McDonald07a}. \sdssiii's \emph{Baryon
  Oscillation Spectroscopic Survey} (\boss; \citealt{Schlegel07a};
\citealt{Eisenstein11a}) aims to do just that by obtaining spectra for
$\approx$ 160,000 quasars ($\approx$ 20 deg$^{-2}$) in the $2.2 \leq z
\leq 3.5$ redshift range. However, only about 40 fibers deg$^{-2}$ are
available to the quasar survey such that \boss\ quasar target
selection must approach an efficiency of 50\,percent.

Quasar target selection in this redshift range is notoriously
inefficient \citep[\eg,][]{Fan99a,Richards02a} because at $z \approx$
2.8 the quasar and stellar loci cross in color space. This problem is
exacerbated by the fact that medium-redshift 2.2 $\leq z \leq 3.5$
quasars are relatively rare, such that to reach a density of 20
deg$^{-2}$, objects must be targeted down to $g \approx 22$ mag, where
photometric uncertainties in the relatively shallow Sloan Digital Sky
Survey (\sdss) imaging can become substantial. Fortunately, since the
\boss\ quasars' primary use is to provide lines of sight through the
intervening intergalactic medium, they do not have to be selected in a
uniform manner. However, in order to allow for ancillary quasar
science such as the measurement of the quasar luminosity function at
the peak of quasar activity, the \boss\ has decided to select half of
the targets in a uniform manner. Therefore, a quasar target-selection
technique that is 50\,percent efficient down to $g \approx 22$ mag at
20 targets deg$^{-2}$ is required.

The last decade has seen the advent of statistical studies with
quasars using purely photometric samples. The $\approx$ 95\,percent
efficient photometric UVX catalog of \citet{Richards04a} was used to
measure the integrated Sachs-Wolfe effect
\citep{Giannantonio06a,Giannantonio08a} and cosmic magnification bias
\citep{Scranton05a}, and to study the clustering of quasars on large
\citep{Myers06a,Myers07a} and small scales
\citep{Hennawi06a,Myers07b,Myers08a}. However, with the latest catalog
of $\approx$ 1,000,000 photometric quasars in \sdss\ Data Release 6
\citep{dr6} brighter than $i = 21.3$ mag \citep{Richards09a}, the
kernel-density-estimation (KDE) technique has reached its limit, since
it does not take into account the photometric uncertainties when
evaluating quasar probabilities. It is already the case that ignoring
the photometric uncertainties results in an incompleteness of the
KDE-selected catalog that is not currently understood. This makes
statistical studies and follow-up of interesting sources \citep[\eg,
  binary quasars;][]{Hennawi10a,Shen10a} difficult. The technique
described in this paper for quasar target selection can also be used
to construct a photometric quasar catalog down to faint magnitudes
that does take photometric uncertainties into account.

Since the first attempts at optical selection of quasars
\citep{Sandage65a}, broad-band quasar target selection has mostly
relied on the ``UV excess'' that characterizes low-redshift quasars
\citep{Schmidt83a,Marshall84a}. However, this selection technique
cannot be used by the \boss, precisely because of the Lyman-$\alpha$
absorption in the optical that the \boss\ aims to study at these
redshifts, which reddens the quasars so that they are no longer UV
excess objects. Quasar target selection in \sdss-\emph{I} and
\sdss-\emph{II} consisted essentially of two color-selection
algorithms using $ugri$ and $griz$ for low- and high-redshift quasars,
respectively \citep{Richards02a}. This color-selection algorithm aimed
to avoid the stellar locus and therefore it cannot be used to
efficiently target the $z \approx 2.5$ to 3 range that is of interest
to the \boss: The \sdss-I/II quasar target selection in this range is
$\lesssim$ 25\,percent efficient and this is at relatively bright
magnitudes, \viz, $i < 19.1$ mag \citep{Richards02a,Richards06a}.

In this \documentname\ we describe a probabilistic target selection
technique that uses density estimation in flux space to assign quasar
probabilities to all \sdss\ point sources. As our density-estimation
tool we use extreme-deconvolution \citep[\exd;][]{Bovy09a}, which uses
the photometric uncertainties of the data and can therefore be used to
obtain reliable target probabilities down to $g \approx 22$ mag. By
targeting the highest probability quasars, an efficient and uniform
medium-redshift quasar survey such as the \boss\ can be performed. We
show, using early \boss\ data, that this approach achieves
approximately 50\,percent efficiency, as required for the \boss's
uniform quasar target-selection algorithm.

We apply this target selection technique to all \sdss\ point sources
and construct the \sdssxdqso\ quasar targeting catalog. This catalog
can be used to define star, low-redshift ($z < 2.2$), medium-redshift,
and high-redshift ($z > 3.5$) target classes. We show that this
selection algorithm performs well at low redshift and that it
outperforms all other flux-based methods for medium-redshift quasar
targeting.

We describe the general density-estimation-based target-selection
technique in \sectionname~\ref{sec:general}. In
\sectionname~\ref{sec:data} we discuss the data used to train our
probabilistic classifier, and in \sectionname~\ref{sec:exdmodel} we
present the specific implementation of the general method that we use
for \boss\ target selection. In \sectionname~\ref{sec:catalog} we
describe the targeting catalog, and in
\sectionname~\ref{sec:properties} we discuss its basic properties. In
\sectionname~\ref{sec:boss}, we show results for medium-redshift
targets based on early \boss\ data. In
\sectionname~\ref{sec:discussion}, we compare our target selection
technique to other methods and describe various extensions to the
basic method described in this \documentname. We conclude in
\sectionname~\ref{sec:conclusion}.

In what follows, AB magnitudes \citep{Oke83a} are used
throughout. Where dereddened fluxes and magnitudes are required we
have used the reddening maps of \citet{Schlegel98a}.

%BOVY: reference point

\section{General considerations}\label{sec:general}

Target selection is essentially a classification problem: Based on a
set of attributes, we must classify an object into one of a discrete
set of classes. This set of classes can be as simple as
target/non-target, but more classes might be beneficial in
circumstances in which different targets are assigned different
priorities (see below). We assume that we have a set of objects with
class assignments available on which we can train the classification
algorithm. This is a classical problem in data analysis/machine
learning, and many classification methods are available (\eg, neural
networks, NNs, \citealt{Bishop95a}; support vector machines,
\citealt{Schoelkopf02a}; Gaussian process classification,
\citealt{Rasmussen06a}). However, object attributes in astronomy are
rarely measured without substantial and heterogeneous measurement
uncertainties, and the training sets are rarely as noisy as the test
sets that are to be classified. Most classification algorithms are
poorly equipped to deal with these complications in a statistically
well-posed manner. They do not naturally degrade the probability of an
object being in a certain class if the measurement uncertainties imply
that the object overlaps several classes.

A simple way to think about classification is to compare the number
densities of the various classes in attribute space: If the number
density of class \classA\ is higher than that of class
\classB\ evaluated at an object's attributes, then the probability
that the object belongs to class \classA\ is higher than that it
belongs to class \classB, with probabilities proportional to the
number densities \citep[\eg,][]{Richards04a}. The advantage of this
density view of classification is that general density-estimation
techniques exist that can handle all of the complications that
necessarily arise with astronomical data, \eg, very noisy attribute
measurements or missing attributes \citep{Bovy09a}. It also provides a
clean probabilistic interpretation of the class assignments.

Consider an object $O$ with attributes \setofaj\ that we wish to
classify into class \classA\ or class \classB, and that these classes
are an exhaustive set. In the context of quasar target selection, the
attributes could be the \ugriz\ fluxes of a point-like object and the
classes \qsoc\ and \starc. We use Bayes' theorem to relate the
probability that object $O$ belongs to class \classA\ to the density
in attribute space
\begin{equation}\label{eq:bayes}
P(O \in \classA | \setofaj) = \frac{p(\setofaj | O \in \classA)\,P(O \in \classA)}{p(\setofaj)}\,,
\end{equation}
where
\begin{equation}\label{eq:normalize}
p(\setofaj) = p(\setofaj | O \in \classA)\,P(O \in \classA) + p(\setofaj | O \in \classB)\,P(O \in \classB)\,,
\end{equation}
since $\classA \cup \classB$ contains all of the possibilities. Note
that we distinguish between discrete probabilities $P(\cdot)$ and
continuous probabilities $p(\cdot)$. The first factor in the numerator
of the right-hand side of \eqnname~(\ref{eq:bayes}) is the density in
attribute space evaluated at the object's attributes \setofaj, the
second factor is the total number of \classA\ objects in an unbiased
sample, and the denominator is a normalization factor. It is easy to
see that this probability is a true probability in the sense that it
always lies between zero and one and that it sums to one.

The common situation in which the attributes of object $O$ are
measured with substantial measurement uncertainty is easily handled in
this framework through marginalization over the ``true'' attributes
\setofajtrue\ based on the observed attributes \setofaj\ and
measurement-uncertainty distribution $p(\setofajtrue|\setofaj)$ (which
can be as simple as a Gaussian distribution for Gaussian
uncertainties)
\begin{equation}\label{eq:errconv}
P(O \in \classA | \setofaj) = \int \dd \setofajtrue\, P(O \in \classA | \setofajtrue)\, p(\setofajtrue|\setofaj)\,,
\end{equation}
where the first factor in the integral is given by
\eqnname~(\ref{eq:bayes}). This is again a well-defined probability.

The target-selection problem then becomes the task of training good
number-density models for both the target population and the
contaminants population to maximize the efficiency and completeness of
the survey. In regions of attribute space where the efficiency is low,
for example, $z \approx 2.8$ quasars where the quasar and stellar loci
cross, the challenge is to develop the best possible density models
for the most efficient target selection.

\section{Training data}\label{sec:data}

For the purpose of the \sdssxdqso\ quasar targeting catalog we need to
classify point-like objects into classes \starc\ and \qsoc. This
assumes that star--galaxy separation is perfect for extended galaxies,
such that the additional class \galaxyc\ can be ignored. If this were
not the case then the class \galaxyc\ would be included and
morphological attributes relevant to star--galaxy separation would be
added to the attribute list \setofaj. Contamination from point-like
galaxies, which is small in the apparent magnitude range of interest,
is automatically handled in what follows since the training set for
the \starc\ class consists of non-varying sources, which includes both
stars and point-like galaxies.

The \sdss\ (\citealt{York:2000gk}) has obtained \emph{u,g,r,i} and
\emph{z} CCD imaging of $\approx$ 10$^4$ deg$^2$ of the northern and
southern Galactic sky
\citep{Gunn:1998vh,Stoughton:2002ae,Gunn06a}. \sdssiii\ has extended
this area by approximately 2,500 deg$^2$ in the southern Galactic cap
\citep{Aihara11a}. All the data processing, including astrometry
\citep{Pier:2002iq}, source identification, deblending and photometry
\citep{Lupton:2001zb}, and calibration \citep{Fukugita:1996qt,
  Hogg01a,Smith:2002pca, Ivezic:2004bf,Padmanabhan08a} are performed
with automated \sdss\ software.

This \sectionname\ describes the data used to \emph{train} the density
classification model. The data used to create the \sdssxdqso\ catalog
are presented in \sectionname~\ref{sec:catalog}.

\subsection{Stellar data}

The stellar training set is generated using the co-added photometry of
objects in 150 deg$^2$ (-$30^\circ < \alpha_{\mathrm{J2000}} <
+30^\circ$ and -$1^\circ.25 < \delta_{\mathrm{J2000}} < +1^\circ.25$)
of \sdss\ Stripe-82 \citep{Abazajian09a}. We use those primary objects
with image processing flags\footnote{See
  \url{http://sdss3.org/dr8/algorithms/bitmask\_flags1.php} and\\
  \url{http://sdss3.org/dr8/algorithms/bitmask\_flags2.php} for a
  description of these flags.} satisfying
\begin{quote}
\flag{!DEBLEND\_TOO\_MANY\_PEAKS} \&\& \flag{!MOVED} \&\&
\flag{STATIONARY} \&\& \flag{BINNED1} \&\& \flag{!SATUR\_CENTER} \&\&
\flag{!BAD\_COUNTS\_ERROR} \&\& \flag{!NOTCHECKED\_CENTER}
\end{quote}
See \citet{Stoughton:2002ae} for a description of the
\sdss\ photometric flags. Objects with a high variability--–where the
$\chi^2$ per degree of freedom of the distribution of $r$-band flux
over successive calibration runs is greater than 1.4--–are removed as
many of these objects are quasars (see J.~A.~Kirkpatrick \etal, 2011,
in preparation).

The training set contains 701,215 objects. We identified 23,540
objects with existing spectra in this set. Only 221 of these are
quasars (120 $z < 2.2$; 84 2.2 $\leq z \leq 3.5$; 17 $z > 3.5$)
indicating that the contamination of the stellar training set with
quasars is small. Because of this, we do \emph{not} remove these known
quasars from the stellar training set.

The fluxes obtained during a single \sdss\ imaging run have typical
uncertainties of $\approx 0.1$ mag ($gri$) and $\approx 0.4$ mag
($uz$) at $i \approx 20$ mag. The co-added photometry in Stripe-82 has
uncertainties that are three to five times smaller, depending on the
number of imaging epochs available for an object. These multi-epoch
observations also show that the single-epoch uncertainties are correct
in that they explain the scatter in multiple observations of
non-variable objects \citep{Ivezic03a,Scranton,Ivezic07a}.

\subsection{Quasar data}\label{sec:qsodata}

We use a sample of 103,601 $z \ge 0.3$ quasars from the \sdss\ DR7
quasar catalog \citep{Schneider10a}. This sample includes 14,063
quasars with $2.2 \le z \le 3.5$ and 3,519 with $z > 3.5$.

Below we model the densities of relative fluxes in a large number of
narrow bins in \iband\ magnitude. Because the number of currently
known quasars is too small to provide sufficient data in each
magnitude bin, we assume that the colors of quasars are independent of
their absolute magnitude. Although there are well-known correlations
between quasar spectral properties and luminosity
\citep[\eg,][]{Baldwin77,Yip2004}, these trends are very subtle and
they mostly affect emission line shapes. These weak luminosity trends
are thus washed out in broad band colors, especially in comparison to
the very large intrinsic scatter. When fitting for the relative-flux
distributions, this approximation allows us to re-scale the fluxes of
all 103,601 quasars to the central $i$-band apparent magnitude of the
bin in question, which is a sufficient number to obtain good
fits. However, the redshift distribution at this magnitude is
different from what would be observed in nature for two reasons.
First, we have taken quasars selected from a broad magnitude range to
be representative of the redshift distribution at the center of the
bin in question. Second, the redshift distribution of these 103,601
quasars is not the true quasar redshift distribution because it has
the \sdss\ quasar selection function imprinted on it
\citep[\eg,][]{Richards06a}.

The luminosity function dictates how the redshift distribution in a
given bin depends on apparent magnitude. To correct the redshift
distributions, we use a model of the quasar luminosity function
\citep{Hopkins07a} to compute a set of relative weights such that the
weighted histogram of the quasars results in the redshift distribution
predicted by the luminosity function for that apparent magnitude
bin. Thus in each magnitude bin, we fit \emph{all} 103,601 quasars in
the training set, re-weighted to produce the right mix of low- and
high-redshift quasars according to the luminosity function.

\section{Extreme-deconvolution density model}\label{sec:exdmodel}

To estimate the density of stars and quasars in flux space we use
\exd\footnote{Code available at
  \url{http://code.google.com/p/extreme-deconvolution/}~.}
\citep{Bovy09a}. At the faint end ($i \gtrsim 20$ mag) of the
magnitude range of interest here the flux uncertainties in the
training set are substantial, even though they are much smaller than
the single-epoch uncertainties used to evaluate quasar probabilities
for the \starc\ class (see below). Deconvolution is therefore
necessary to avoid adding in uncertainties twice at the
density-evaluation phase. While \exd\ can handle missing data as well,
we do not need this feature here. However, this capability of \exd\ is
crucial when we want to extend the current framework to include
near-infrared (\nir), ultraviolet (\uv), or variability information,
since these data will not be available for every object in the
training set (see \sectionname~\ref{sec:extensions}).

\exd\ models the underlying, deconvolved, distribution as a sum of $K$
Gaussian distributions, where $K$ is a free parameter that needs to be
set using an external objective. It assumes that the flux
uncertainties are known, as is the case for point-spread function
(PSF) fluxes for point sources in
\sdss\ \citep{Ivezic03a,Scranton,Ivezic07a}. \exd\ consists of a fast
and robust algorithm to estimate the best-fit parameters of the
Gaussian mixture. For example, we were able to fit the color
distribution of the full stellar catalog of 701,215 objects in only a
few hours using 30 four-dimensional Gaussians. It is robust in the
sense that even a poor initialization quickly leads to an acceptable
solution. We are interested not so much in the true underlying
distribution function as in finding a good fit to the observed density
(after convolving the model with the data uncertainties) without
overfitting, so it is not absolutely necessary to find the exact best
fit in the complicated likelihood surface.  Therefore, we use the
simplest version of \exd\ that does not use the heuristic search
extension or priors on the parameters \citep{Bovy09a}.

The \exd\ method works by iteratively increasing the likelihood of the
underlying, $K$ Gaussian model given the data. It is an extension of
the Expectation Maximization (EM) algorithm for fitting mixtures of
Gaussians in the absence of noise \citep{Dempster77a} to the case
where noise is significant or there are missing data. The algorithm
basically iterates through an expectation (E) and a maximization (M)
step. During the expectation step the data and the current estimate of
the underlying density are used to calculate the expected value of (a)
indicator variables that for each data point indicate which Gaussian
it was drawn from and (b) the true, noiseless value of each data point
and its second moment. In the maximization step, these expected values
are used to optimize the amplitude, mean, and variance of each of the
$K$ Gaussians. The E and M steps are iterated until the likelihood
ceases to change substantially. The algorithm has the property that
after each EM iteration the likelihood of the model is increased.

\subsection{Construction of the model}

The full model consists of fitting the flux density in a number of
bins in \iband\ magnitude for the various classes of objects. We
describe the model in a single bin first for a single example
class. Throughout we will use the \starc\ class as the example.

We opted for a binning approach because the true five-dimensional
distribution of fluxes has a dominant power-law shape corresponding to
the number counts as a function of apparent magnitude. However, most
of the information for discriminating between quasars and stars is not
in this power-law behavior, but in the behavior of colors or relative
fluxes. While the latter can be represented well by mixtures of
Gaussian distributions (see below), the power-law behavior cannot
without using large numbers of Gaussians. For this reason we chose to
take out the power-law degree of freedom, \ie, the overall behavior of
the density as a function of apparent magnitude. Neither the color
distributions of quasars or that of stars is a strong function of
apparent magnitude, such that the binning described below does not
introduce strong assumptions about the behavior of the color
distributions. Any weak magnitude dependence of the color
distributions is captured in our model since we use narrow bins in
\iband\ magnitude (see below).

The model only includes the fluxes of an object and not its position
on the sky. Since quasars are uniformly distributed on the sky, while
stars are concentrated near the Galactic plane, the position on the
sky of an object is a potential discriminant. Because the fluxes are
much more informative about an object's type than its position, we do
not include the position in the model. We discuss below how the model
could be extended to include the position of an object as part of the
model.

In a single bin in \iband\ magnitude, we separate the absolute flux
from the flux relative to $i$ in the likelihood in
\eqnname~(\ref{eq:bayes}) as follows
\begin{equation}\label{eq:exdmodel}
p(\setoffj|O \in \mathrm{\starc})= p(\setofrfj | \fluxi, O \in
\mathrm{\starc})\,p(\fluxi | O \in \mathrm{\starc})\,,
\end{equation}
where we now specify that the attributes \setofaj\ are the
\ugriz\ fluxes \setoffj, \setofrfj\ are the fluxes relative to $i$, and
\fluxi\ is the \iband\ flux. We model these two factors separately.

We model the first factor using \exd\ in narrow bins in
\iband\ magnitude described in detail below. We use relative fluxes
rather than colors---logarithms of relative fluxes---since the
observational uncertainties are closer to Gaussian for relative fluxes
than they are for colors. Except for the absence of a logarithm,
relative fluxes are similar to colors and have the same number of
degrees of freedom, \viz, four. The fact that fluxes must be larger
than zero while the Gaussian mixture model does not contain any such
constraints, which is one reason to model the logarithm of the fluxes
rather than the fluxes themselves, does not matter greatly here
because most of the objects in the training set are at least
five-sigma detections. However, for $z > 3$ quasars the $u$-band has
zero flux and the flux measurement can be negative, in which case
magnitudes are badly behaved; relative fluxes remain well-behaved in
this case. To evaluate the \exd\ probabilities during training, we
always convolve the underlying model with the objects'
uncertainties. We assume that the relative-flux uncertainties are
Gaussian---which is a good assumption because the \iband\ magnitude is
always measured at a reasonably signal--to--noise ratio---such that
the convolution of the Gaussian mixture with the uncertainties results
in a Gaussian mixture, with an object's uncertainty variance added to
the model variance for each of the components.

We model the four-dimensional relative fluxes \setofrfj, which each
come with an individual, non-diagonal, four by four uncertainty
covariance, using \ngauss\ Gaussian components. The number
\ngauss\ was decided upon as follows. For a few bins we performed
\exd\ fits with 5, 10, 15, 20, 25, and 30 components. While fits with
less than \ngauss\ components overly smoothed the observed
distribution, models with more than \ngauss\ components used the extra
components to fit extremely low significance features in the observed
distribution. Because of the scale of the full model (see below) no
bin-by-bin objective method for setting the number of components was
pursued, although we did verify that all of the resulting fits were
reasonable.

We model the second factor, $p(\fluxi | i \in \mathrm{\starc})$, by
first combining it with the number factor $P(i \in
\mathrm{\starc})$. This combined factor becomes the number density as
a function of apparent magnitude. This quantity will always be
expressed in units of deg$^{-2}$. For the \starc\ class we model the
number density directly using the number counts of the training data,
by spline interpolating the histogram of \iband\ magnitude number
counts per square degree. For the \qsoc\ class we use a model for the
quasar luminosity function \citep{Hopkins07a} to calculate the number
density of quasars as a function of apparent \iband\ magnitude; we
multiply these theoretical number densities with a simple model for
the \sdss\ incompleteness of point sources
\begin{equation}
I(i) = \left(1 + \exp\left(\frac{i-21.9}{0.2}\right)\right)^{-1}\,,
\end{equation}
designed to reproduce the incompleteness as given in
\citet{Abazajian03a}. The $p(\fluxi | i \in \mathrm{class})$ factors
for the various target classes are shown in
\figurename~\ref{fig:numcounts}. The total number densities for the
various classes are given in \tablename~\ref{table:numcounts}.

The full model consists of \nbins\ bins of width \binwidth\ mag
between $i = \ilow$ and $i = \ihigh$, spaced \binspacing\ mag apart
such that adjacent bins overlap. We further divide the \qsoc\ class
into three subclasses corresponding to low-redshift ($z < \zmin$),
medium-redshift ($\zmin \leq z \leq \zmax$; the \boss\ quasar redshift
range), and high-redshift ($z > \zmax$) quasars. The \exd\ fits for
all but the brightest bin for a given class are initialized using the
best-fit parameters for the previous bin. There are typically $\approx
20,000$ objects in each bin for the stellar training data; for the
quasars there are 85,998 low-redshift, 14,060 medium-redshift, and
3,519 high-redshift quasars in each bin.

In each of 4 $\times$ 47 bins we fit 20 four-dimensional Gaussians,
yielding a total of $4 \times 47 \times (20 \times 15 - 1) = 56,212$
parameters.

\subsection{Comparison of the model and observations}

In this \sectionname\ we assess the performance of the \exd\ technique
for modeling the relative flux distributions of stars and quasars, and
provide examples which demonstrate that the algorithm produces
excellent fits to the data.

The \exd\ method implicitly produces an error-deconvolved model of the
relative-flux distribution, which would correspond to the true parent
distribution of our training set in the limit of infinite
signal-to-noise ratio. We refer to this as the `underlying' model of
the data. An important test of our approach is to apply the
\exd\ technique to the noisier single-epoch data of the stars in
Stripe-82, and compare our determination of the underlying model to
the much higher signal-to-noise ratio Stripe-82 co-added photometry
\emph{of the same sources}. This is a stringent test since the
underlying model fit is based on much noisier single-epoch data than
the co-added photometry.

If the co-added photometry were perfect, then the co-added data would
represent samples of the underlying model with zero noise.  In
practice, even the Stripe-82 co-adds are noisy at the faint end, thus
the relative-flux distribution of the co-adds will be correspondingly
smoothed. To make a sensible comparison we sample the underlying model
at the same number of points as are in the co-added catalog in a given
magnitude bin, and convolve these samples with uncertainties of the
co-added data.  As we do not have a model for the noise in the
co-added data, and our samples are in general at different locations
in relative flux space, we simply match each sample from the
underlying distribution to the closest object in the co-added catalog
and use that co-add object's uncertainties.

\begin{figure}
\includegraphics[width=0.32\textwidth,clip=]{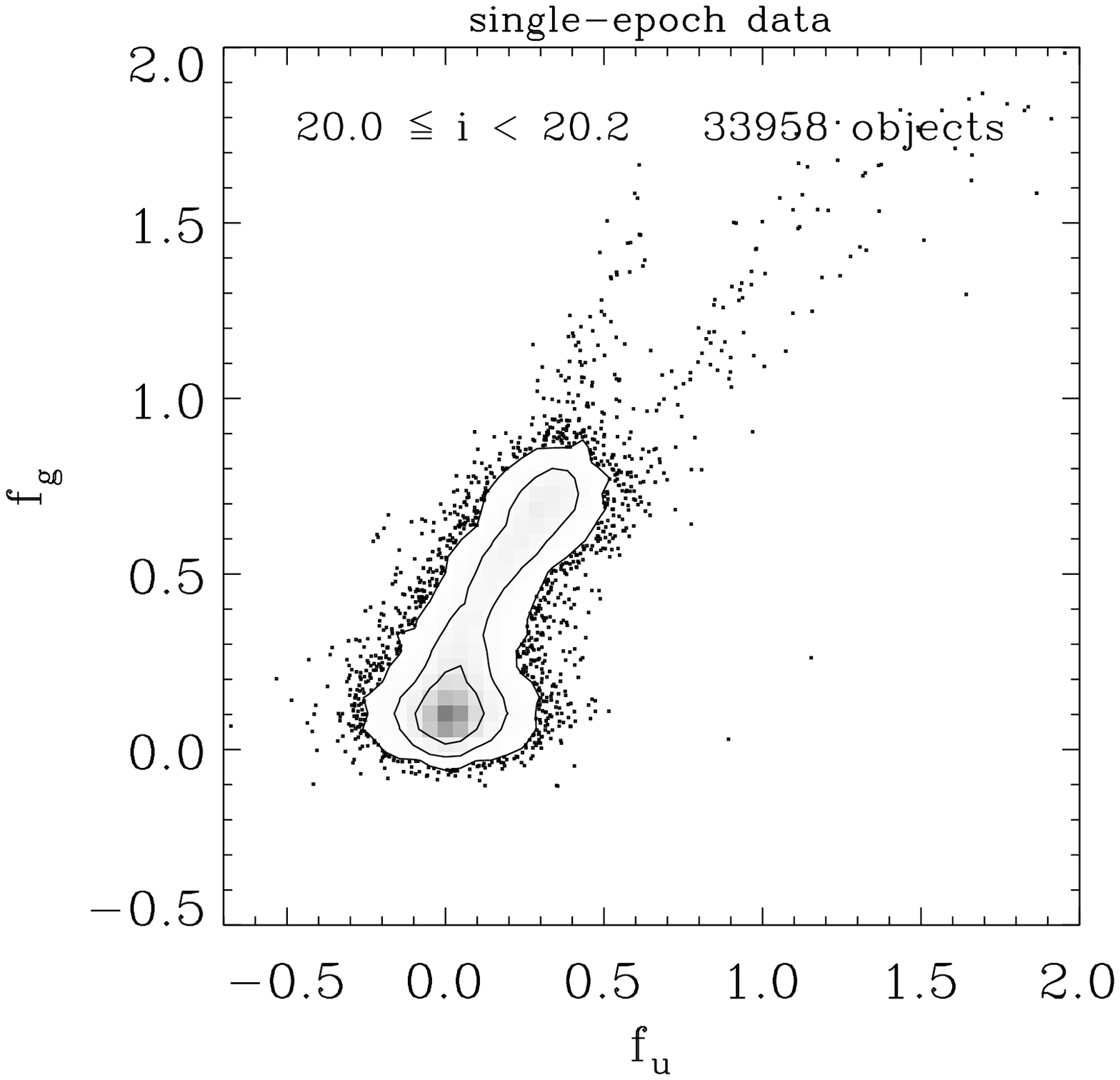}
\includegraphics[width=0.32\textwidth,clip=]{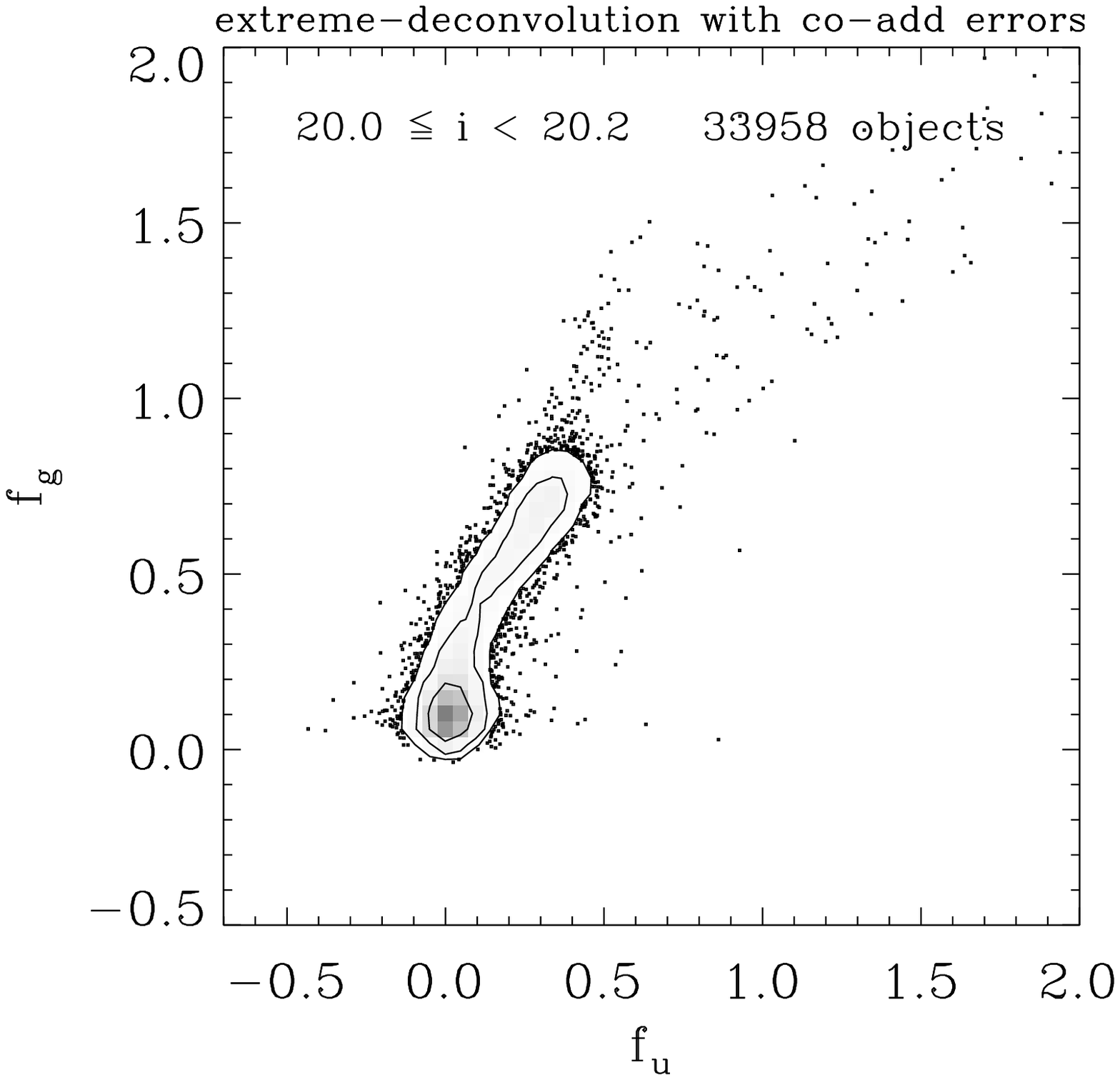}
\includegraphics[width=0.32\textwidth,clip=]{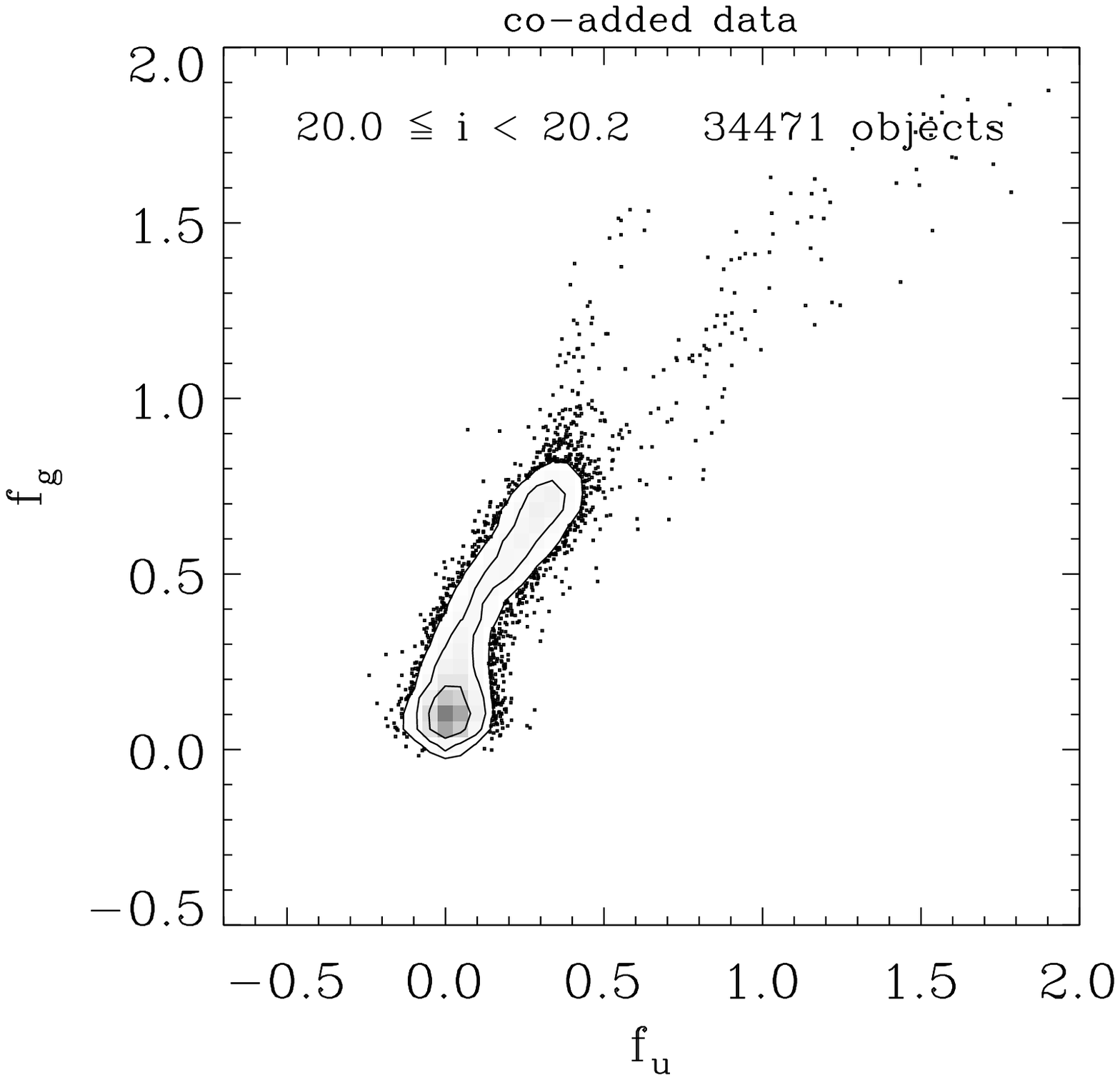}\\
\includegraphics[width=0.32\textwidth,clip=]{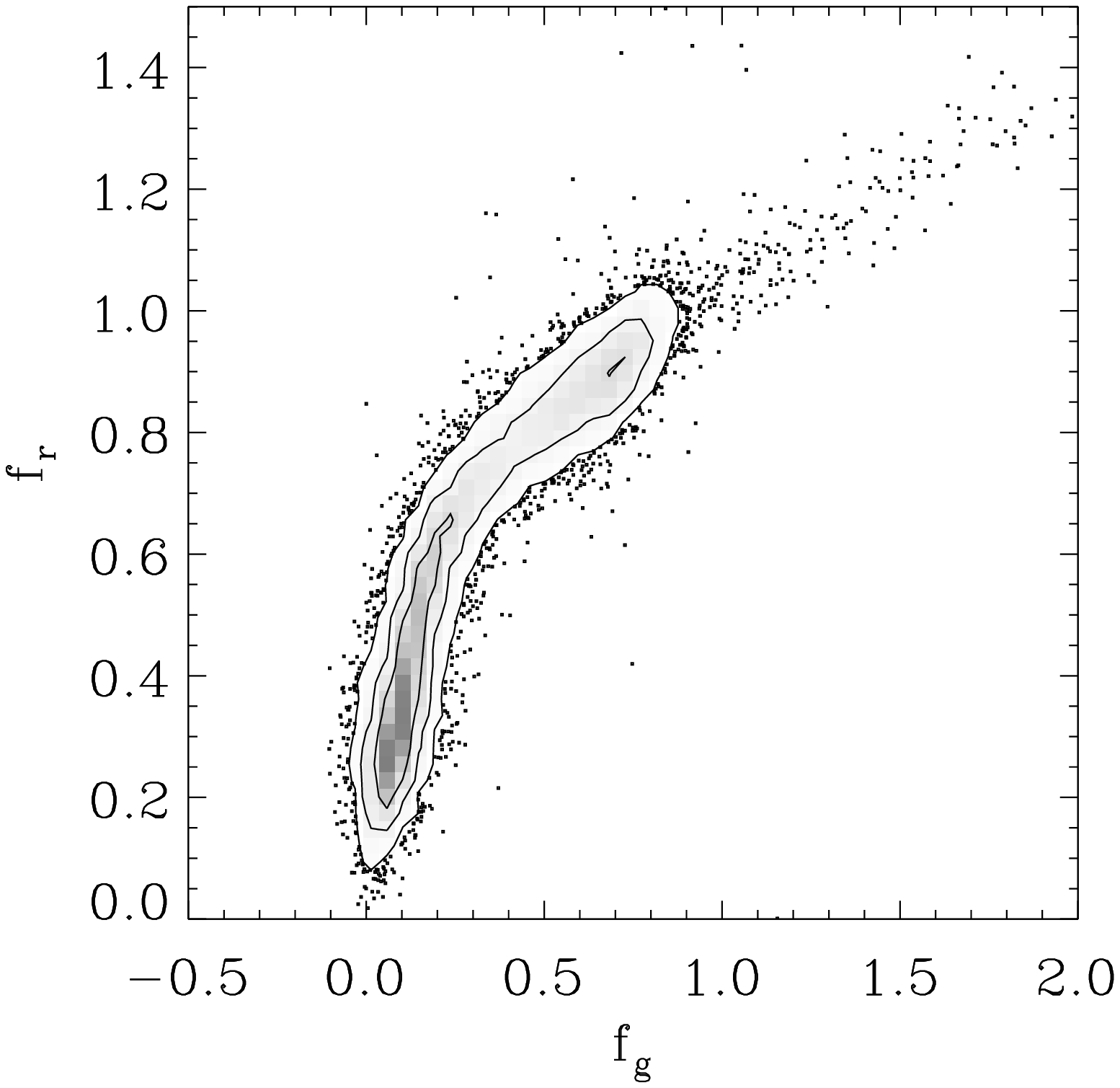}
\includegraphics[width=0.32\textwidth,clip=]{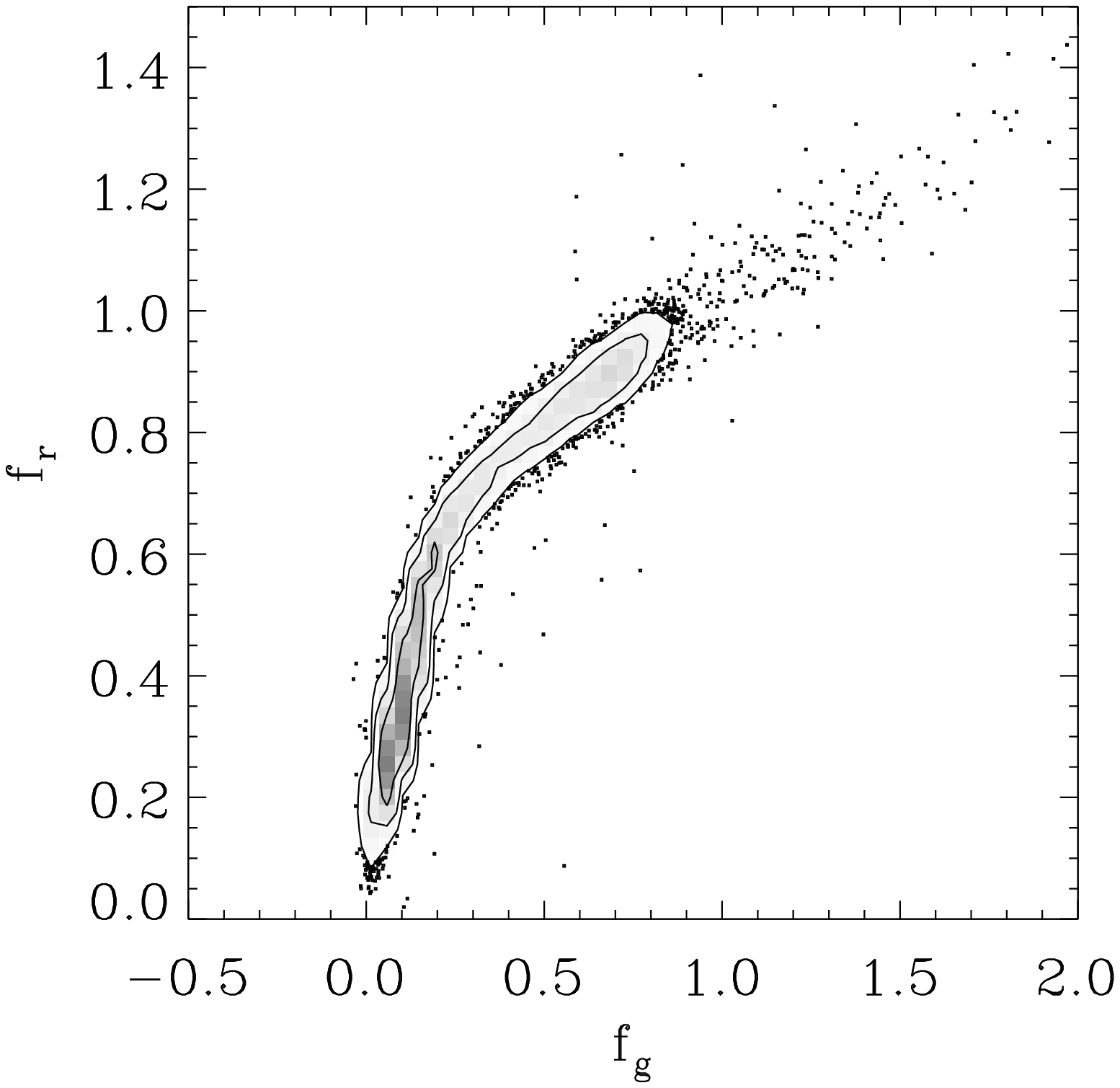}
\includegraphics[width=0.32\textwidth,clip=]{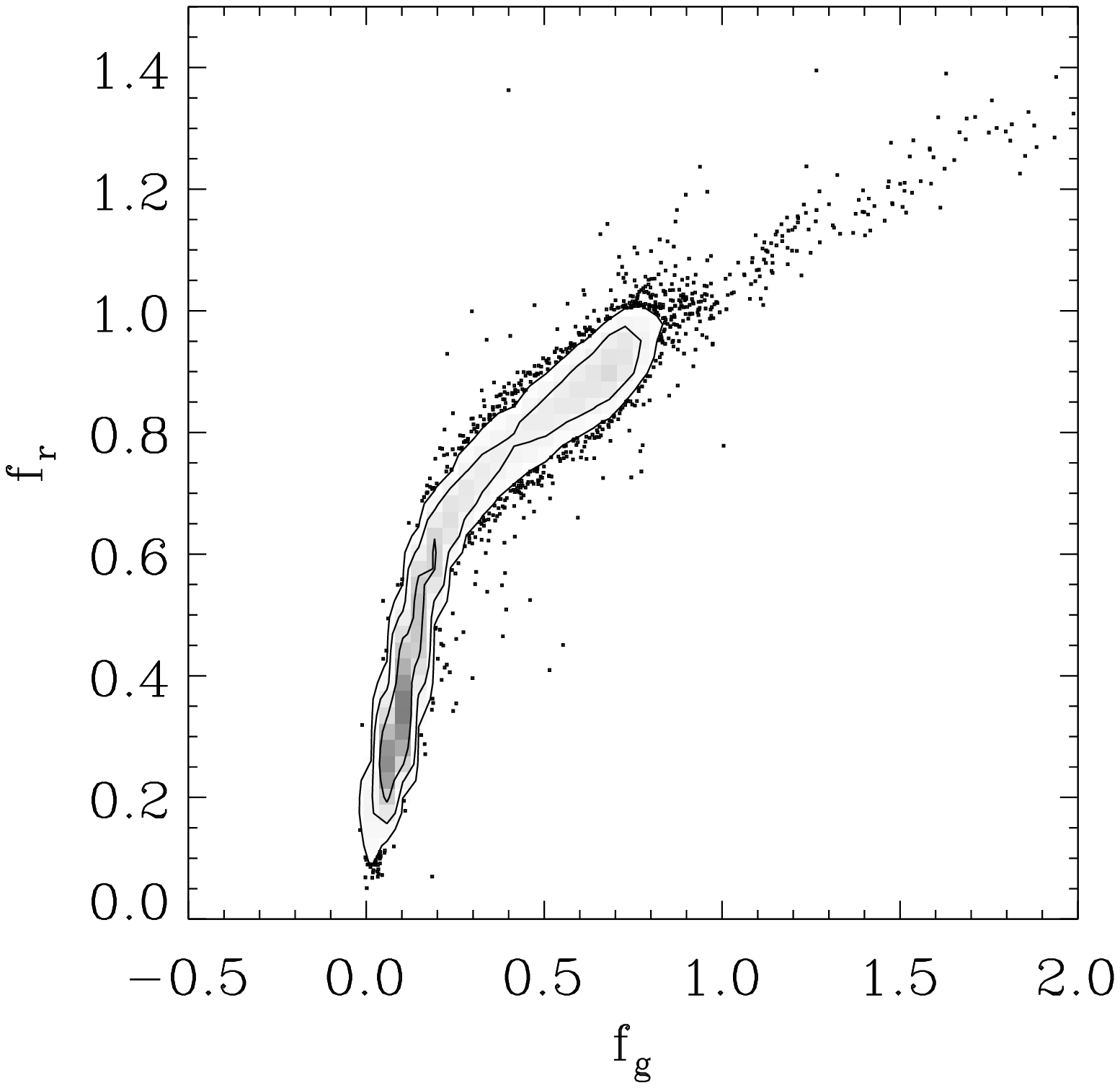}\\
\includegraphics[width=0.32\textwidth,clip=]{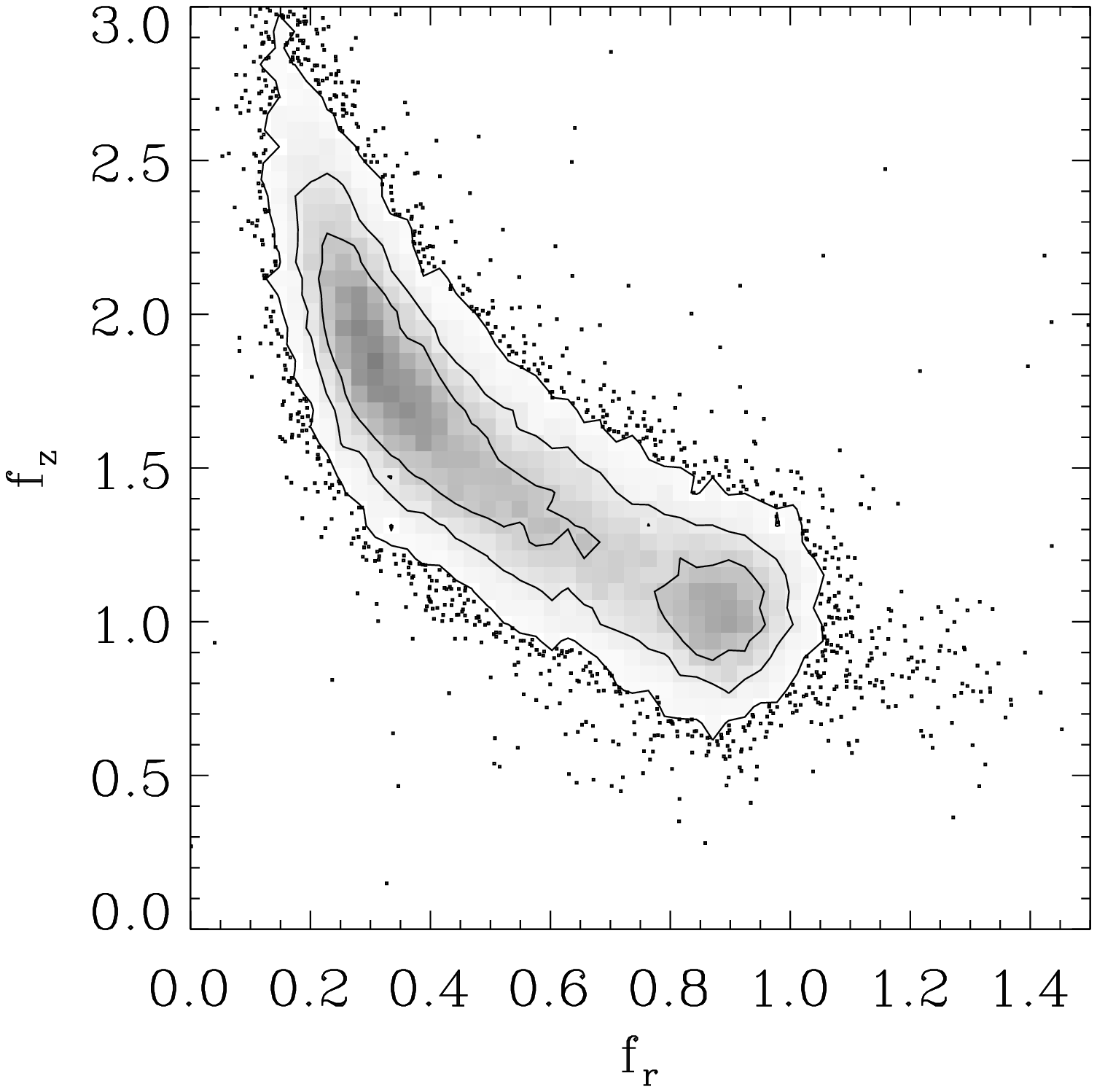}
\includegraphics[width=0.32\textwidth,clip=]{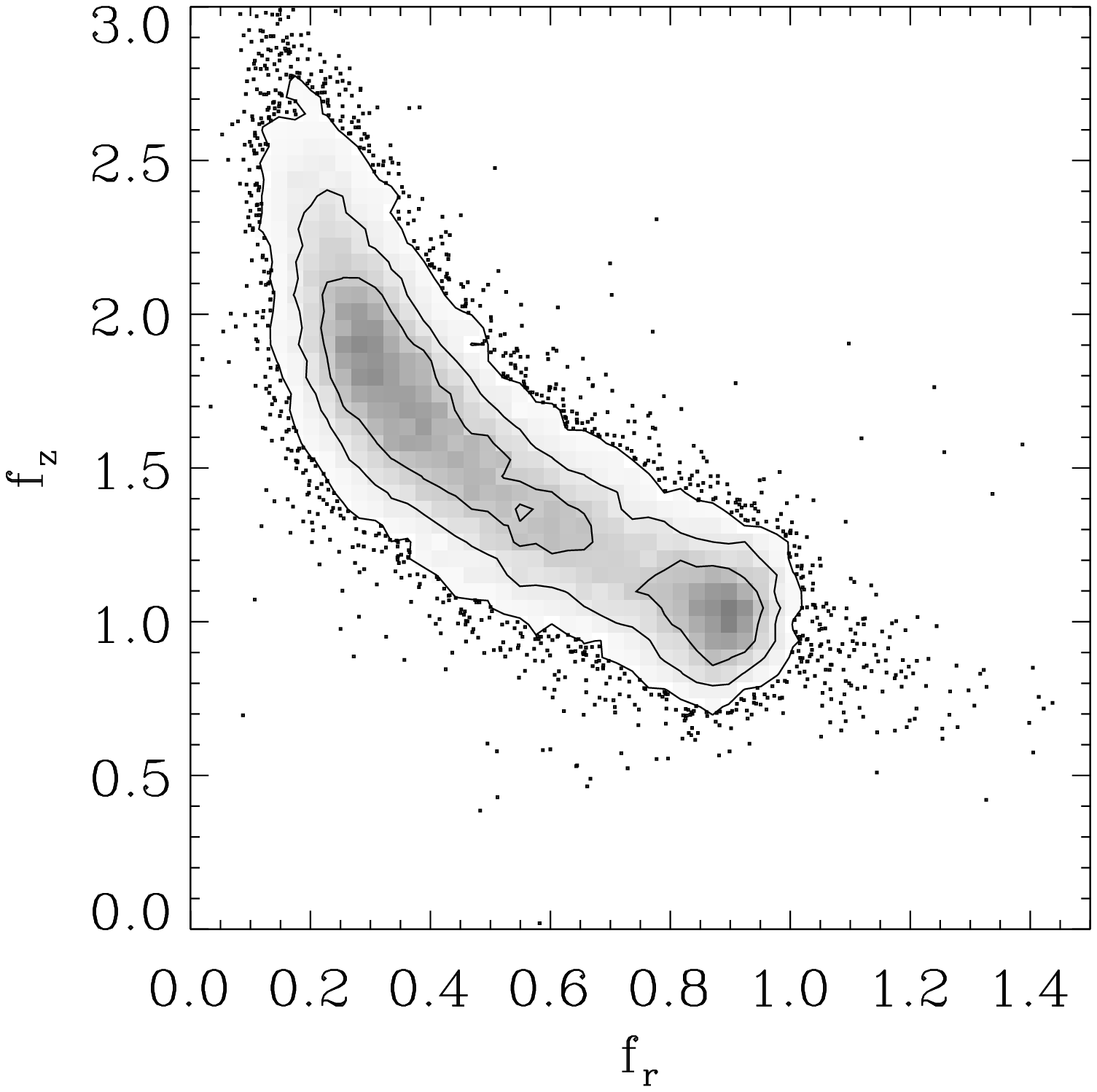}
\includegraphics[width=0.32\textwidth,clip=]{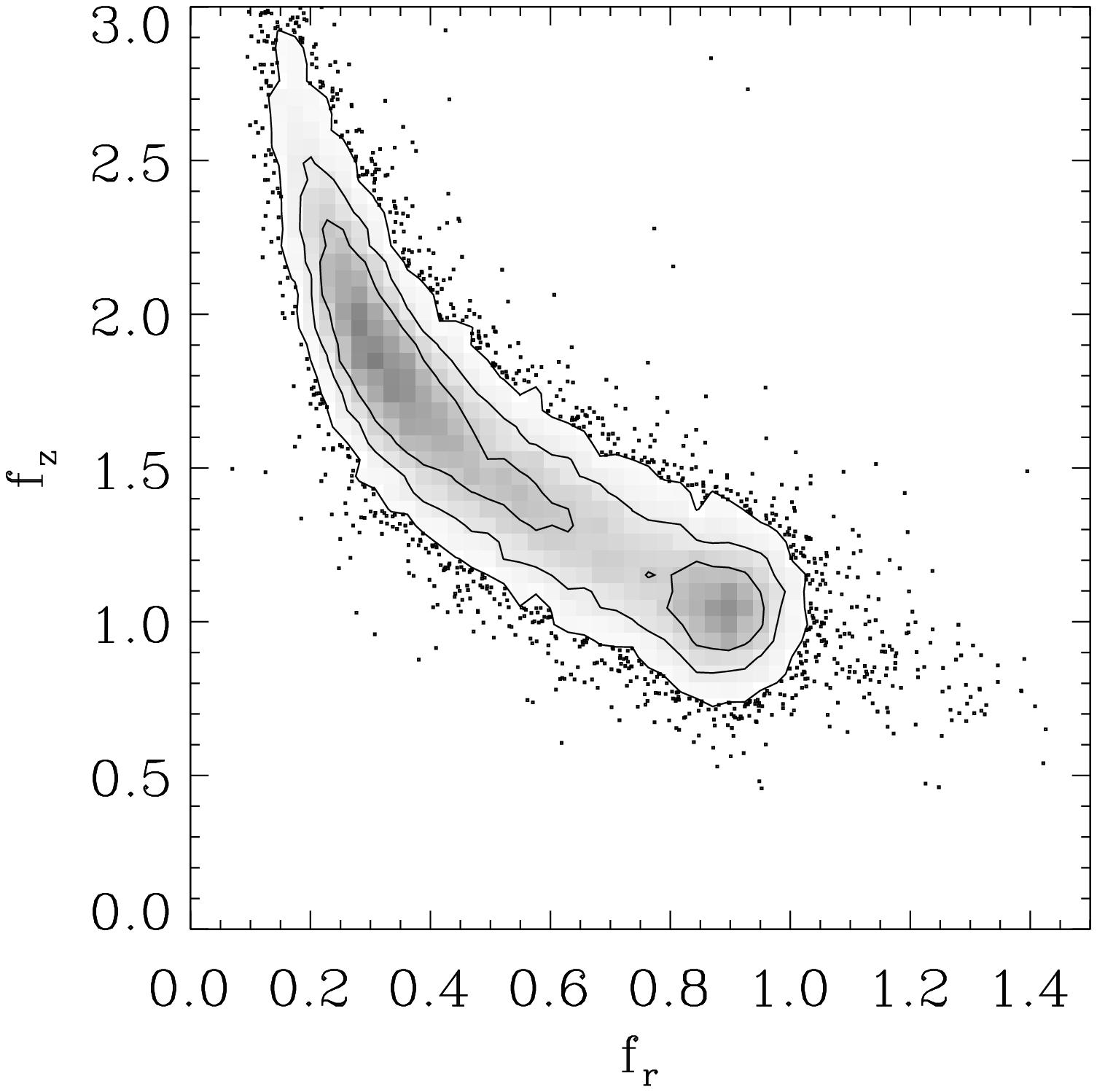}
\caption{Flux-flux diagrams for a bin in $i$-band magnitude from the
  single-epoch \starc\ catalog. The first column shows the
  single-epoch data, the second column shows a sampling from the
  extreme-deconvolution fit to the single-epoch relative fluxes with
  the errors from the co-added data added, and the third column shows
  the co-added data. The relevant goodness-of-fit comparison is
  between the second and third column. The grayscale is linear in the
  density and the contours contain 68, 95, and 99\,percent of the
  distribution. Samples falling outside the outermost contour are
  individually shown. This \figurename\ shows that the \exd\ technique
  recovers the underlying error-deconvolved distribution given noisy
  training data.}\label{fig:singlezexfit}
\end{figure}

\begin{figure}
\includegraphics[width=0.32\textwidth,clip=]{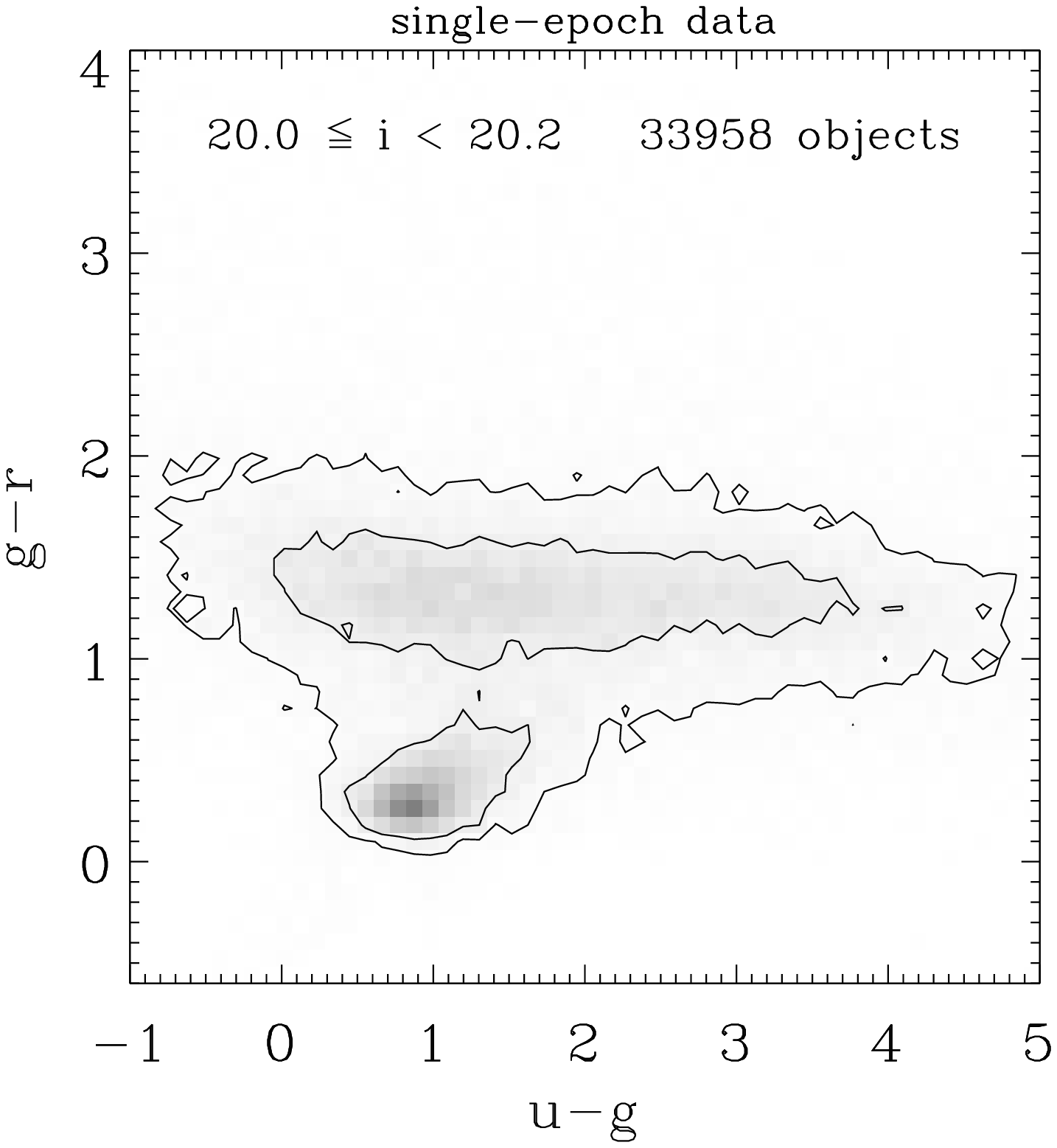}
\includegraphics[width=0.32\textwidth,clip=]{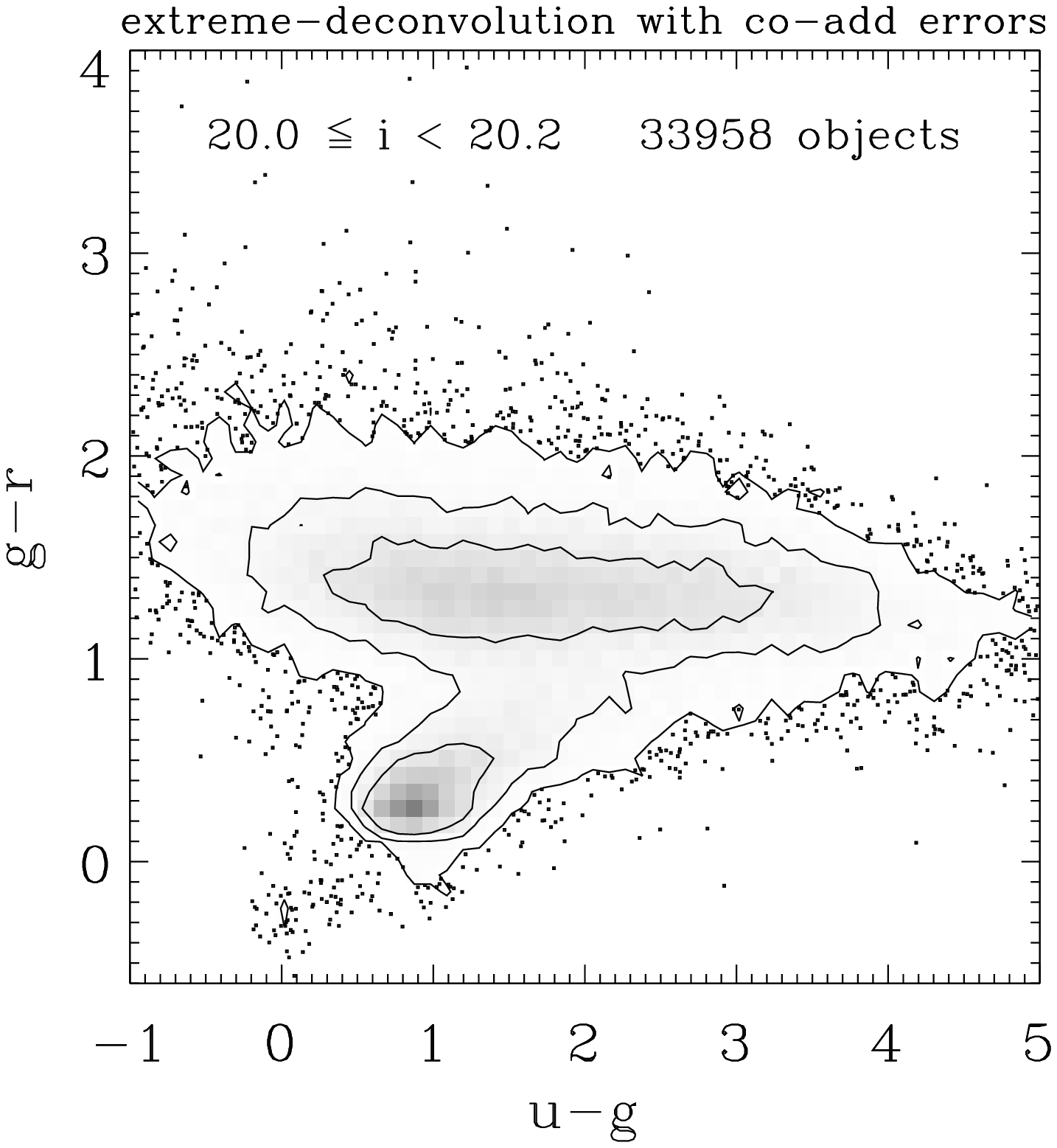}
\includegraphics[width=0.32\textwidth,clip=]{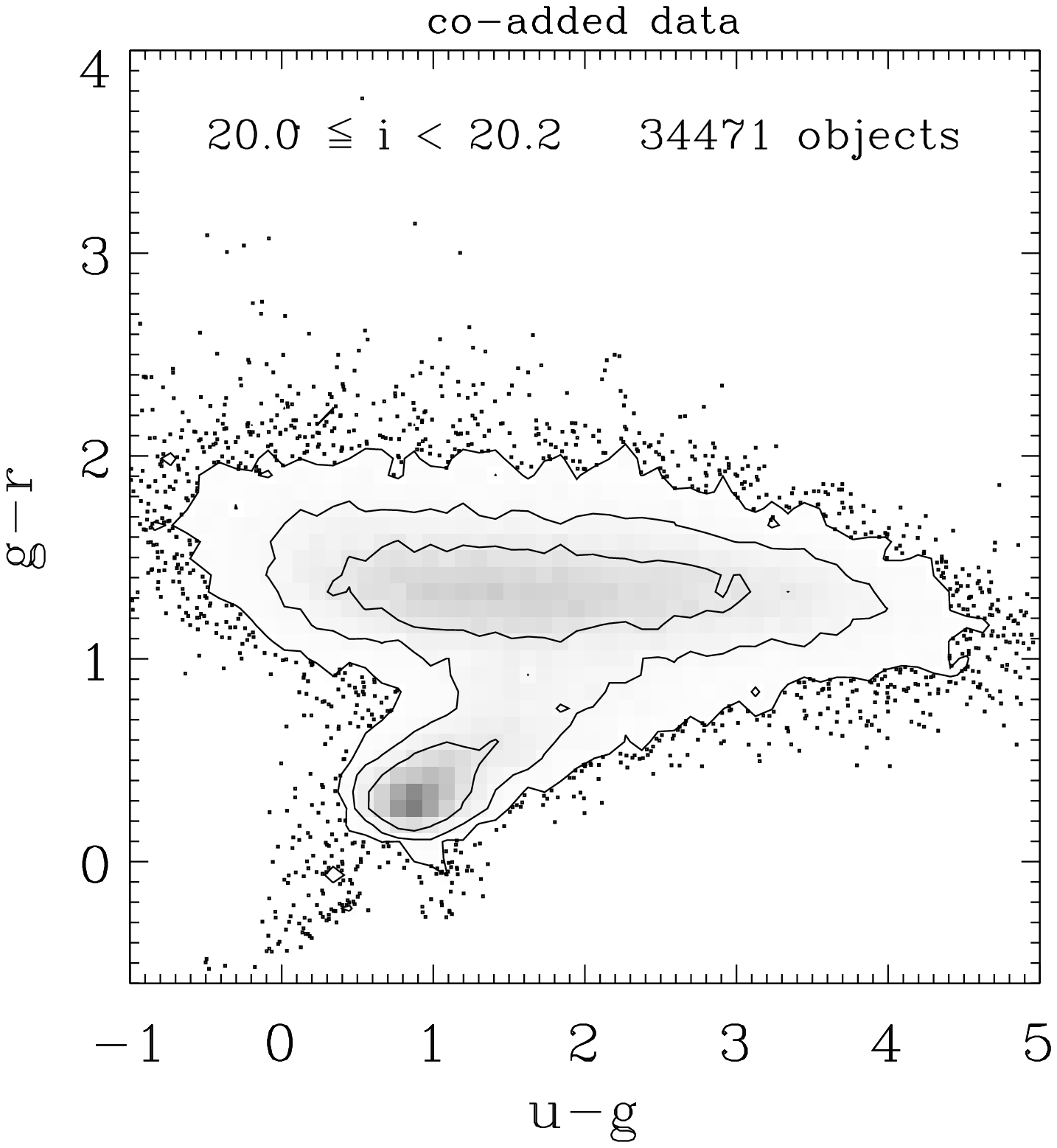}\\
\includegraphics[width=0.32\textwidth,clip=]{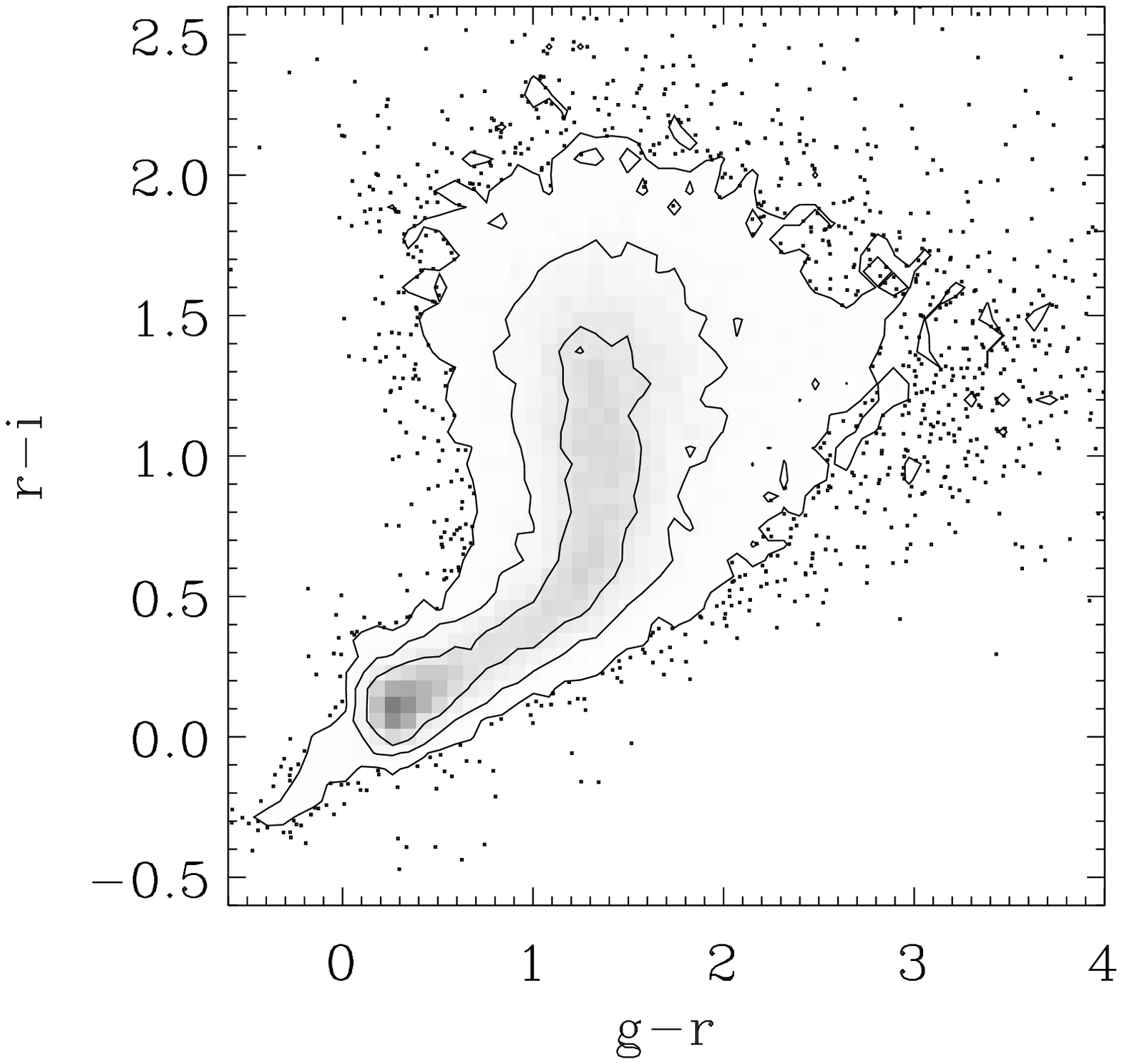}
\includegraphics[width=0.32\textwidth,clip=]{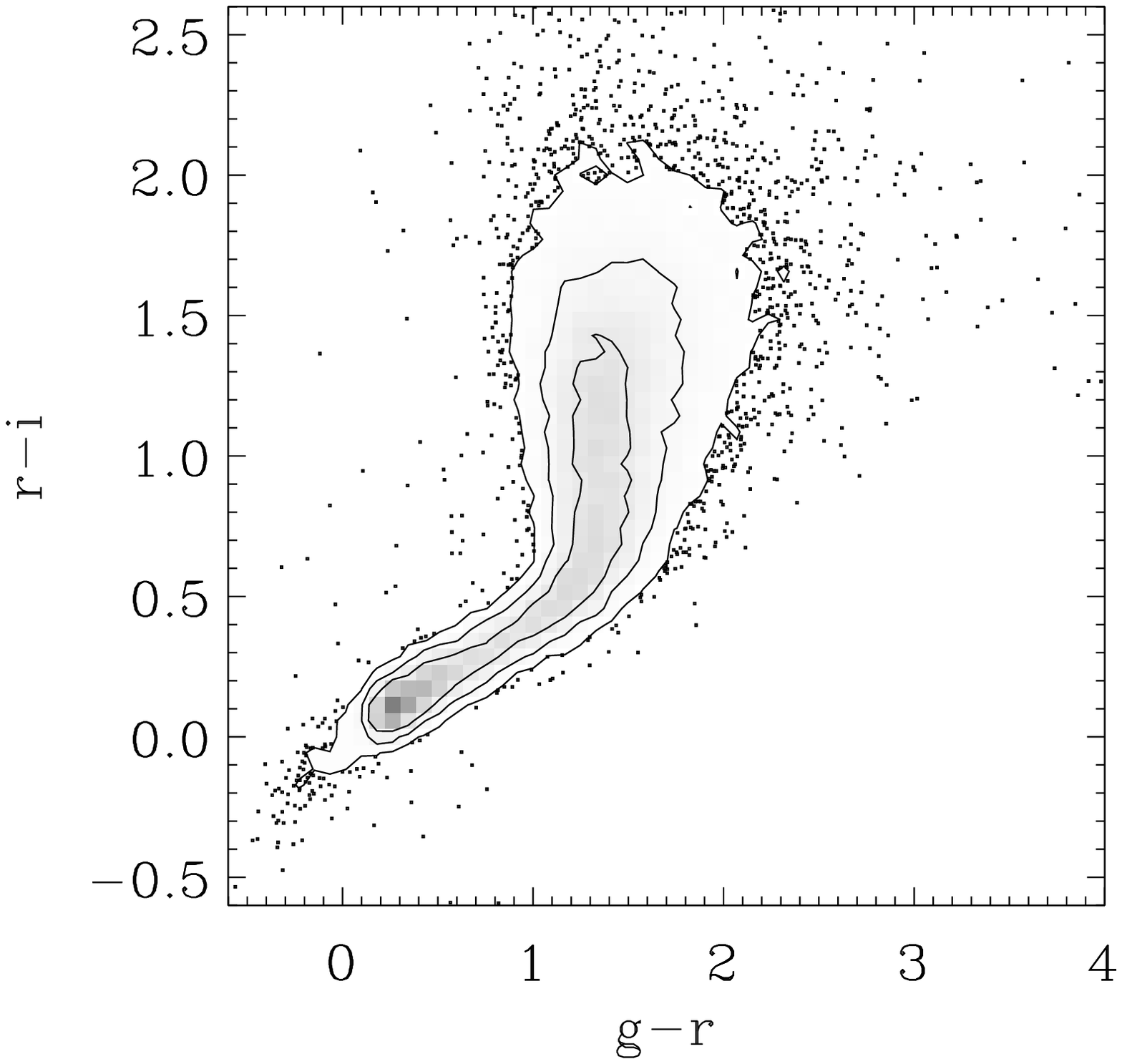}
\includegraphics[width=0.32\textwidth,clip=]{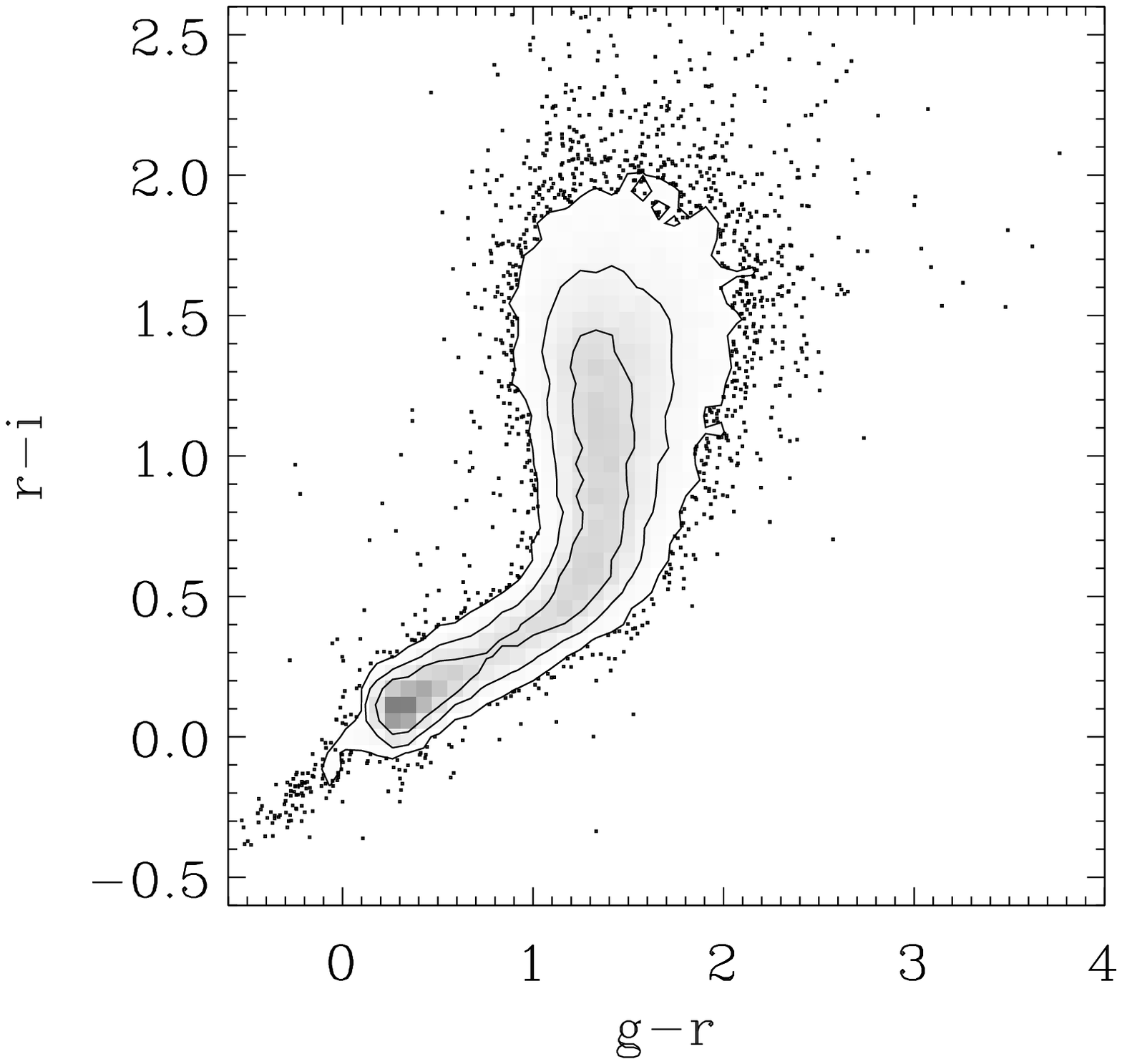}\\
\includegraphics[width=0.32\textwidth,clip=]{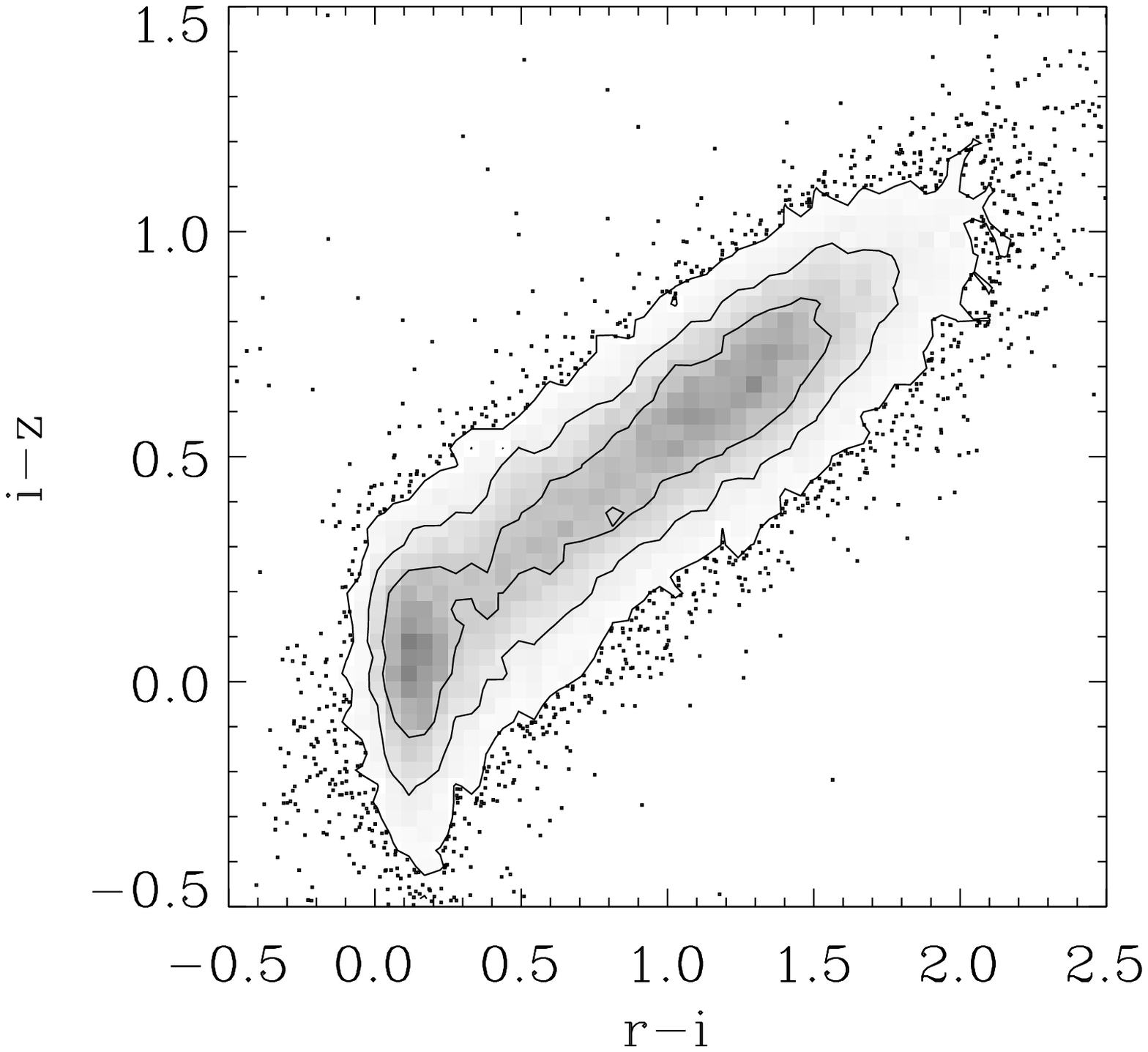}
\includegraphics[width=0.32\textwidth,clip=]{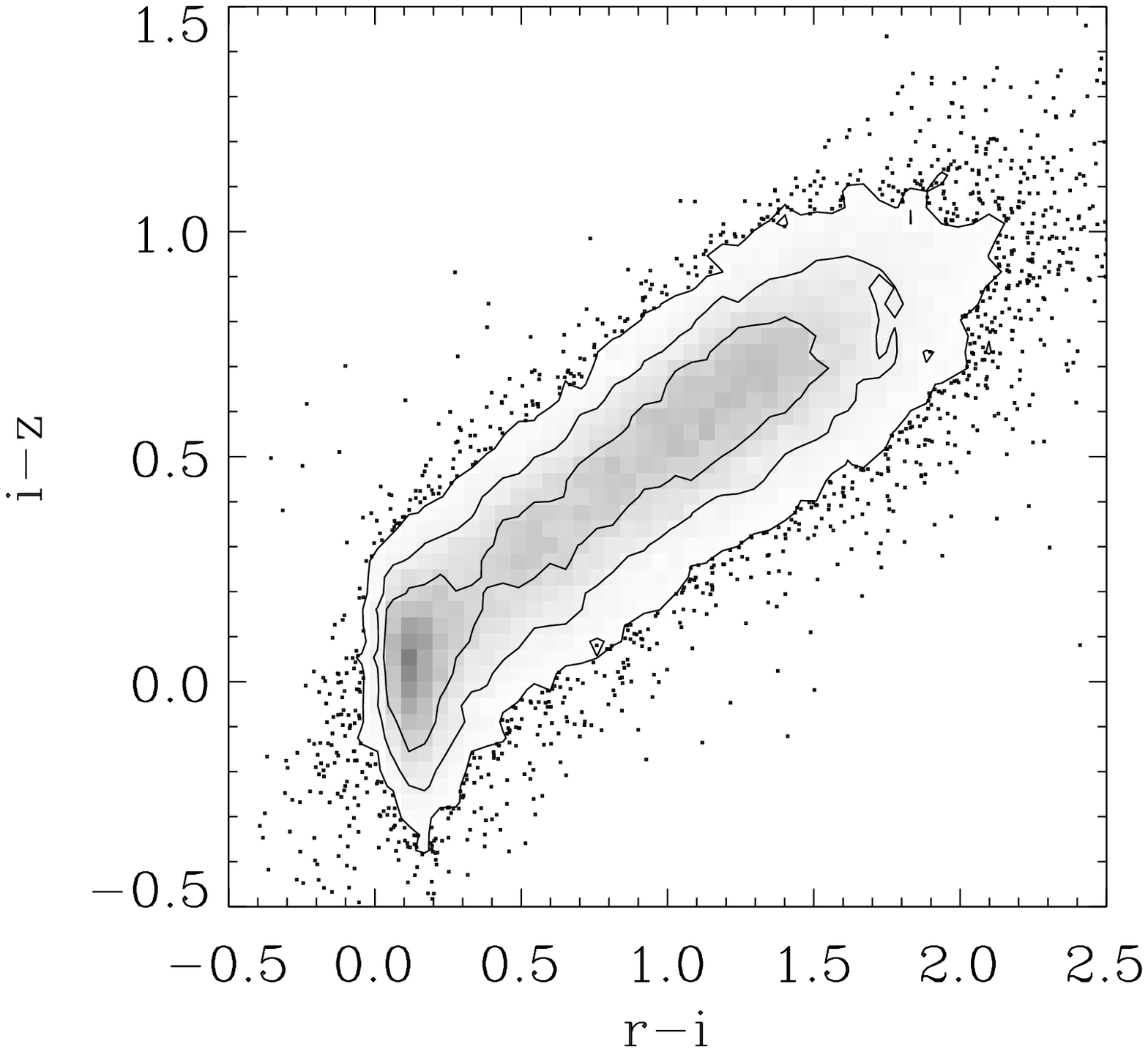}
\includegraphics[width=0.32\textwidth,clip=]{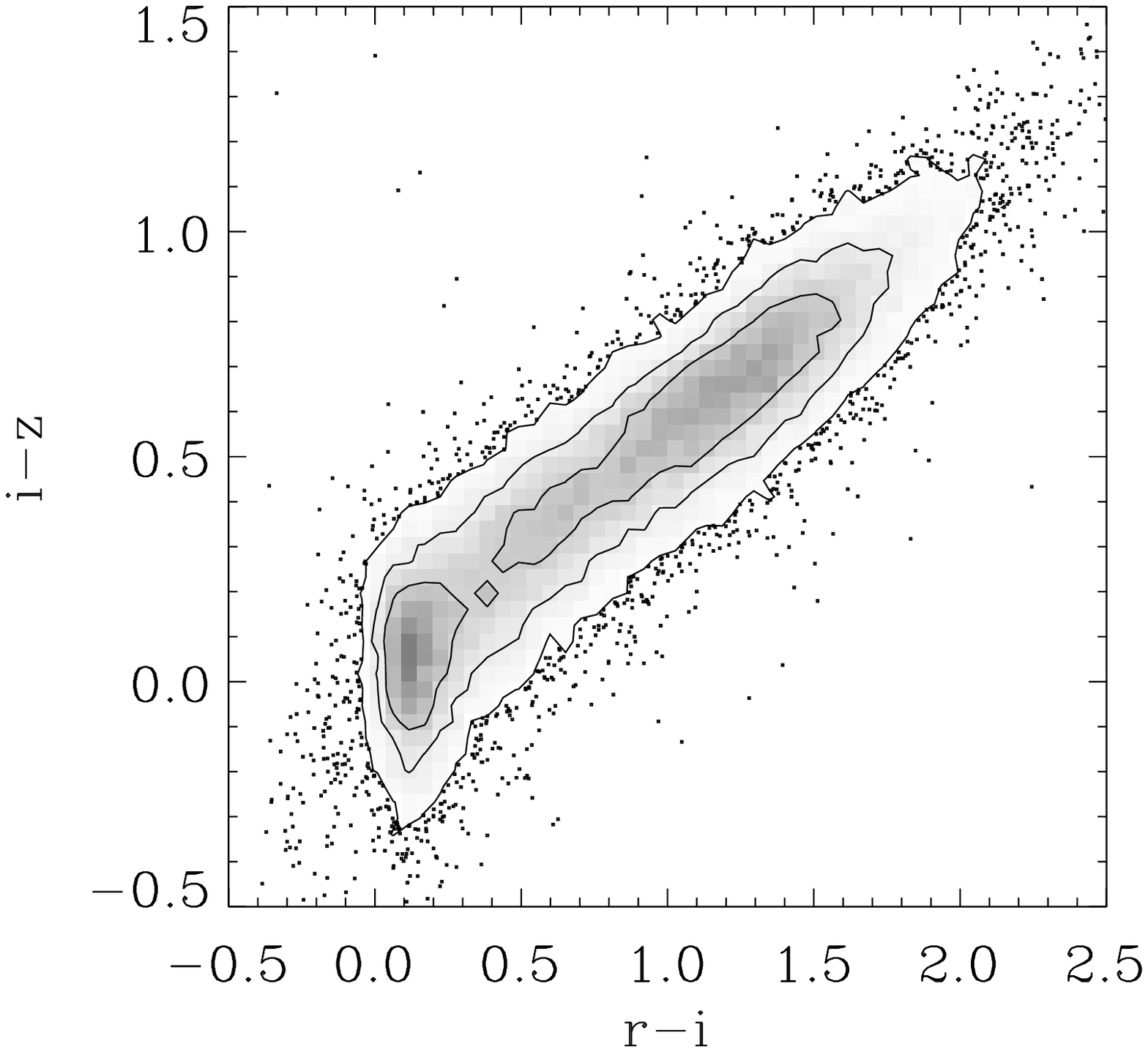}
\caption{Same as \figurename~\ref{fig:singlezexfit} but the color-color
  diagrams.}\label{fig:singlezexfitcolor}
\end{figure}

The results for this exercise for a single bin are shown in
\figurename s~\ref{fig:singlezexfit} and
\ref{fig:singlezexfitcolor}. We see that the agreement between the
model and the co-add truth is excellent.

\begin{figure}
\includegraphics[width=0.24\textwidth,clip=]{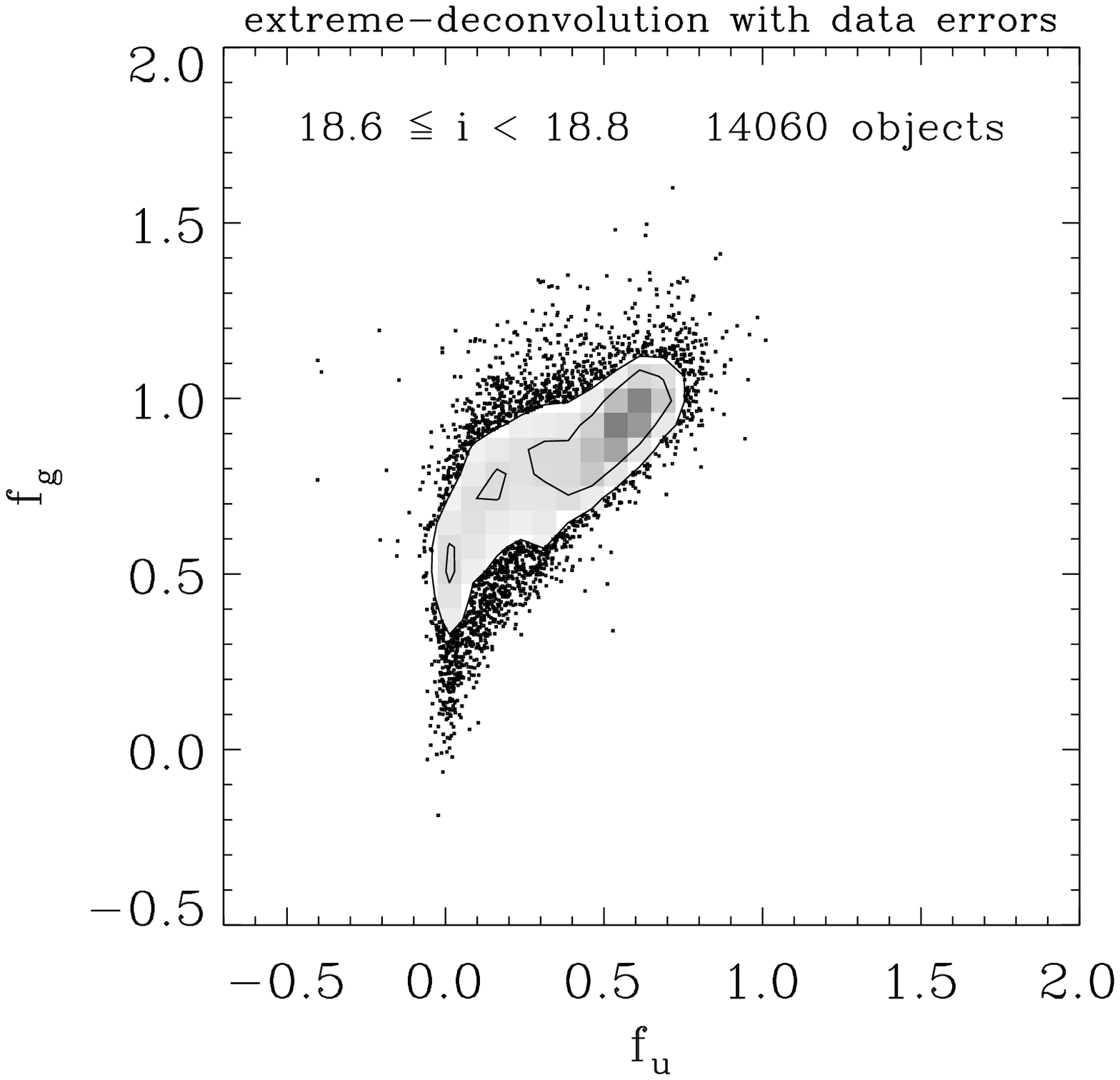}
\includegraphics[width=0.24\textwidth,clip=]{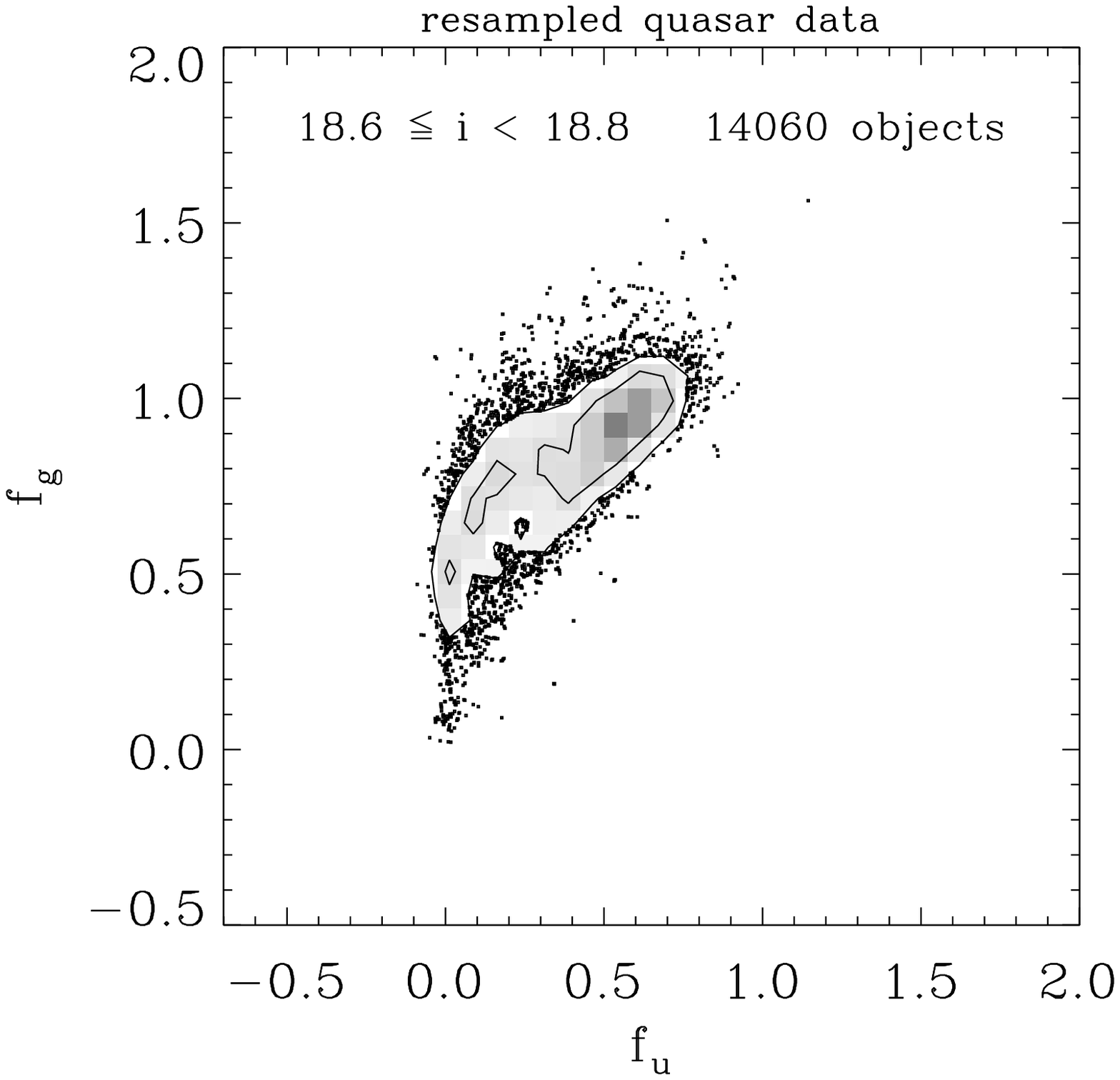}
\includegraphics[width=0.24\textwidth,clip=]{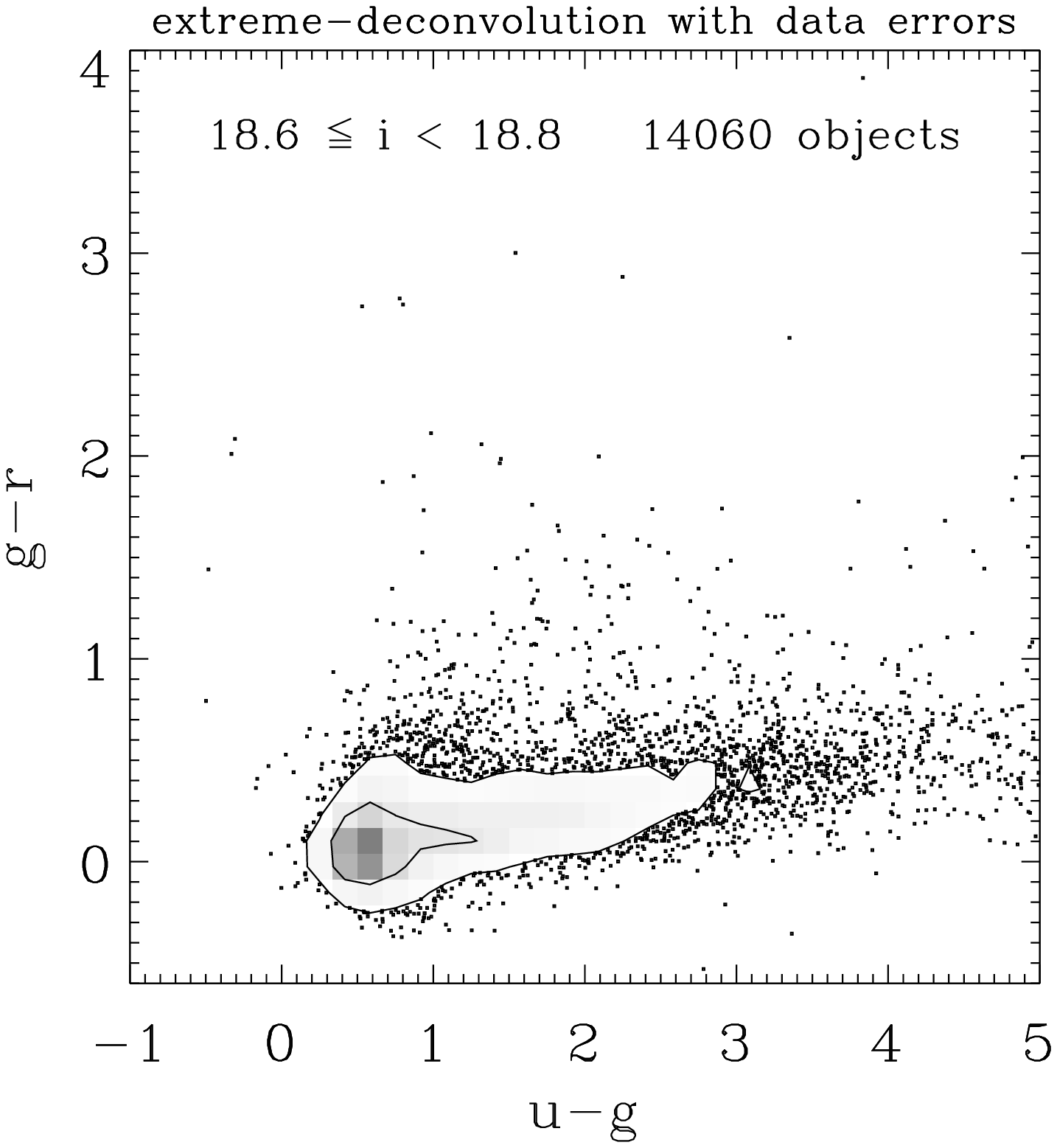}
\includegraphics[width=0.24\textwidth,clip=]{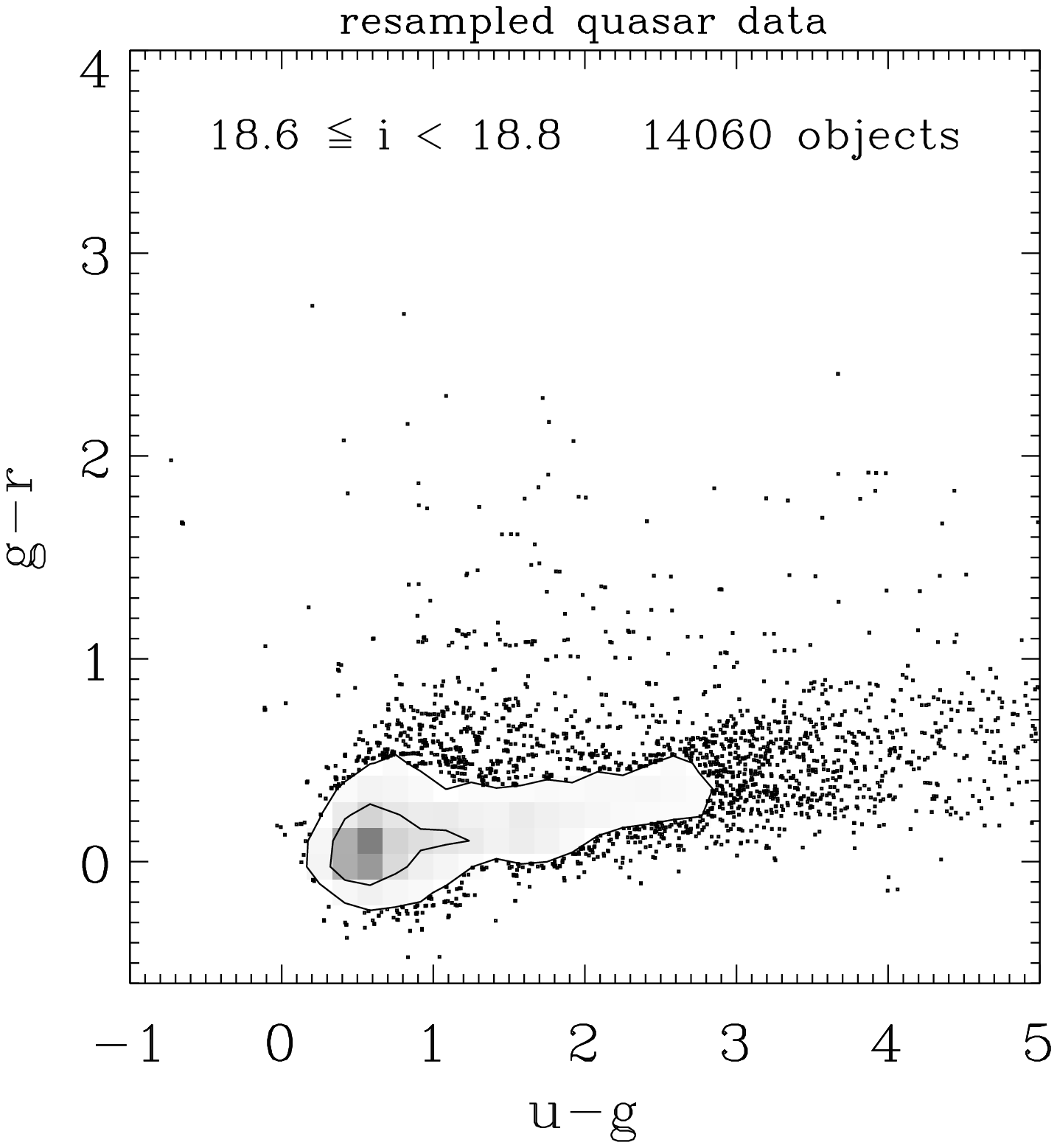}\\
\includegraphics[width=0.24\textwidth,clip=]{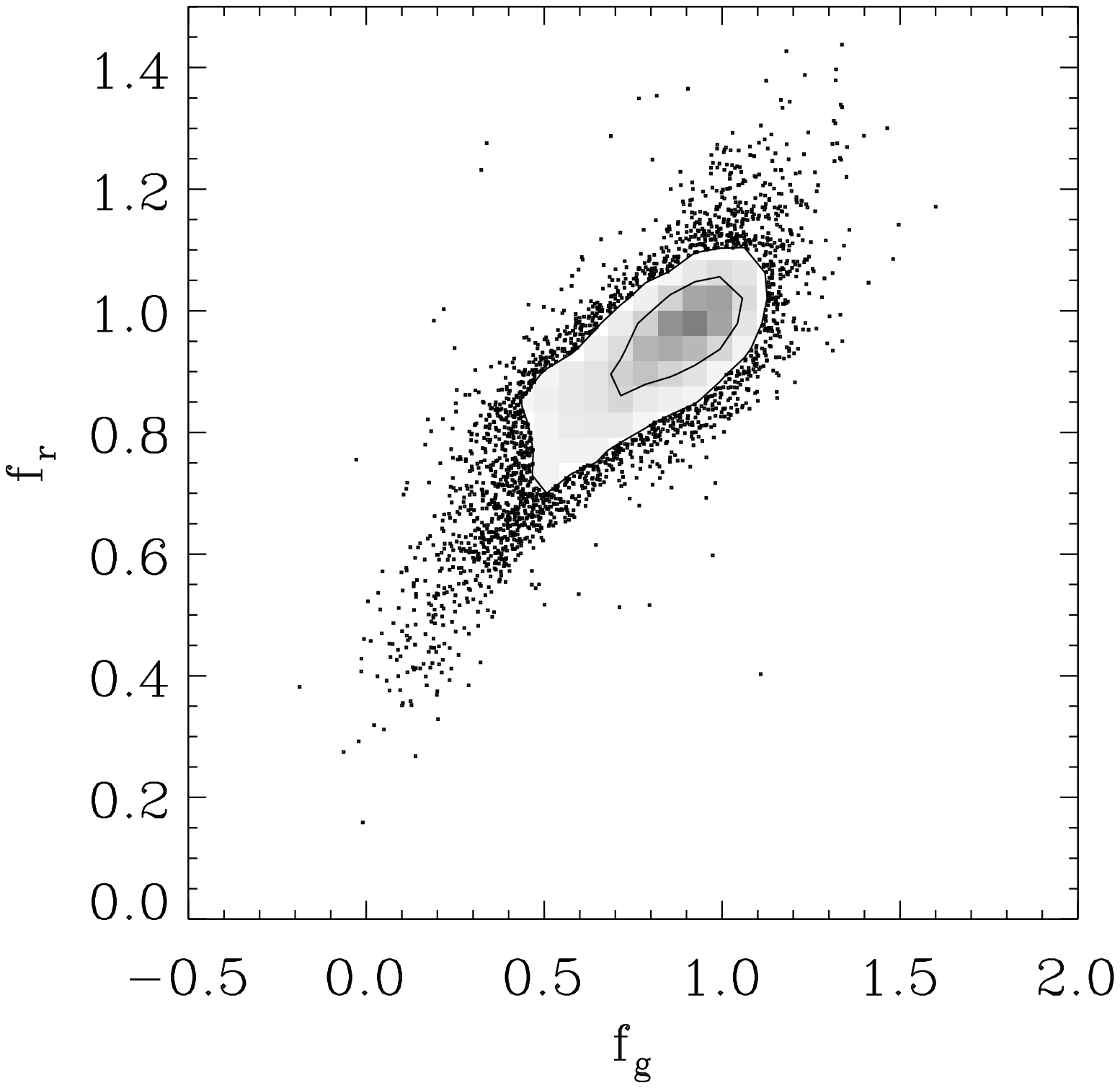}
\includegraphics[width=0.24\textwidth,clip=]{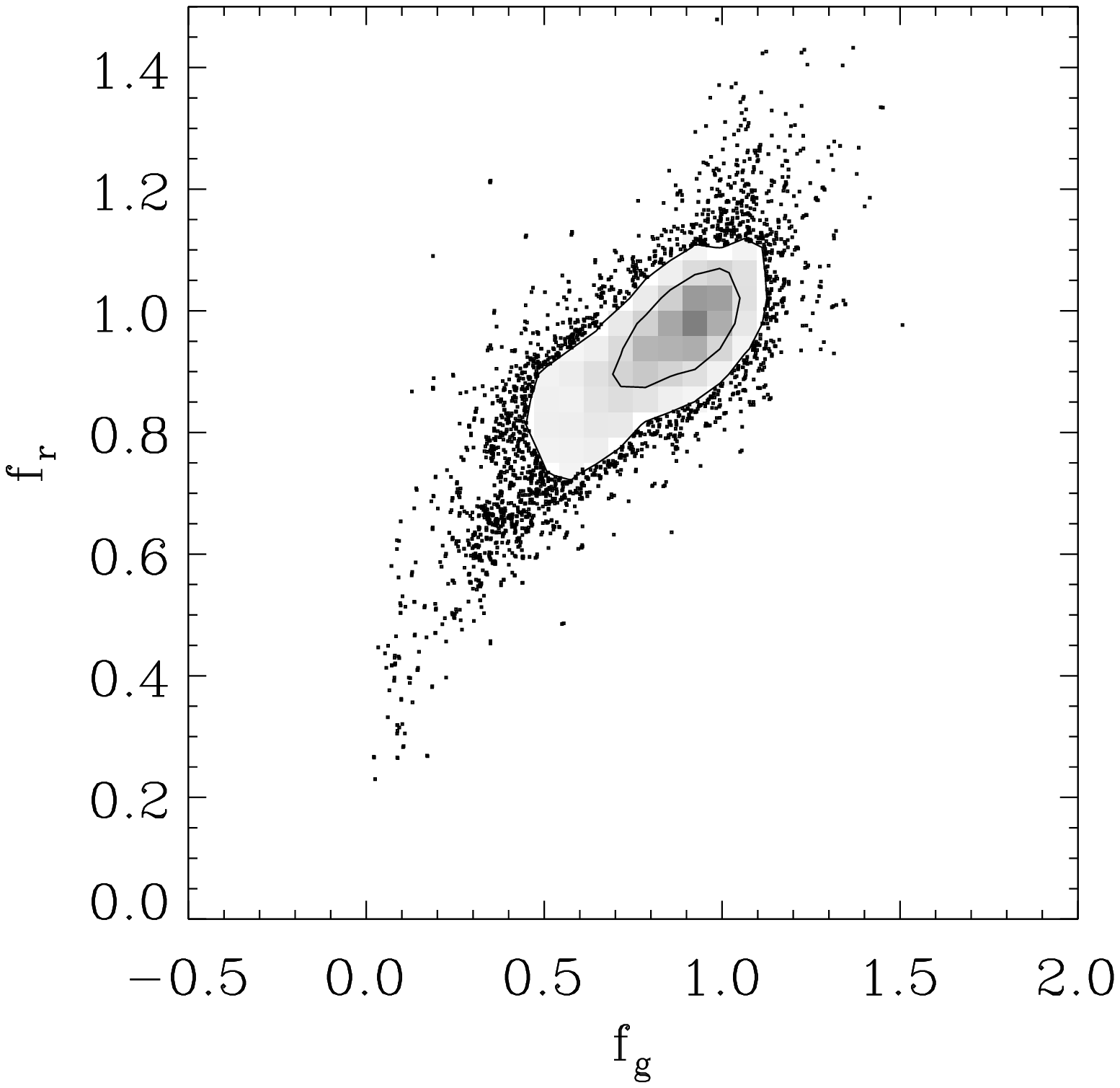}
\includegraphics[width=0.24\textwidth,clip=]{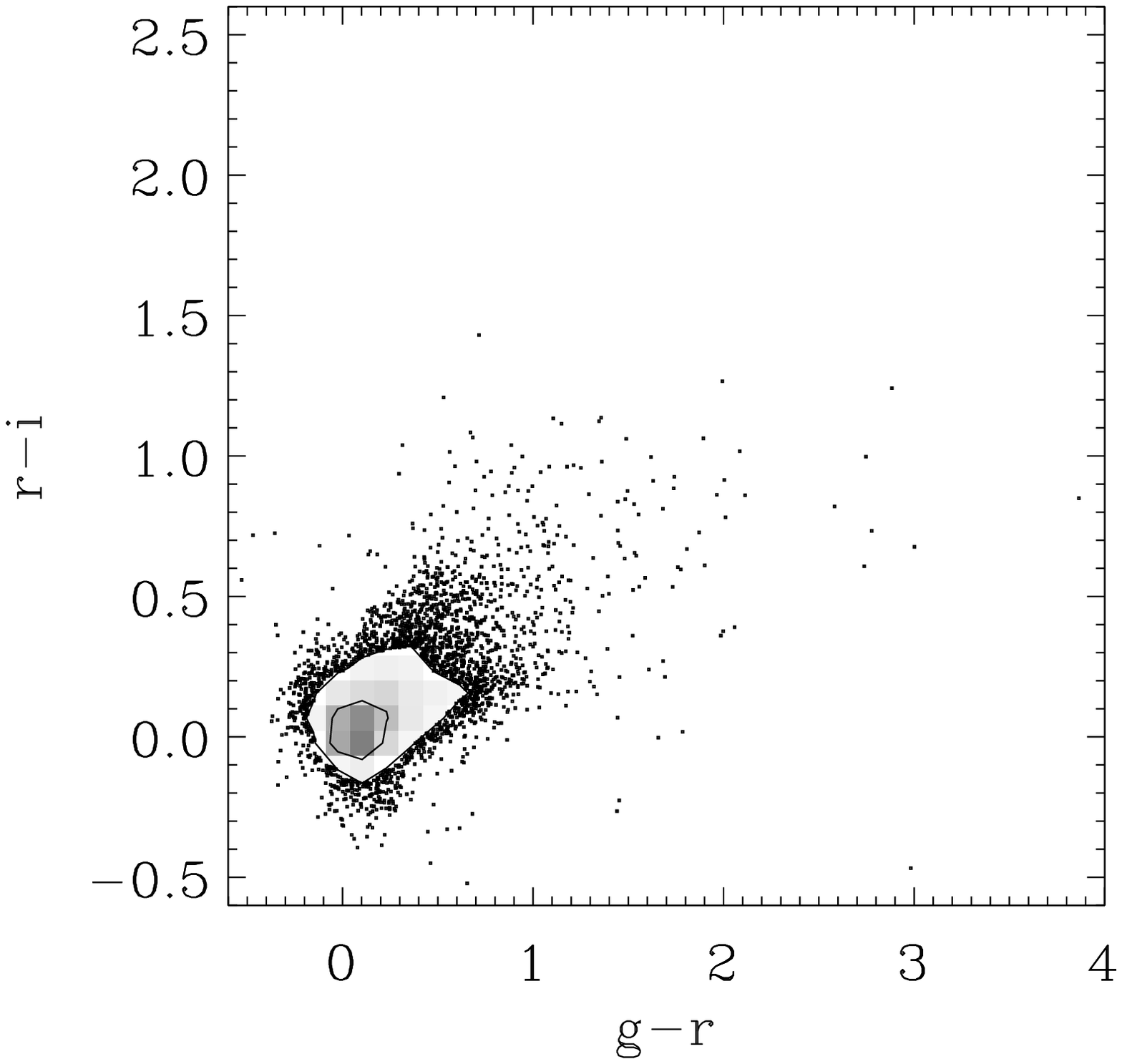}
\includegraphics[width=0.24\textwidth,clip=]{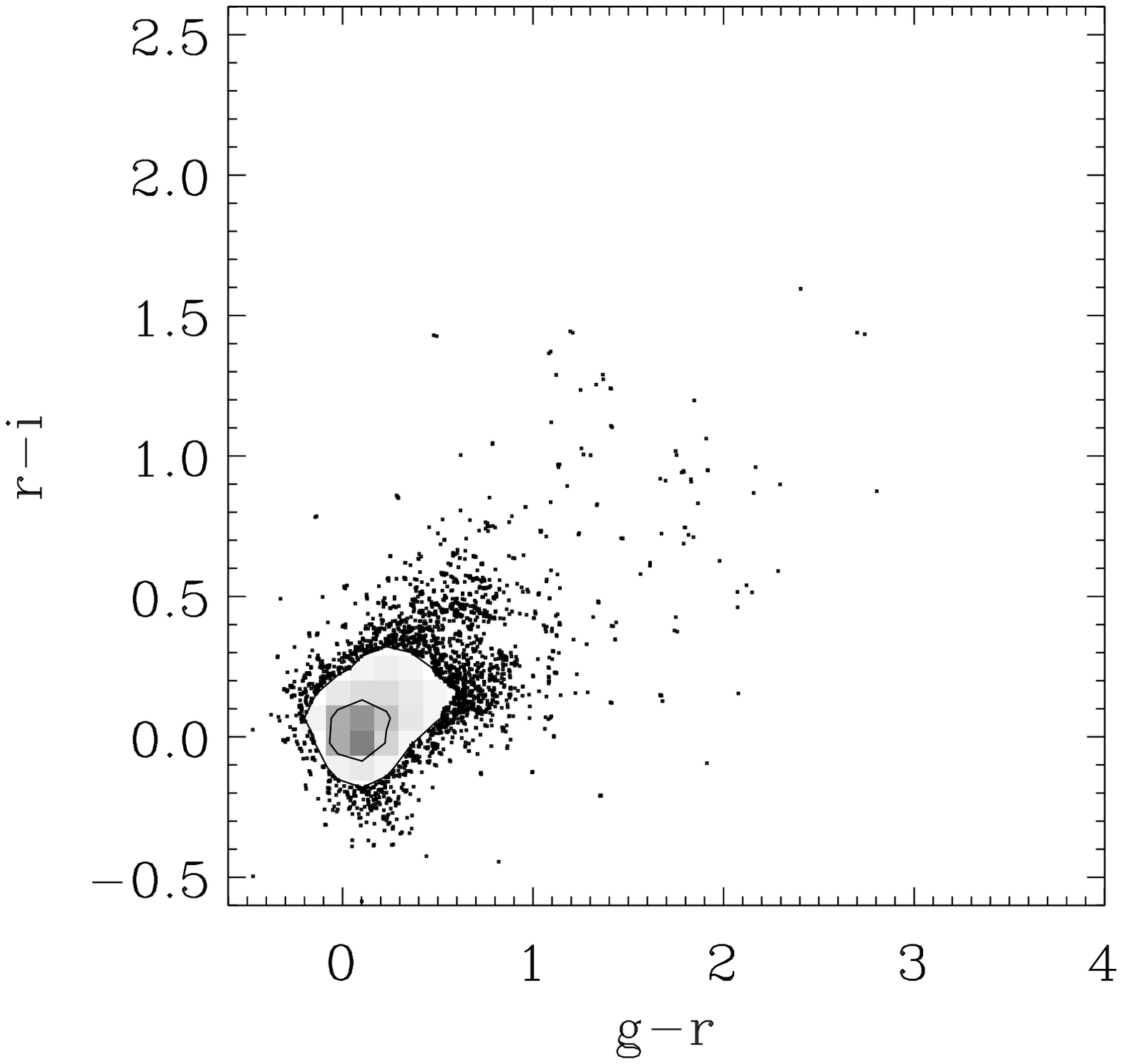}\\
\includegraphics[width=0.24\textwidth,clip=]{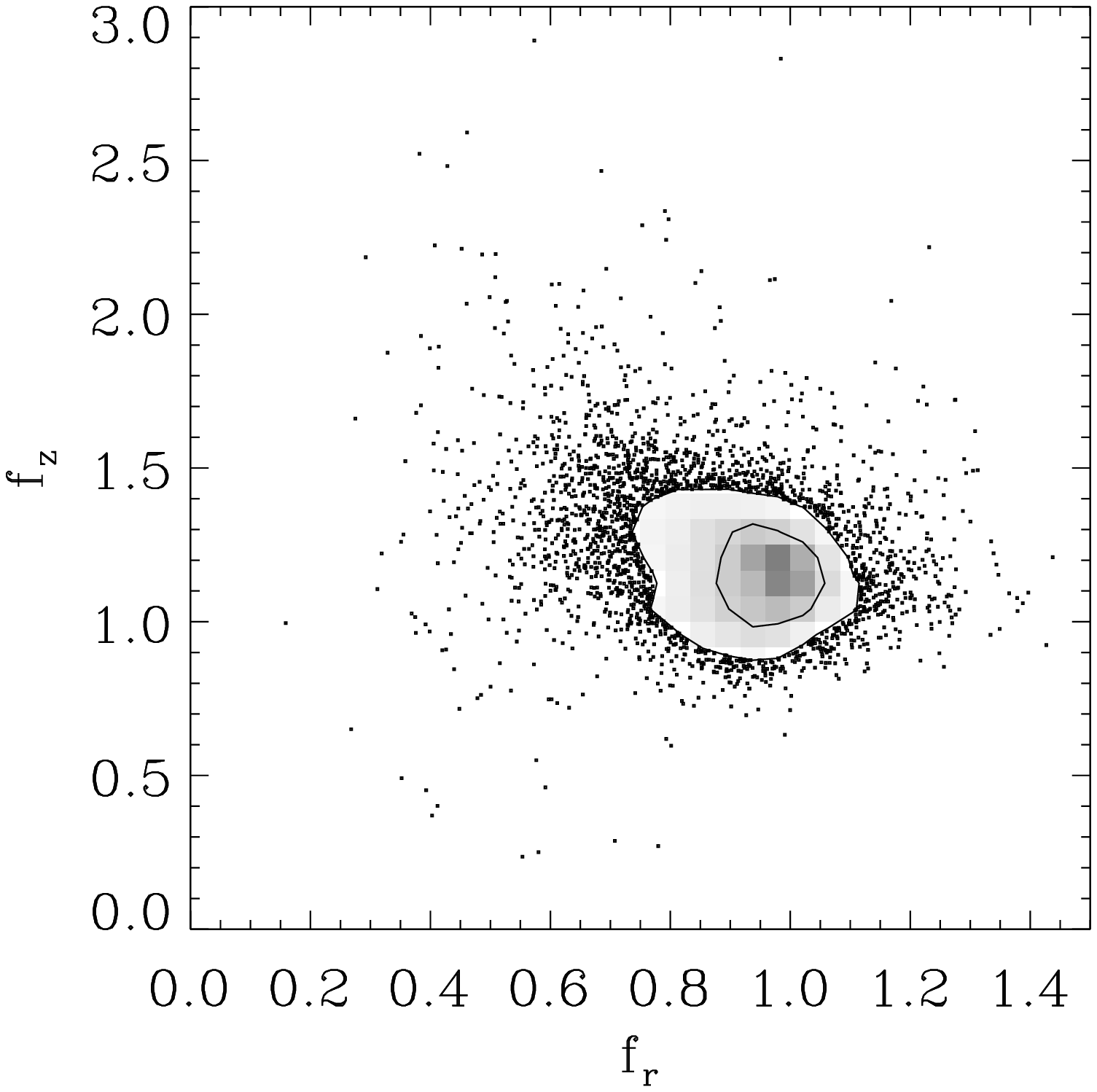}
\includegraphics[width=0.24\textwidth,clip=]{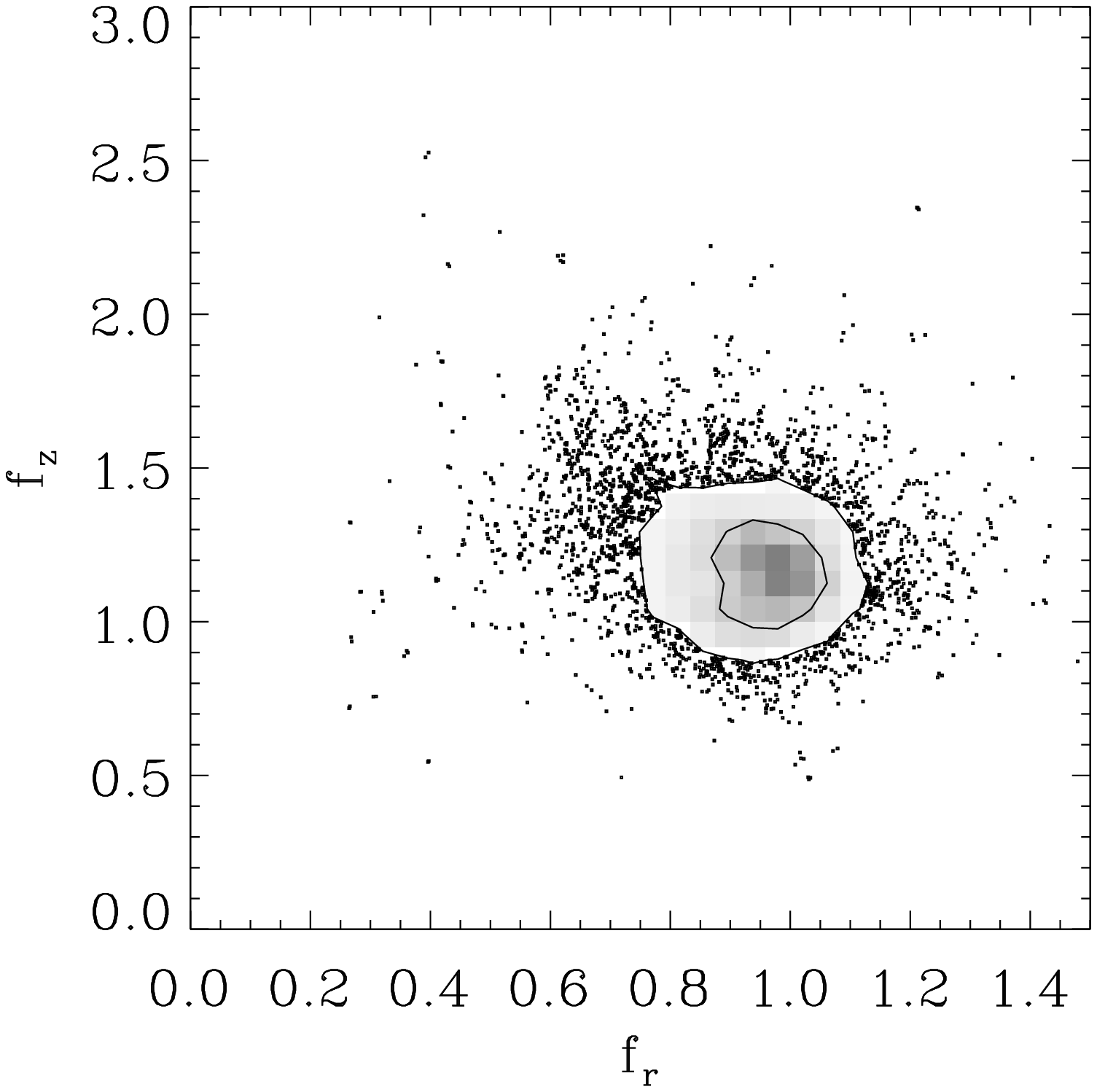}
\includegraphics[width=0.24\textwidth,clip=]{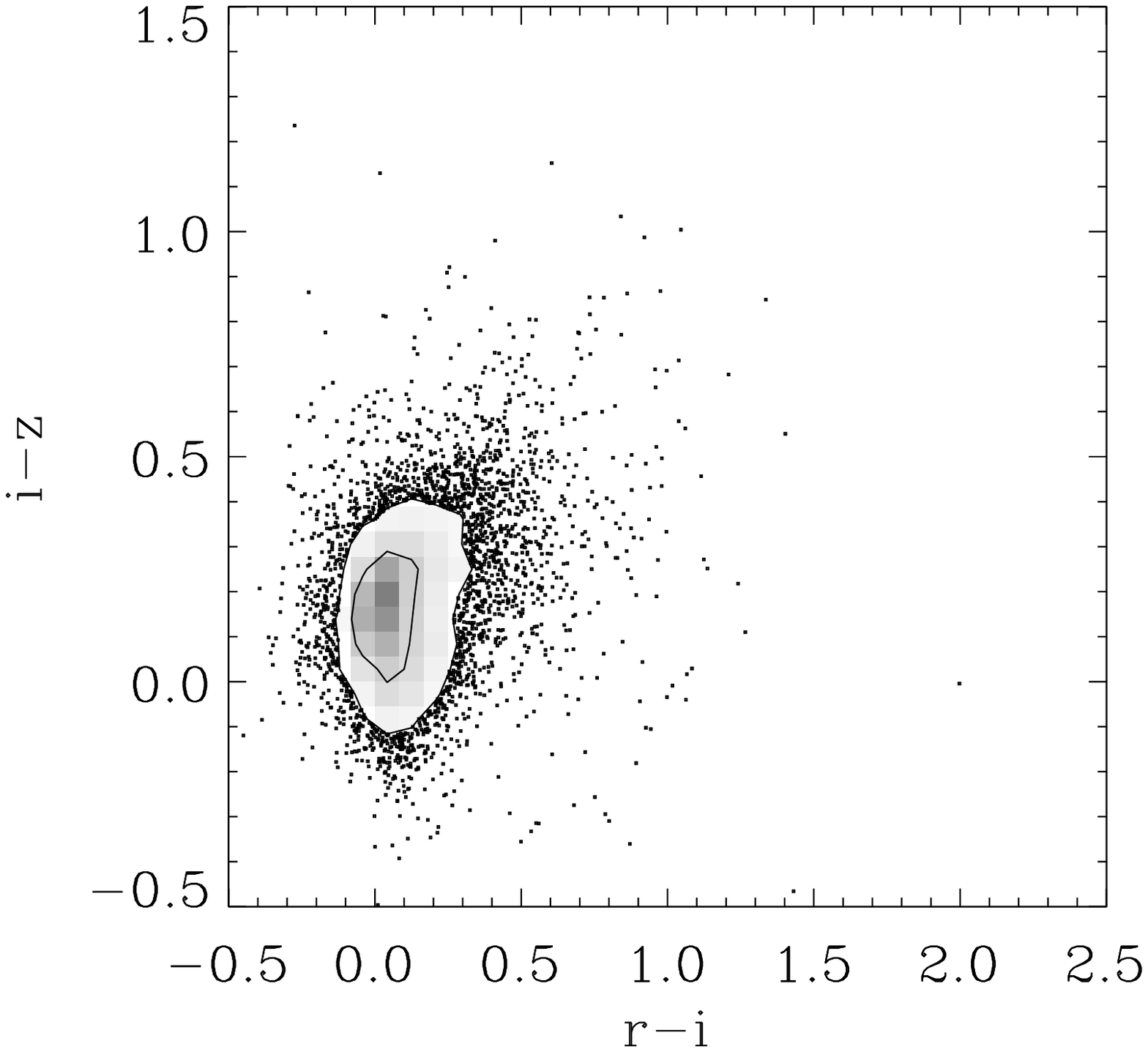}
\includegraphics[width=0.24\textwidth,clip=]{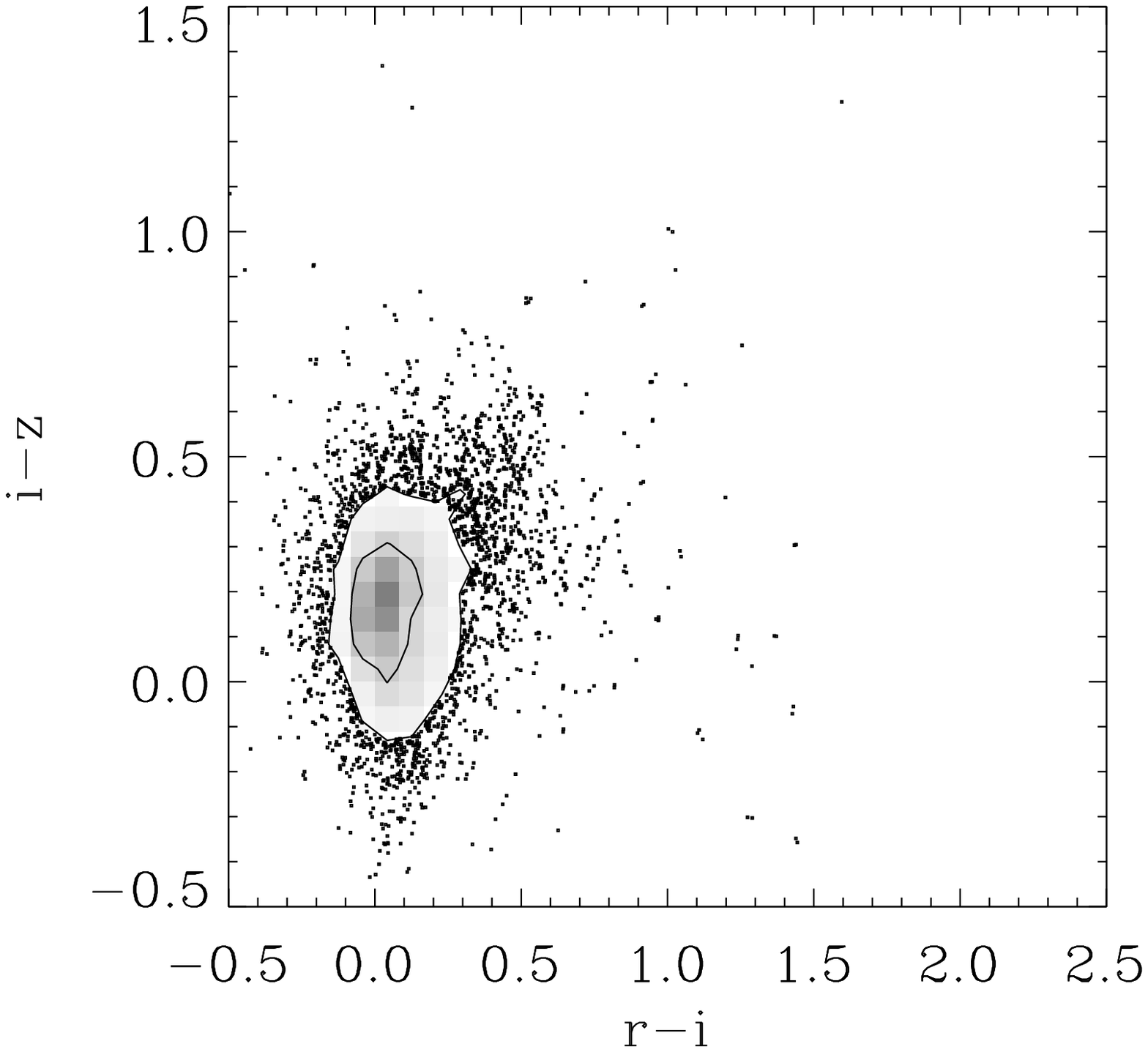}
\caption{Flux-flux and color-color diagrams for a bin in $i$-band
  magnitude from the quasar ($2.2 \leq z \leq 3.5$) catalog. The first
  column shows a sampling from the extreme-deconvolution fit with the
  errors from the quasar data added and the second column shows the
  quasar data resampled according to the quasar luminosity function as
  described in \sectionname~\ref{sec:qsodata}. The third and fourth
  columns show the same as the first and second columns, but for
  colors.}\label{fig:bosszexfit}
\end{figure}

\begin{figure}
\includegraphics[width=0.24\textwidth,clip=]{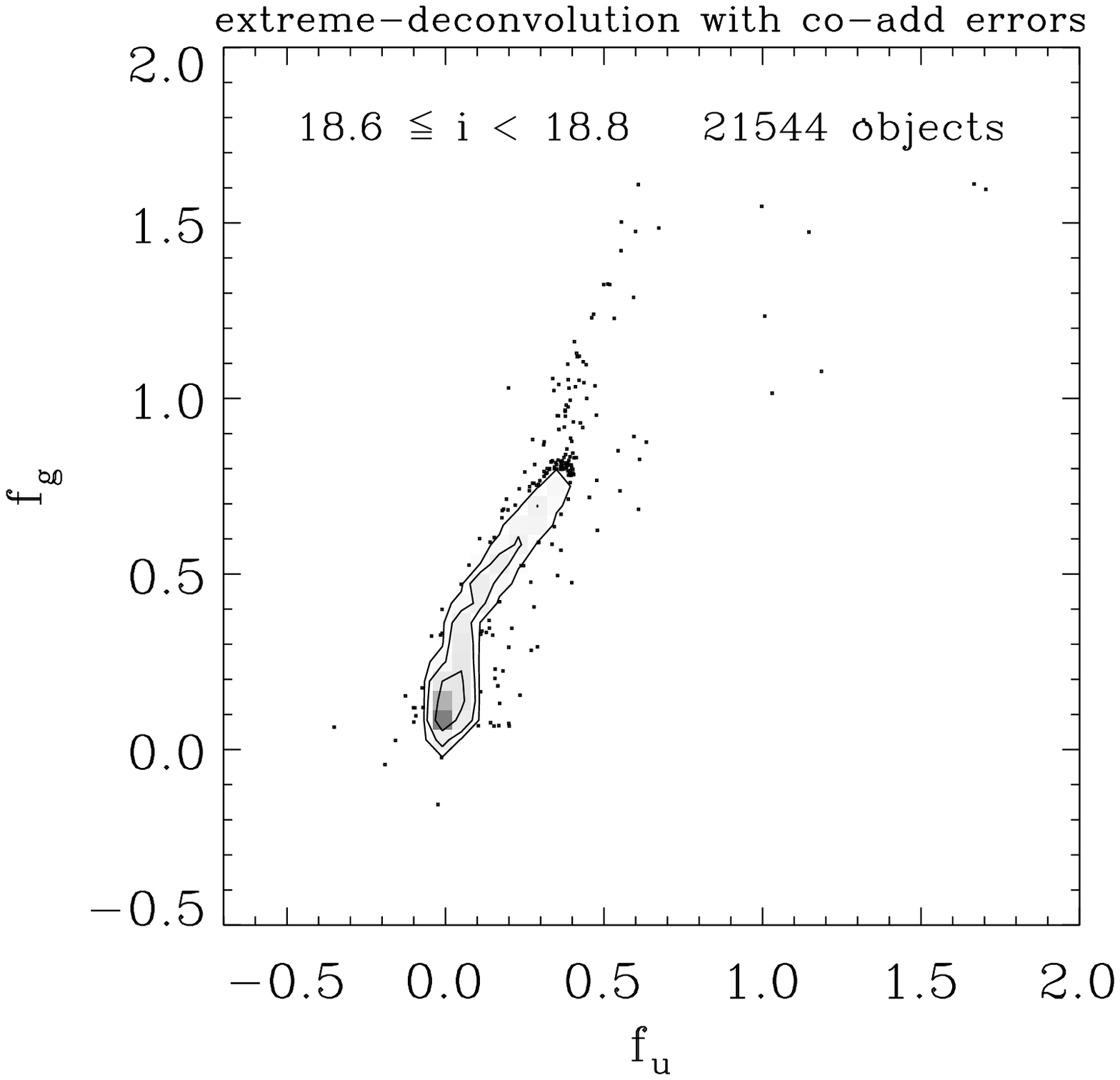}
\includegraphics[width=0.24\textwidth,clip=]{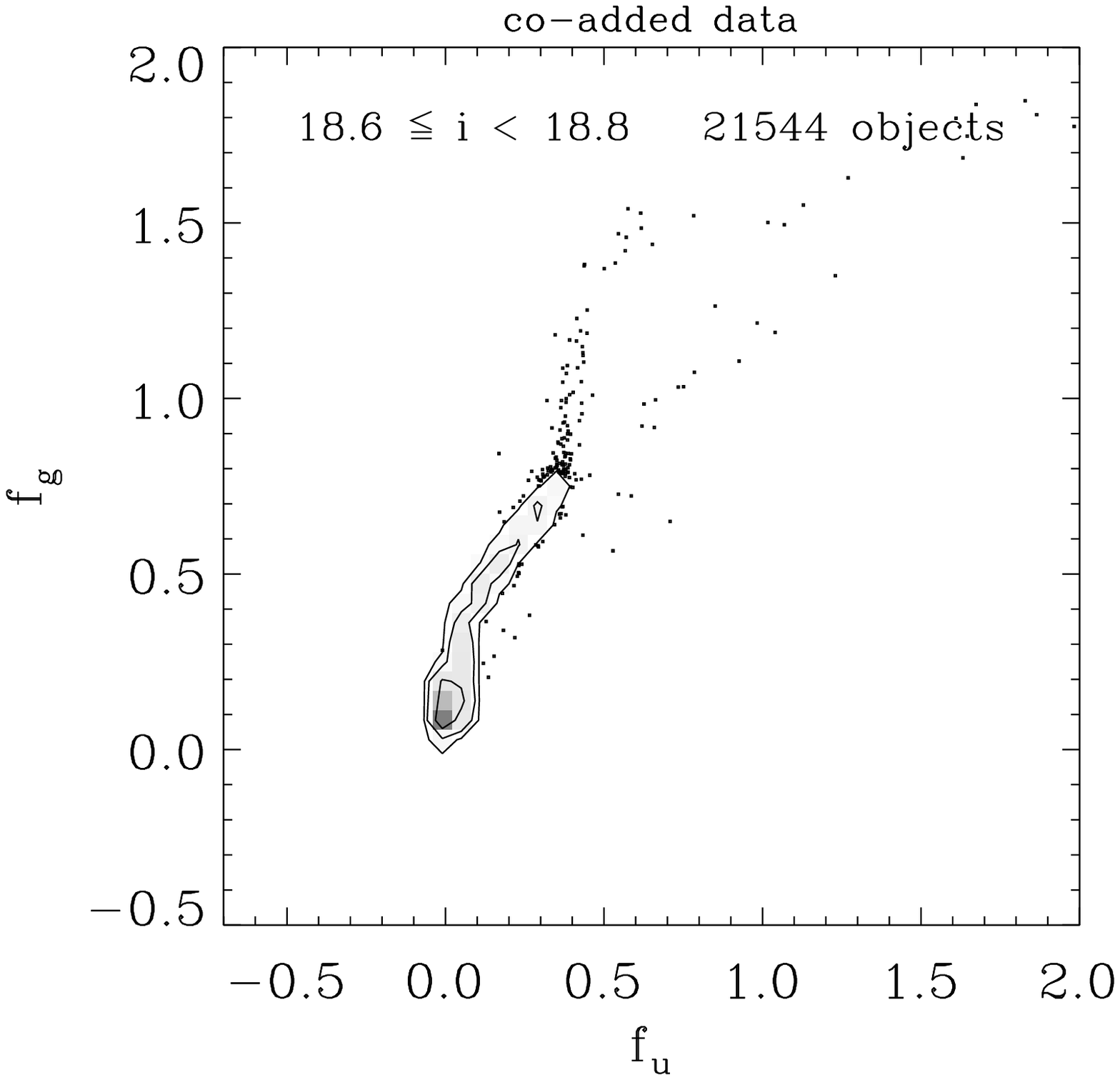}
\includegraphics[width=0.24\textwidth,clip=]{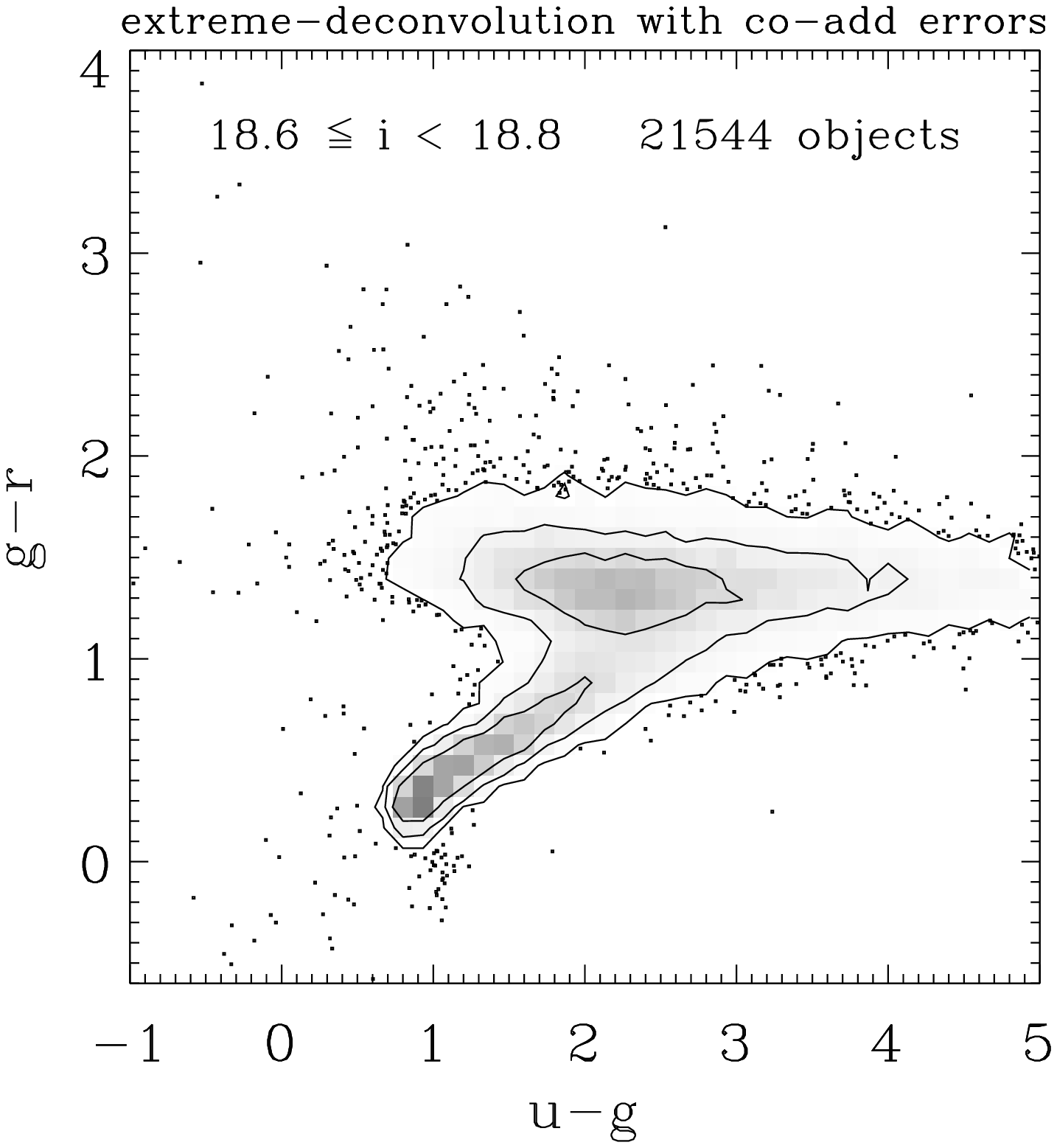}
\includegraphics[width=0.24\textwidth,clip=]{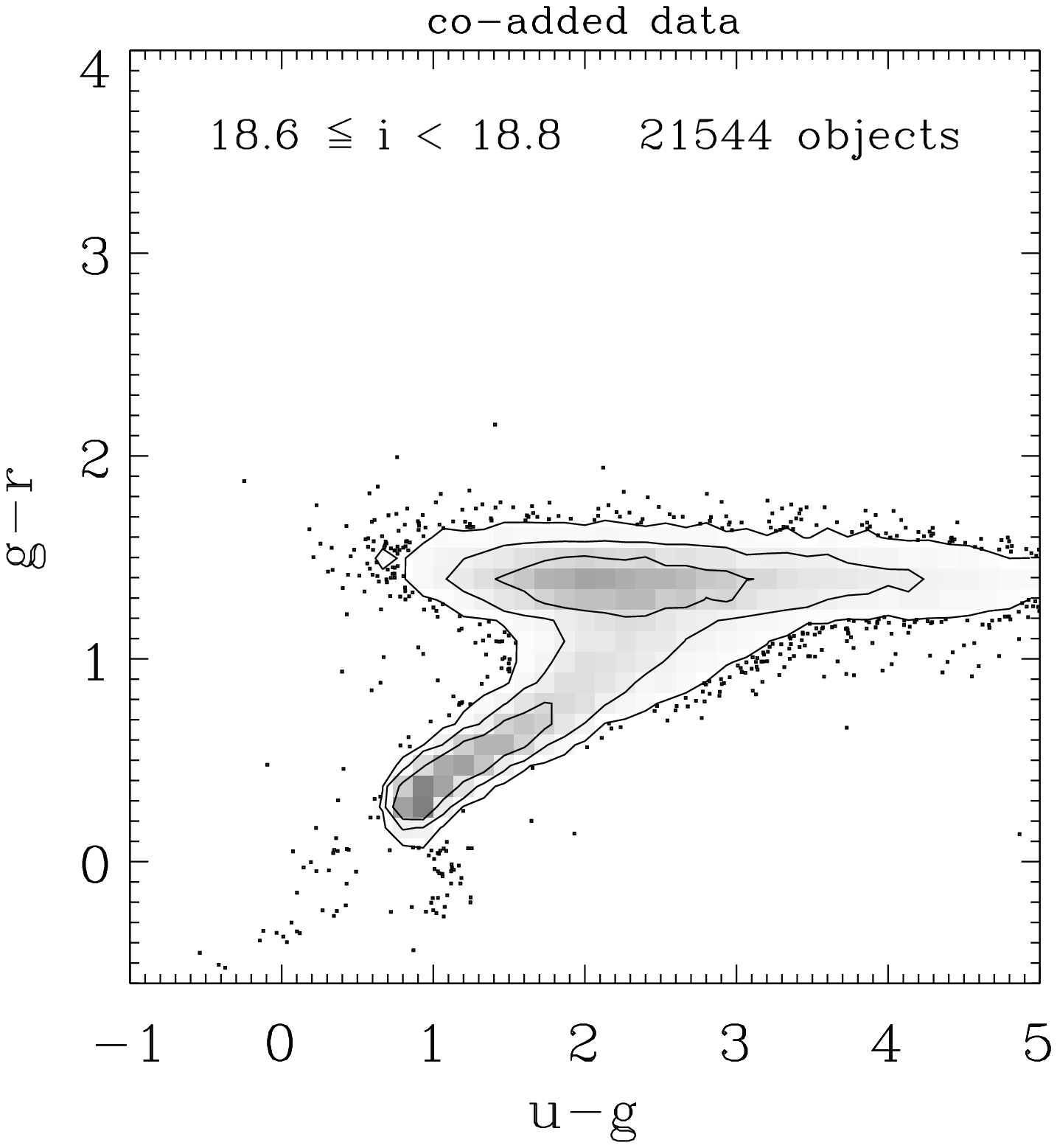}\\
\includegraphics[width=0.24\textwidth,clip=]{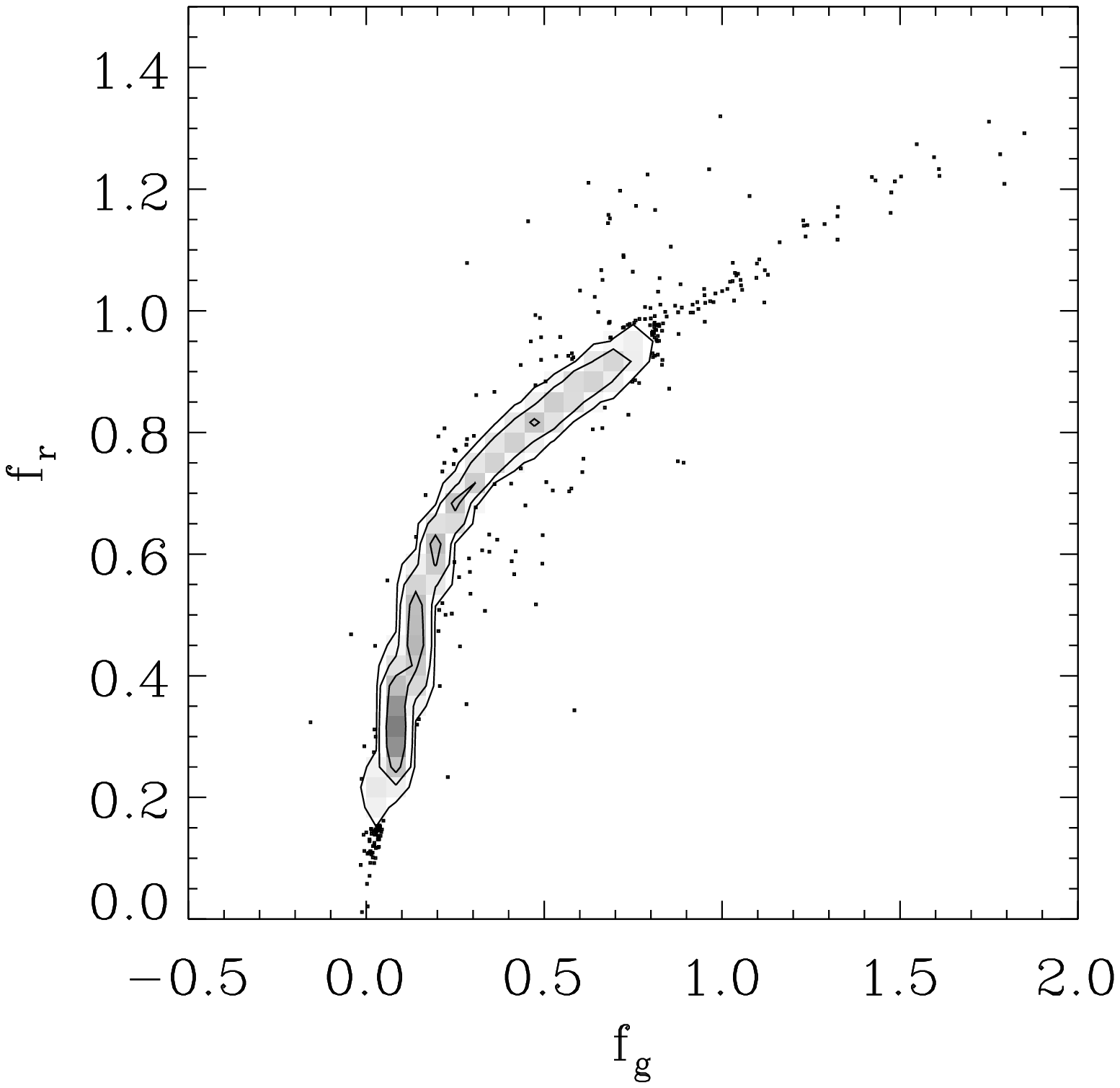}
\includegraphics[width=0.24\textwidth,clip=]{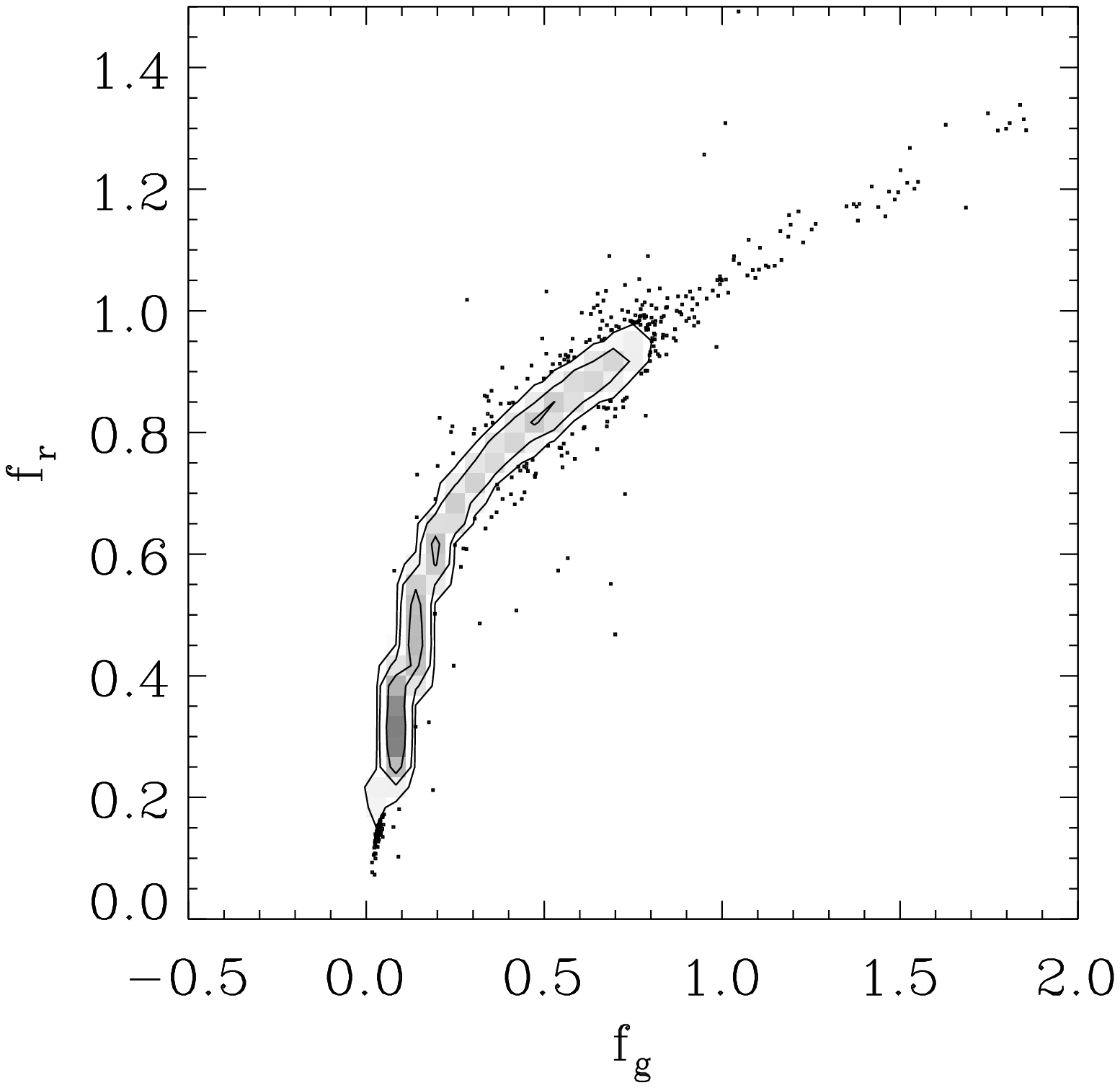}
\includegraphics[width=0.24\textwidth,clip=]{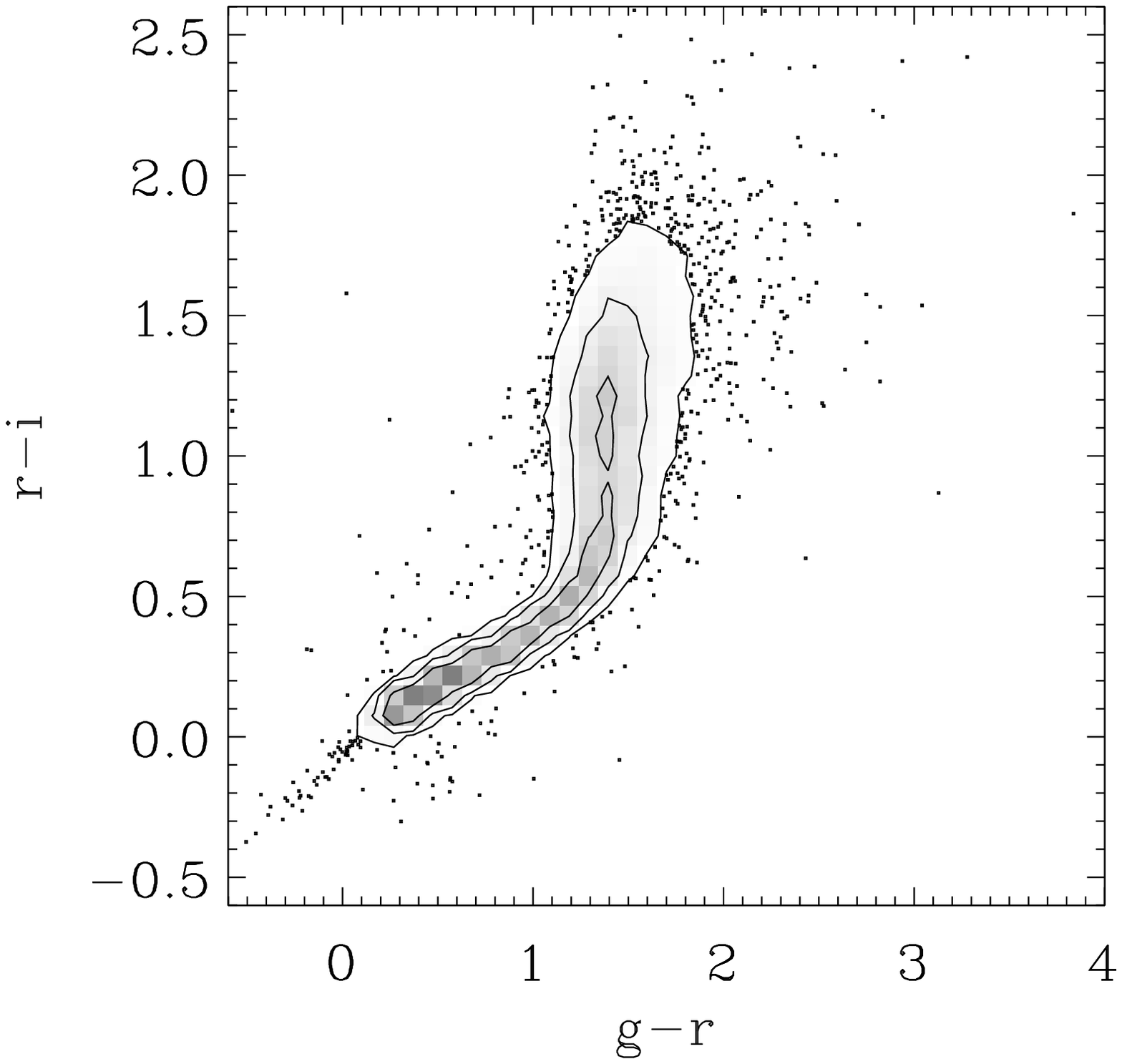}
\includegraphics[width=0.24\textwidth,clip=]{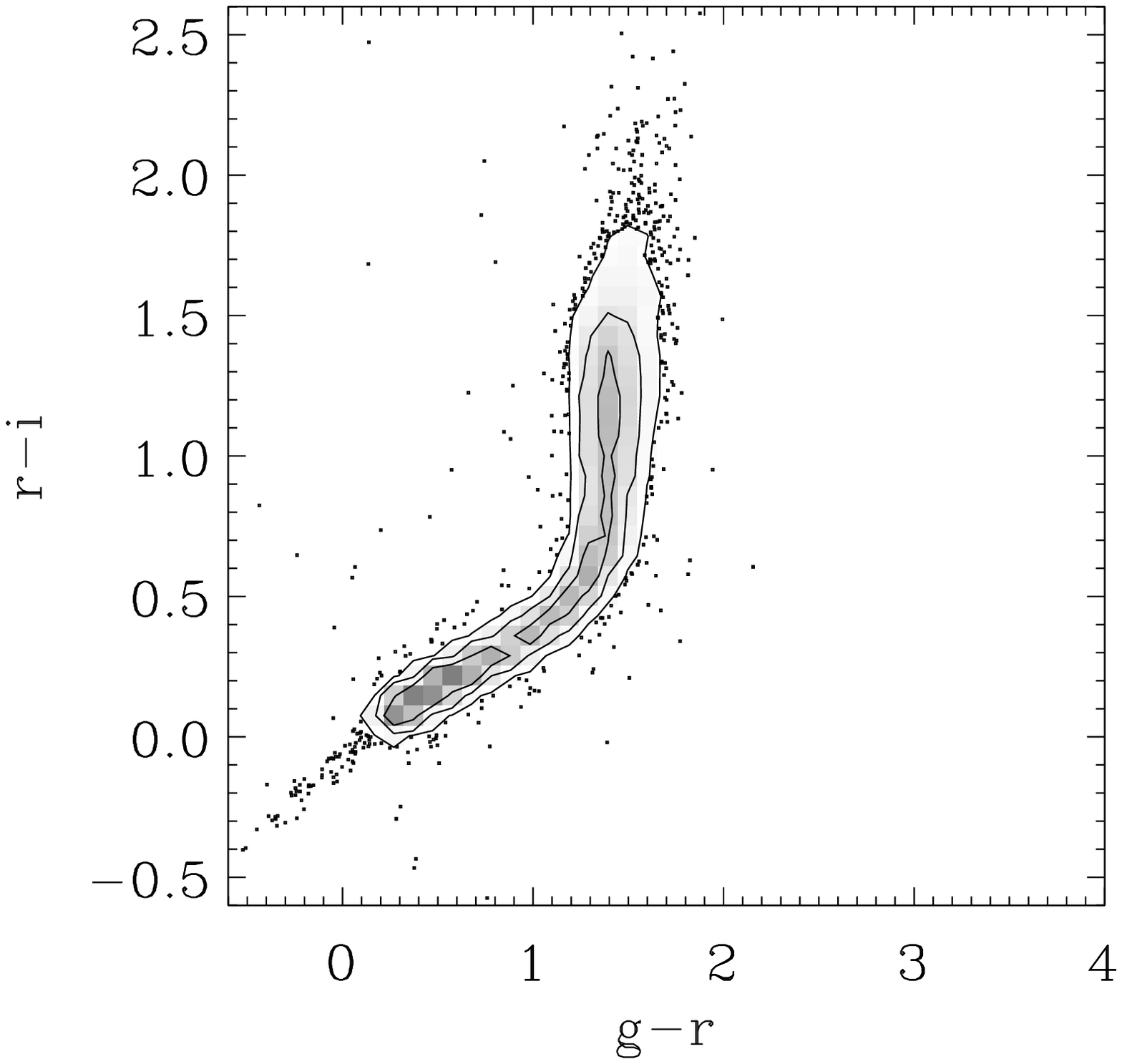}\\
\includegraphics[width=0.24\textwidth,clip=]{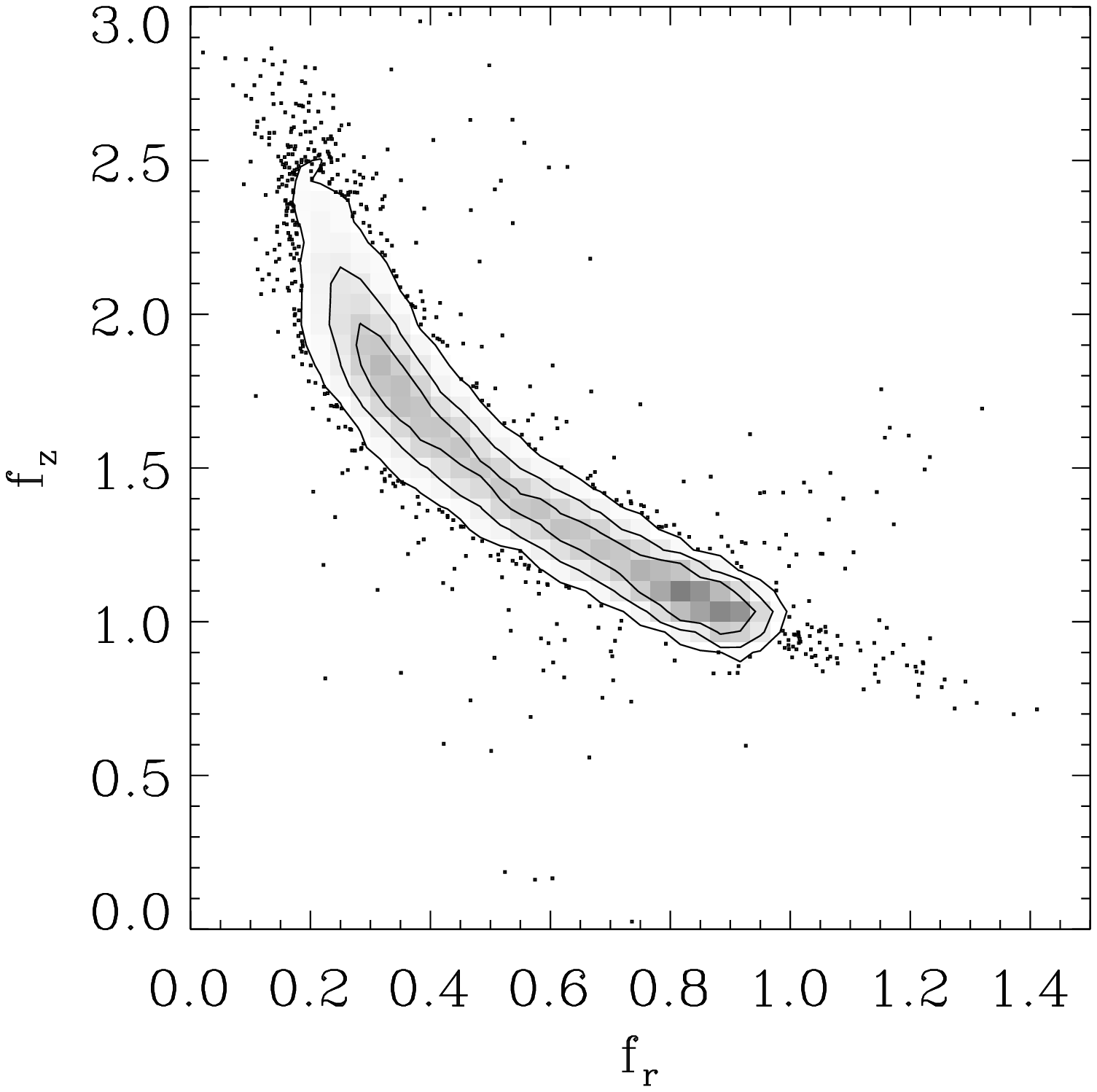}
\includegraphics[width=0.24\textwidth,clip=]{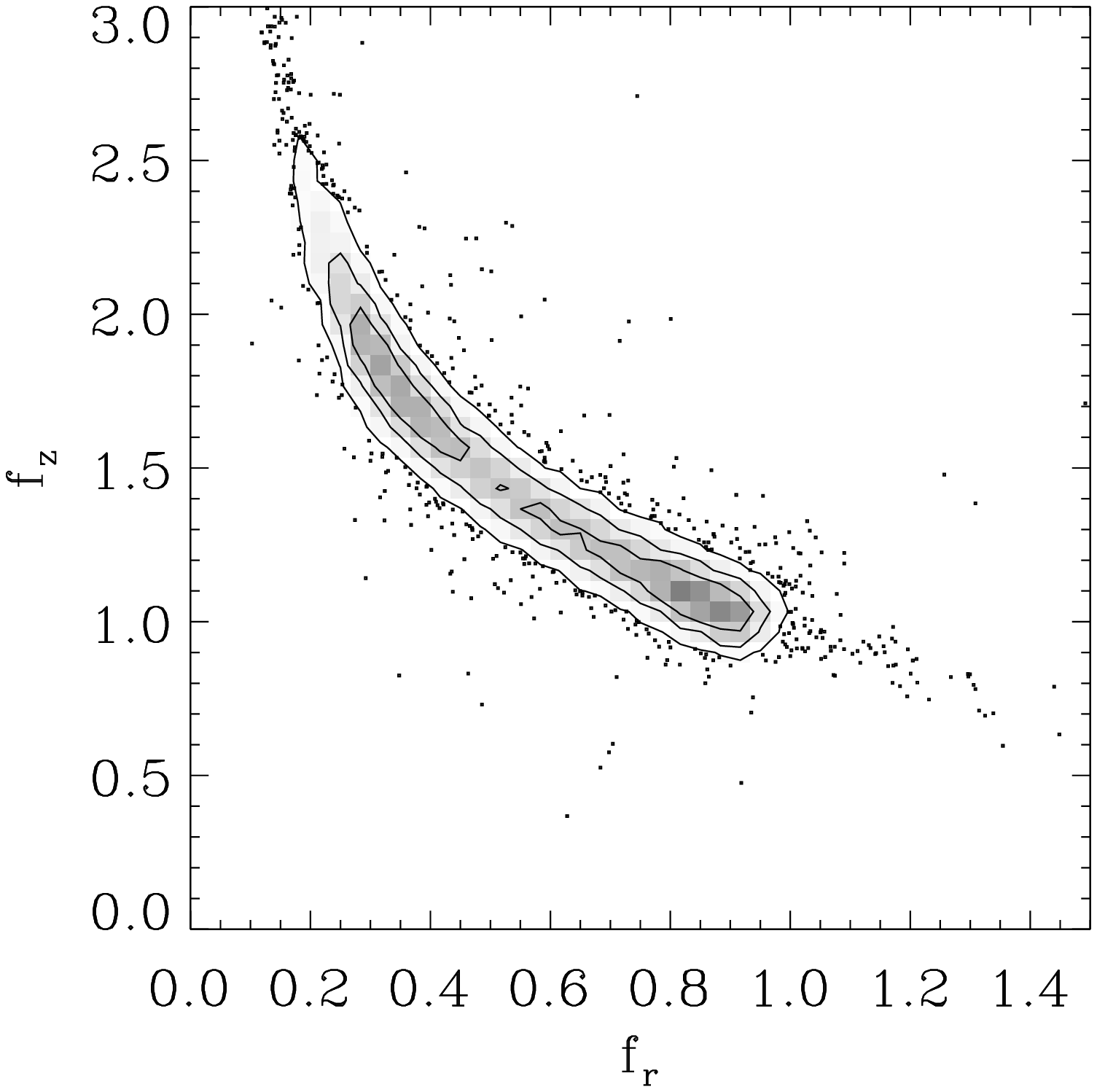}
\includegraphics[width=0.24\textwidth,clip=]{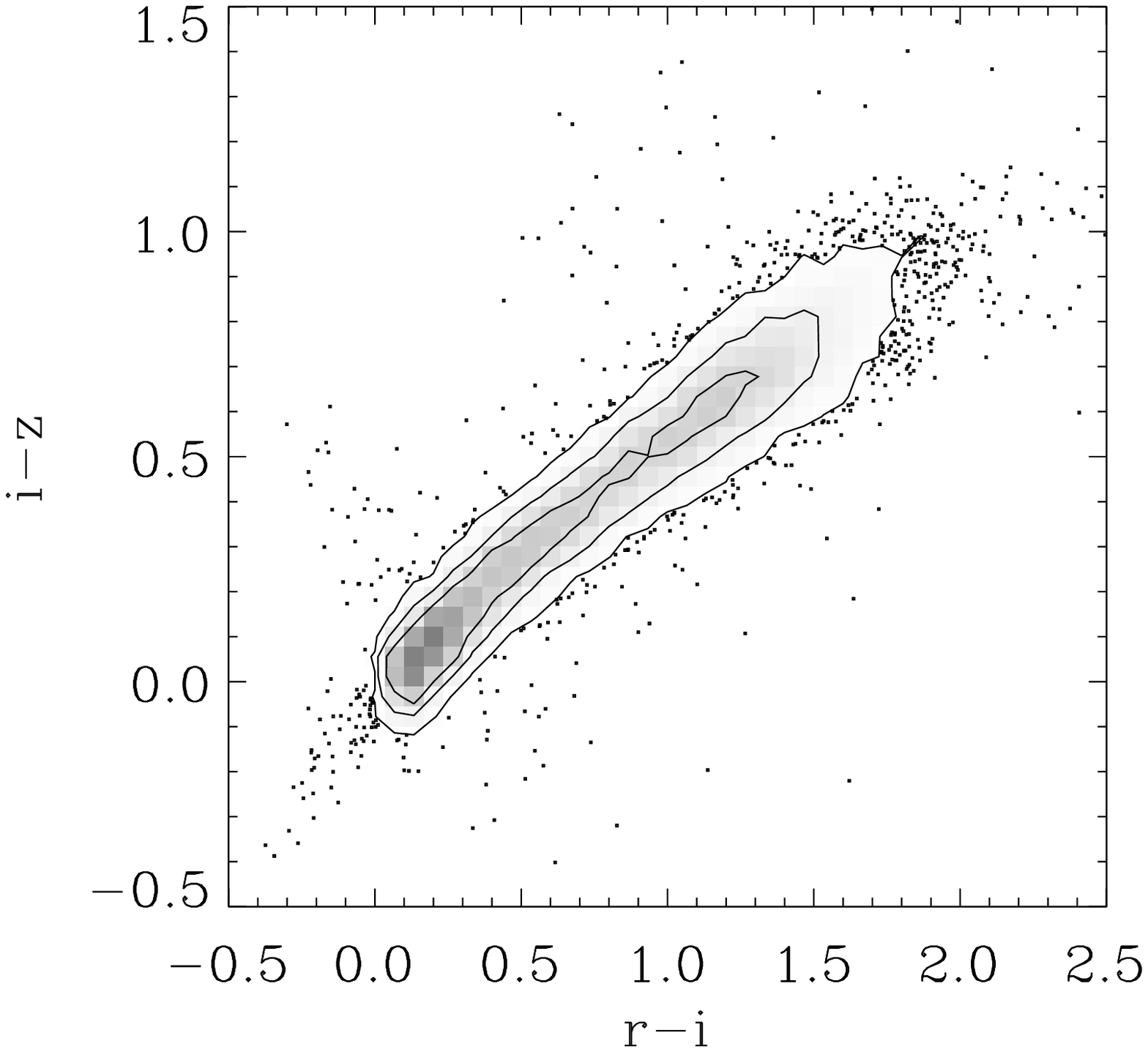}
\includegraphics[width=0.24\textwidth,clip=]{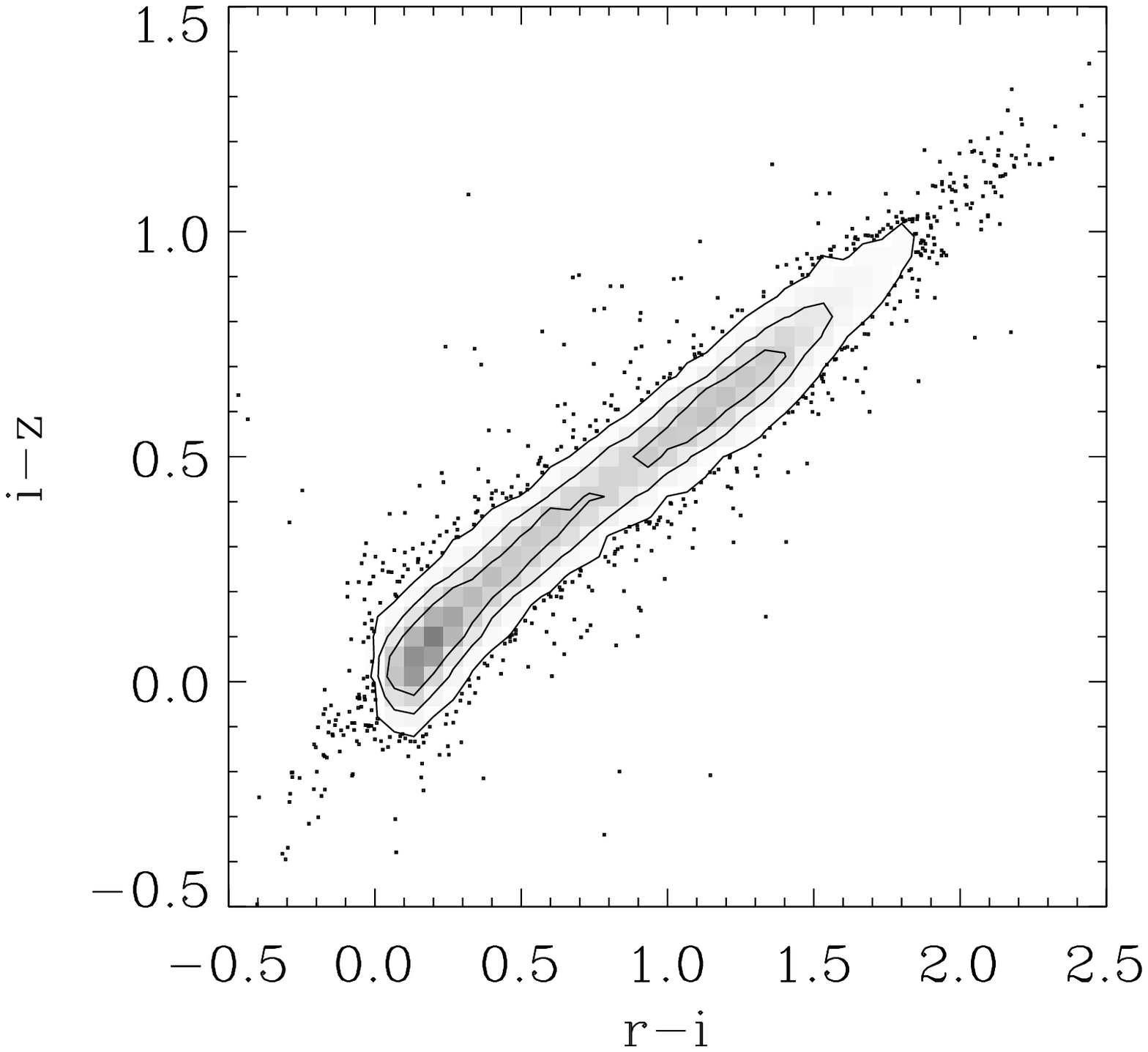}
\caption{Flux-flux and color-color diagrams for a bin in $i$-band
  magnitude from the stellar training catalog of co-added,
  non-variable \sdss\ Stripe-82 point-source data. The first column
  shows a sampling from the extreme-deconvolution fit with the errors
  from the stellar data added and the second column shows the stellar
  training data. The third and fourth columns show the same as the
  first and second columns, but for colors.}\label{fig:exfitstar1}
\end{figure}

\begin{figure}
\includegraphics[width=0.24\textwidth,clip=]{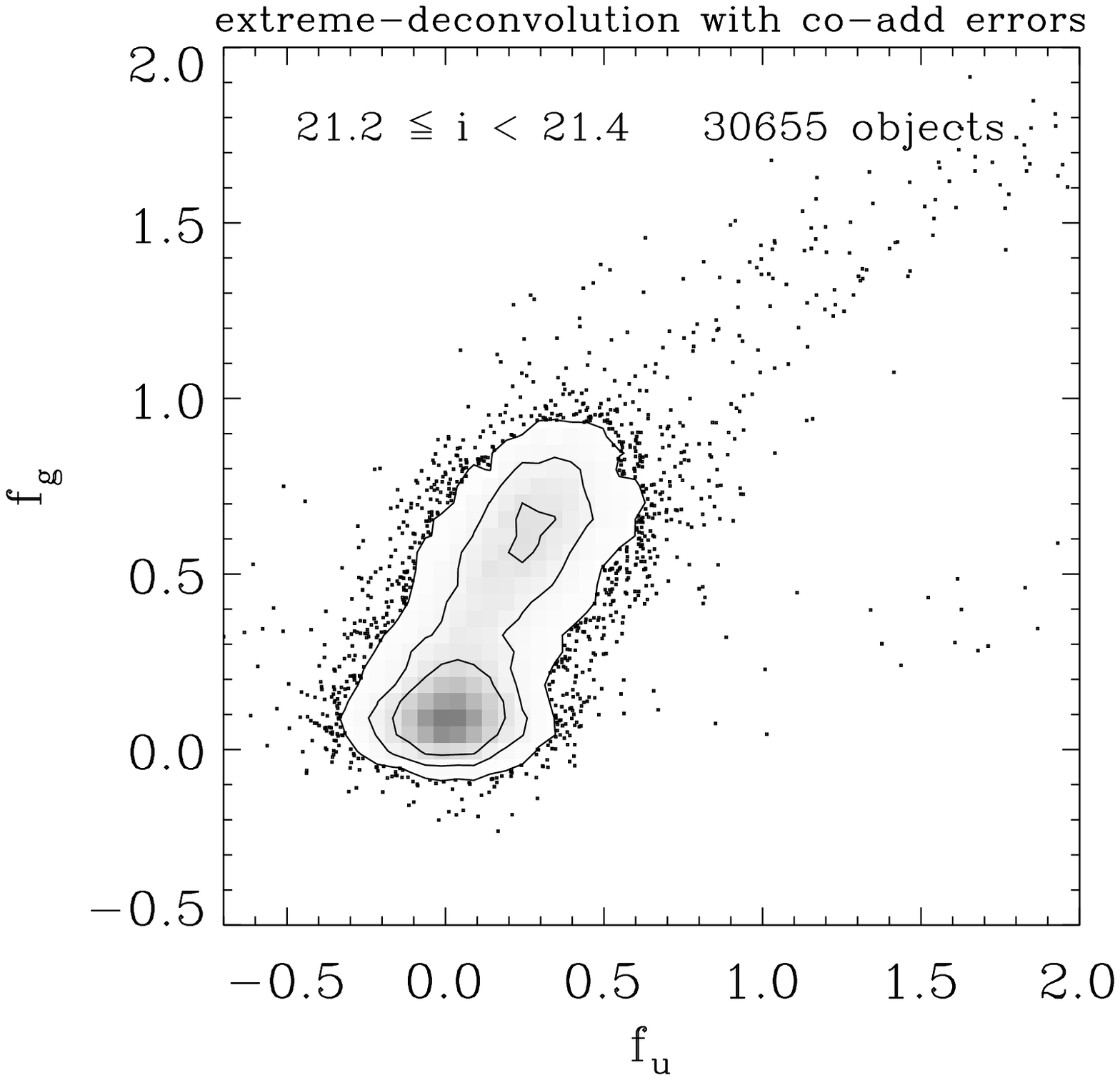}
\includegraphics[width=0.24\textwidth,clip=]{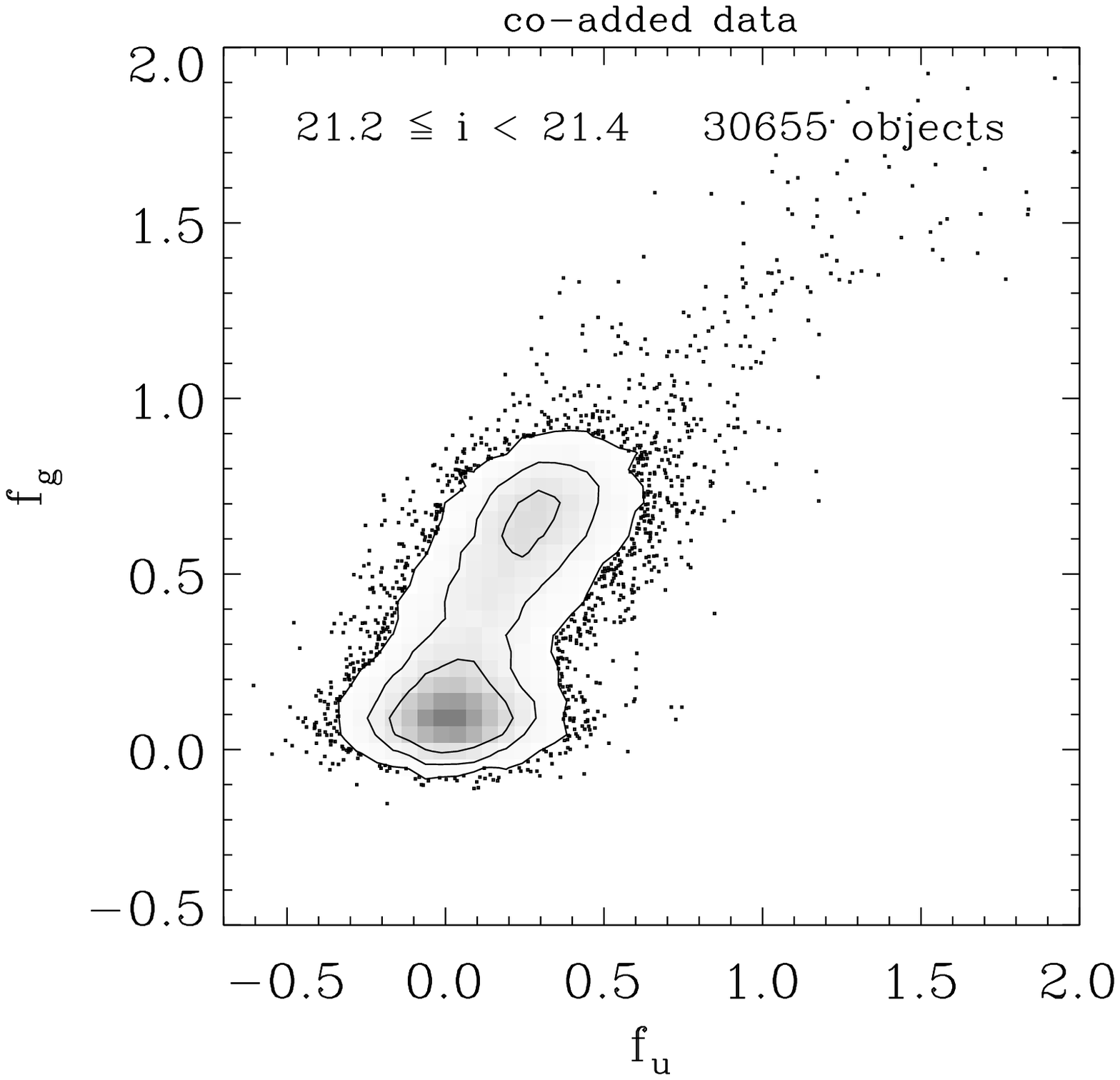}
\includegraphics[width=0.24\textwidth,clip=]{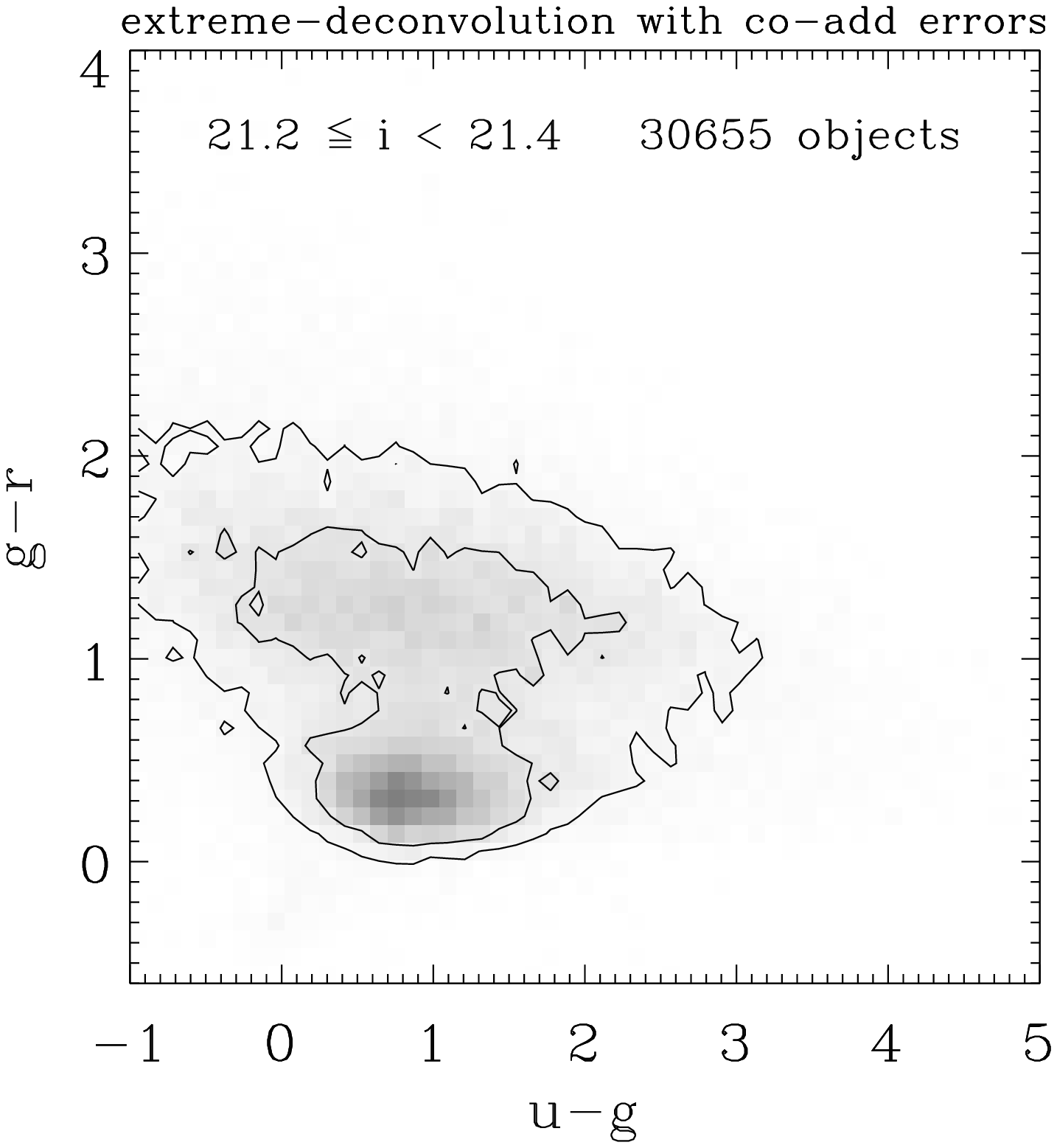}
\includegraphics[width=0.24\textwidth,clip=]{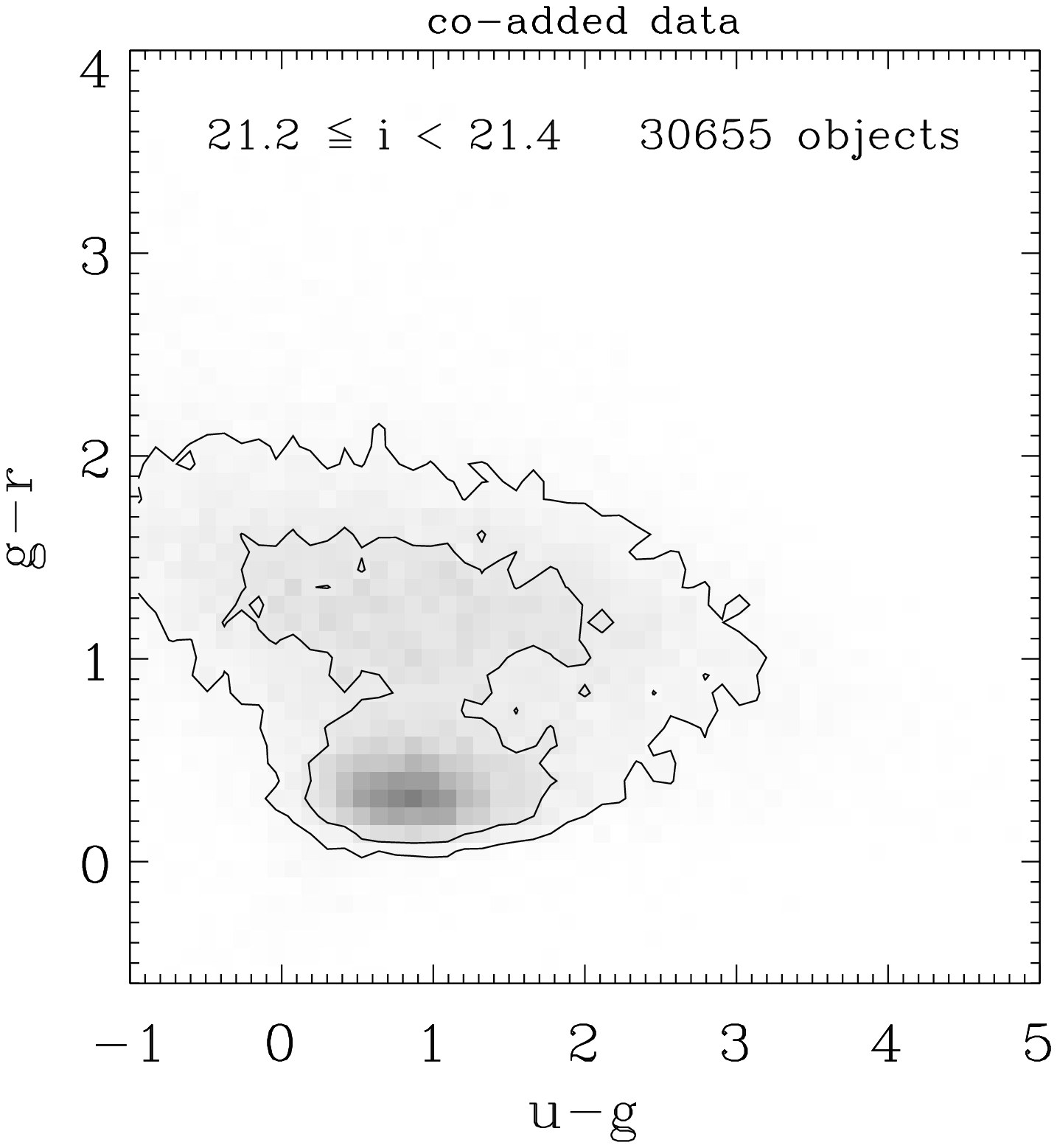}\\
\includegraphics[width=0.24\textwidth,clip=]{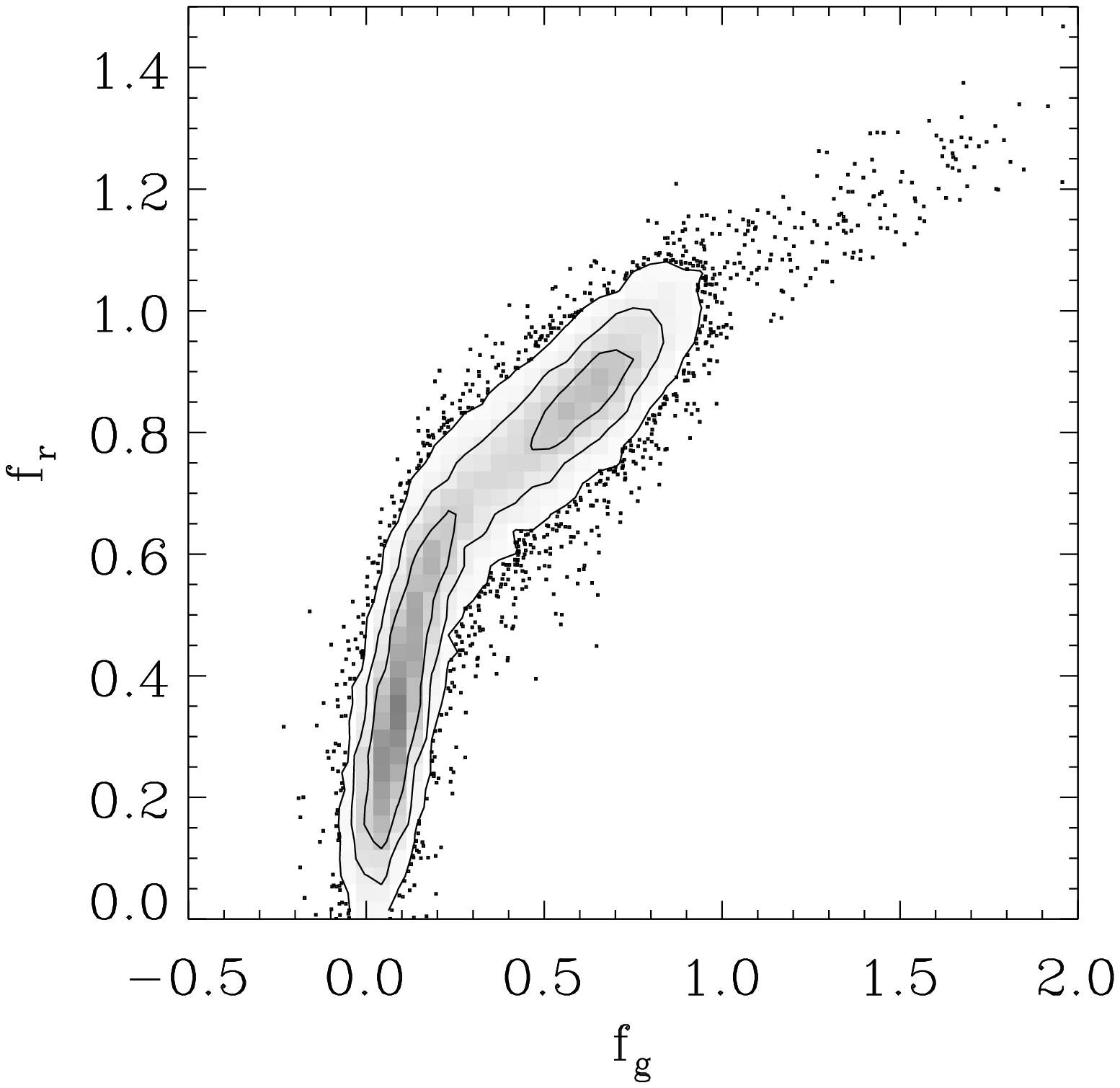}
\includegraphics[width=0.24\textwidth,clip=]{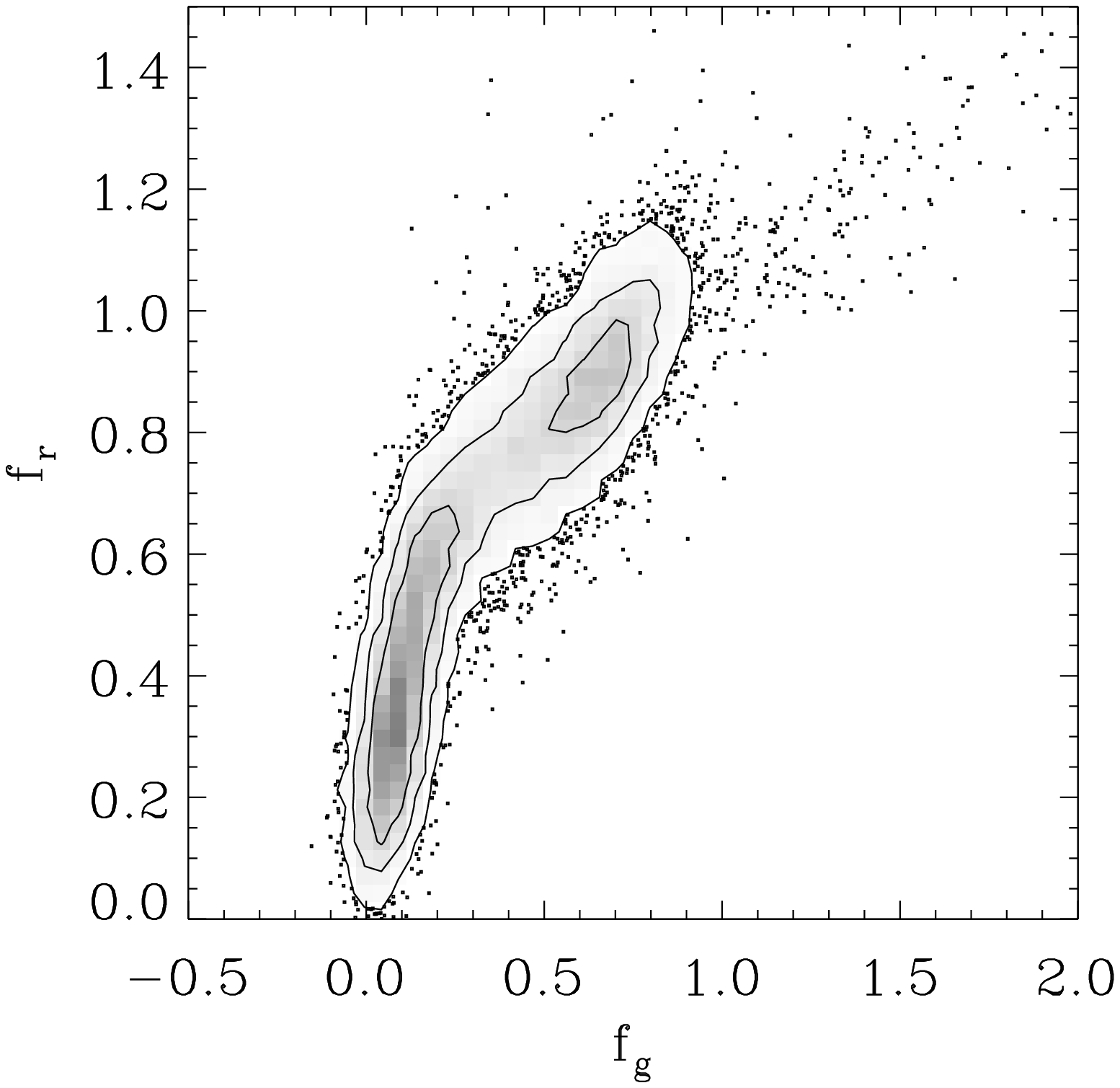}
\includegraphics[width=0.24\textwidth,clip=]{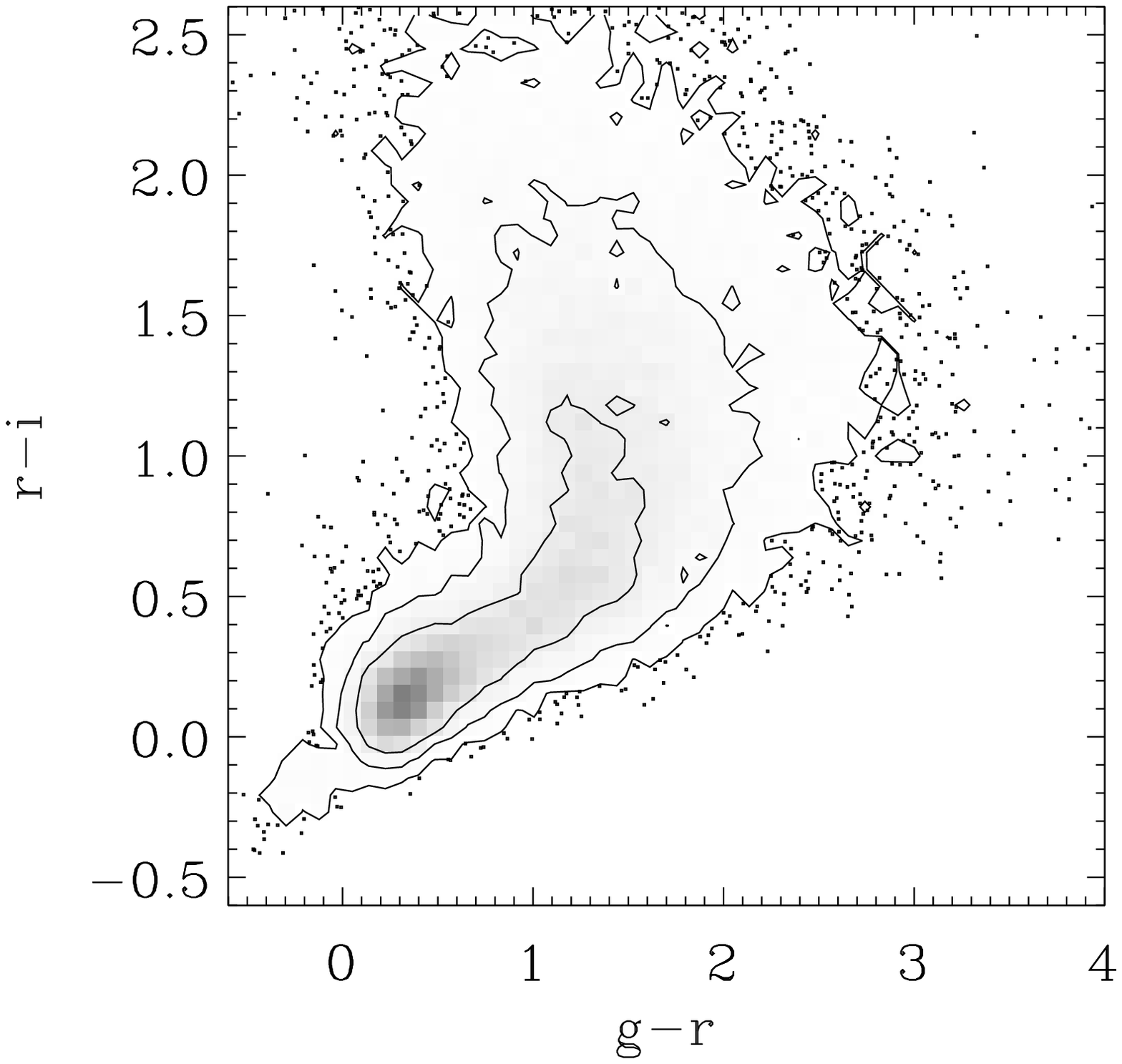}
\includegraphics[width=0.24\textwidth,clip=]{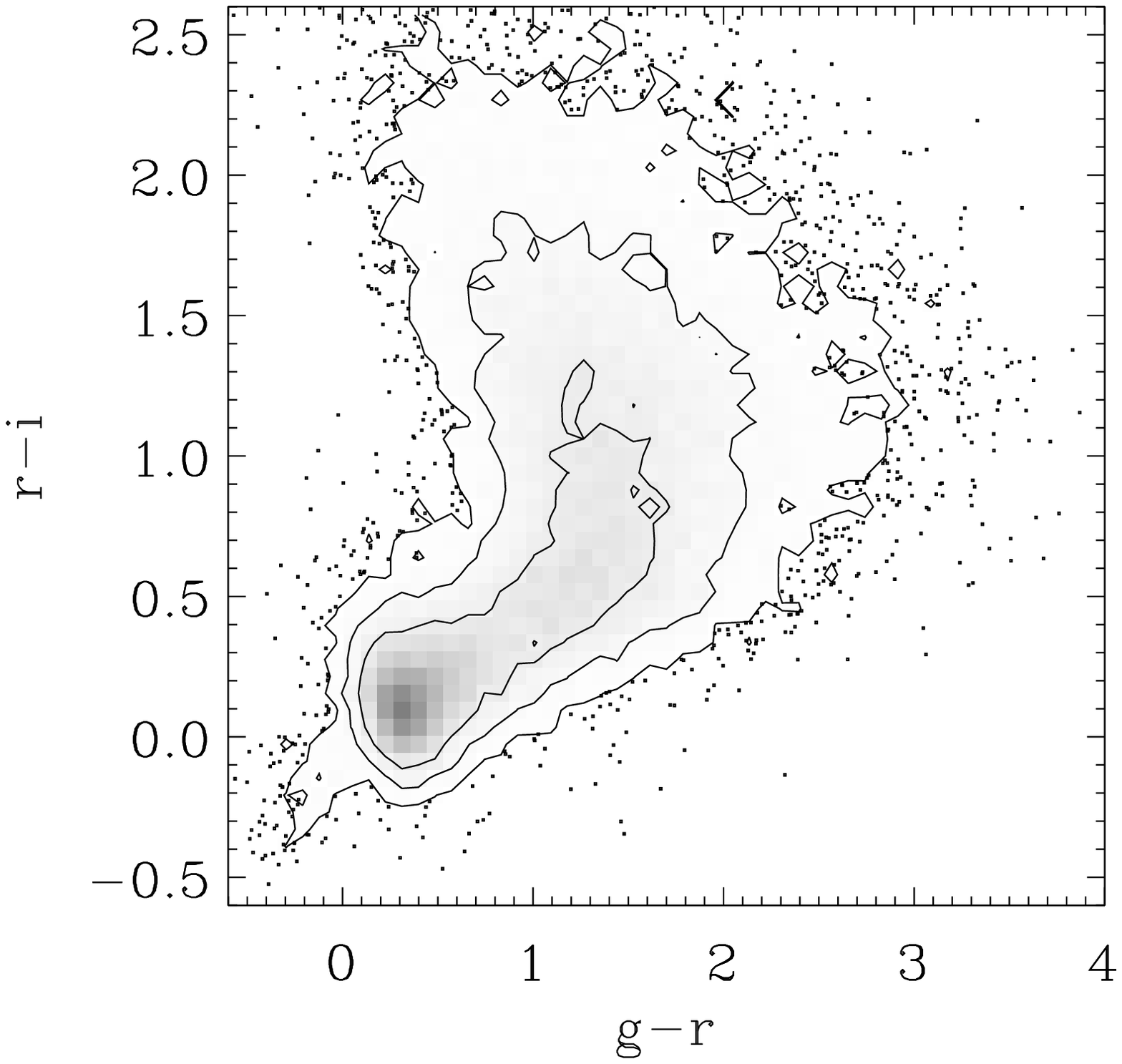}\\
\includegraphics[width=0.24\textwidth,clip=]{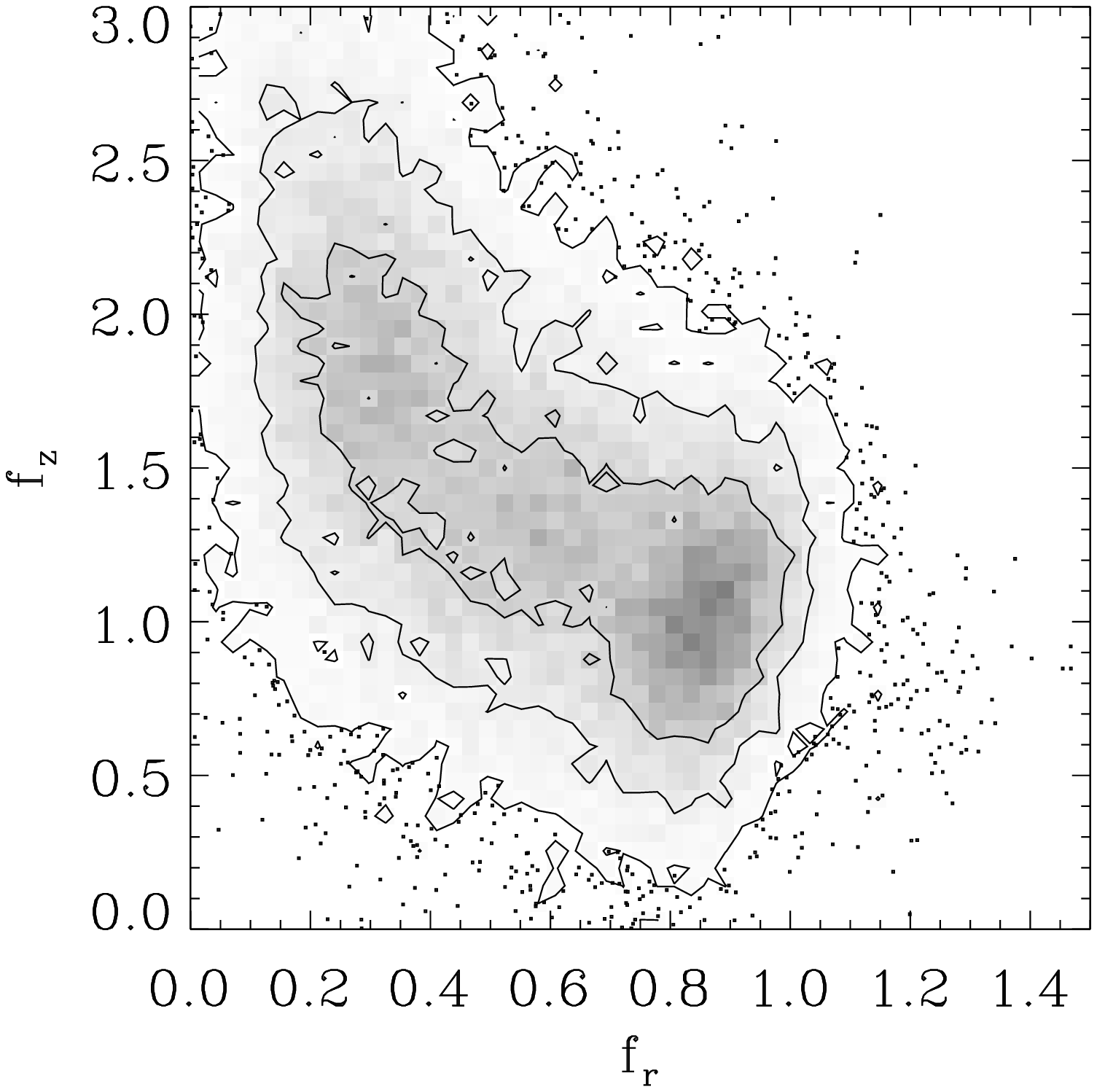}
\includegraphics[width=0.24\textwidth,clip=]{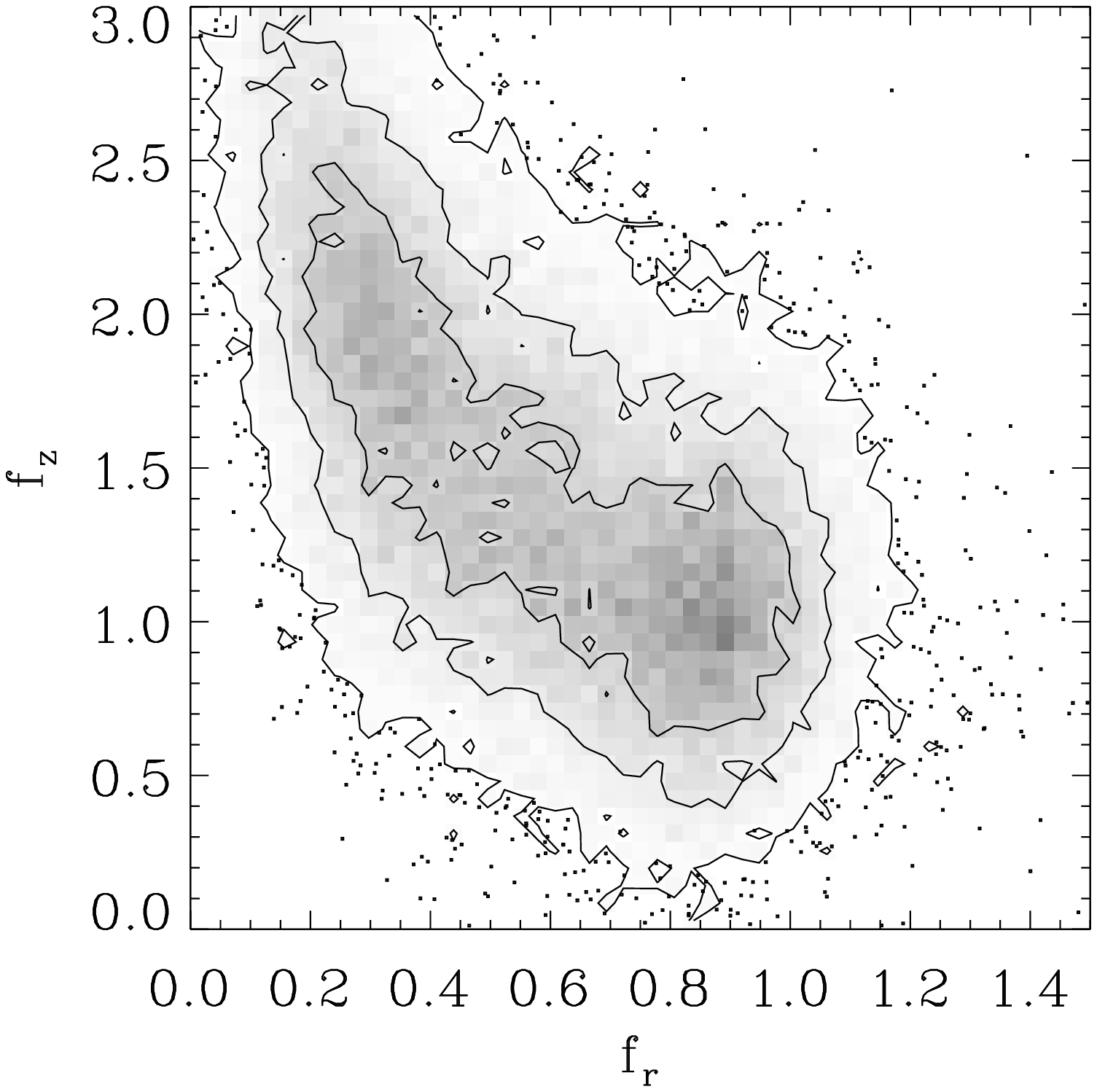}
\includegraphics[width=0.24\textwidth,clip=]{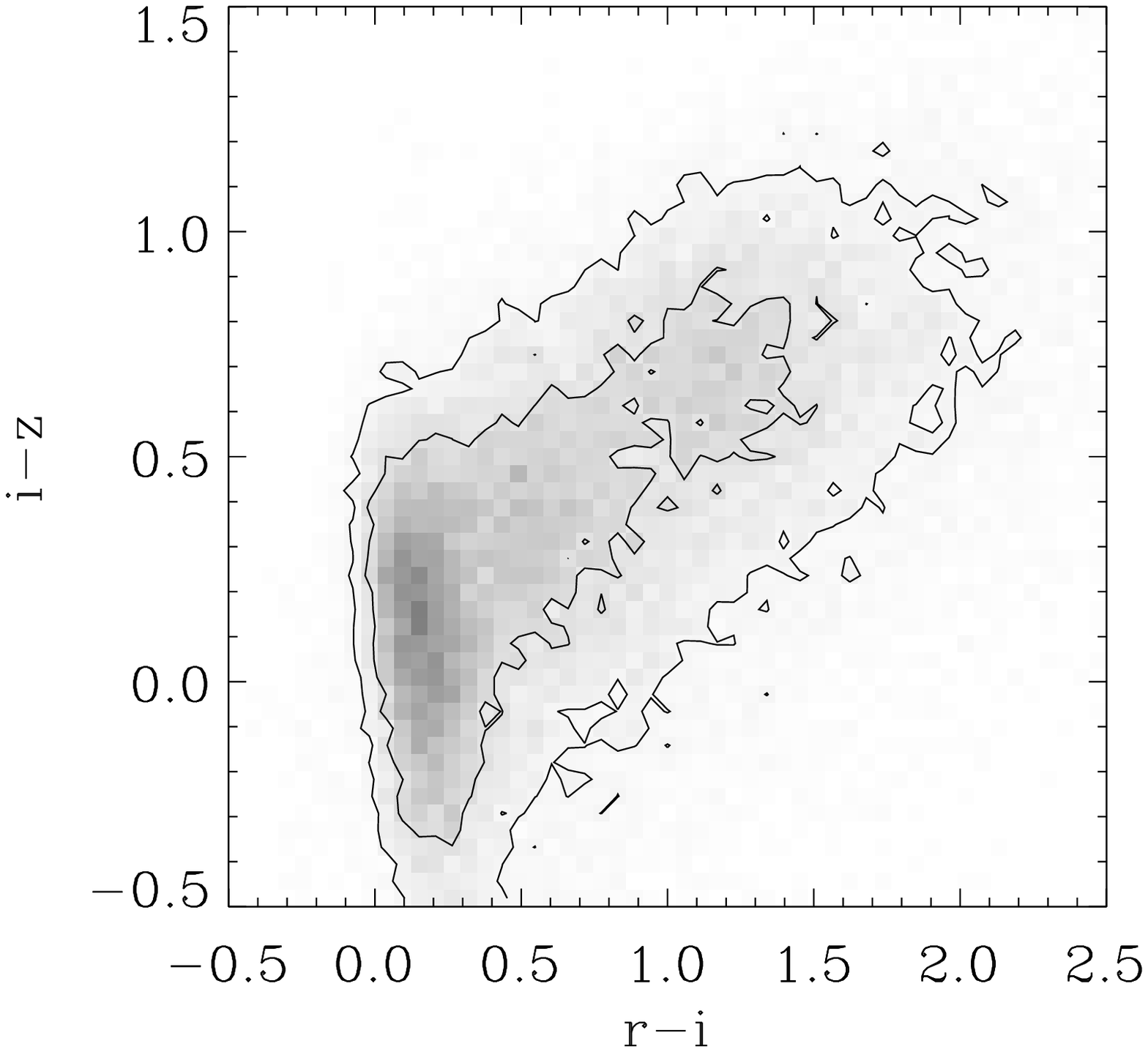}
\includegraphics[width=0.24\textwidth,clip=]{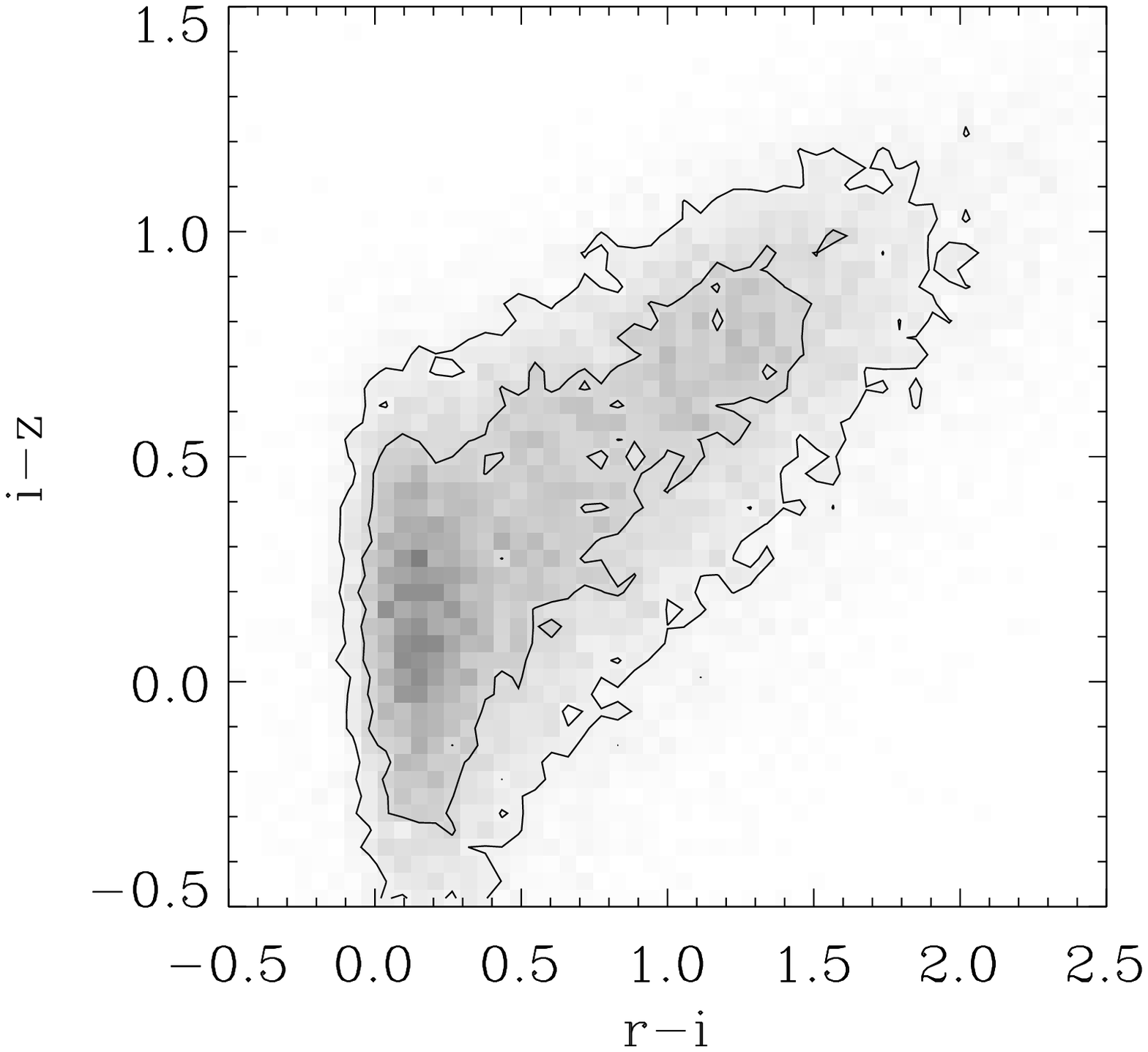}
\caption{Same as \figurename~\ref{fig:exfitstar1}, but for a fainter
  \iband\ apparent magnitude bin.}\label{fig:exfitstar}
\end{figure}

After fitting the relative flux distribution of the training data we
can also compare the best-fit model with the training data by
convolving the model with the data uncertainties. This
uncertainty-convolution is again performed by matching a sampled point
to the nearest object in the training set and drawing from that
object's uncertainty distribution. Some results of these tests are
shown in \figurename s~\ref{fig:bosszexfit} to \ref{fig:exfitstar}.

\figurename~\ref{fig:bosszexfit} shows flux-flux and color-color
diagrams in one of the bins in \iband\ magnitude for medium-redshift
quasars. The first and third columns show a sampling from the best-fit
\exd\ model to the relative fluxes. The second and fourth columns show
the data that these fits were based on: These data are quasars from
the \sdss\ DR7 quasar catalog resampled according to the quasar
luminosity function as explained in
\sectionname~\ref{sec:qsodata}. Similar figures for the other apparent
magnitude bins are very similar as the distribution of quasar colors
does not vary much with apparent magnitude. The same comparison for
low- and high-redshift quasars was performed for quality assurance,
but these figures are not shown here.

\figurename~\ref{fig:exfitstar1} shows the comparison between the
\exd\ fit to the co-added point-source data from \sdss\ Stripe-82 and
the data itself for a relatively bright apparent magnitude
bin. \figurename~\ref{fig:exfitstar} shows the same for a fainter bin
in \iband\ apparent magnitude. In this bin the photometric
uncertainties are significant such that the stellar locus is severely
smoothed. We see that the underlying deconvolved \exd\ model is good
in the sense that after convolution with the data uncertainties it
reproduces the observed distribution.

The resampled error-convolved fits to the data in all of these cases
are essentially indistinguishable from the real data.

\section{The \sdssxdqso\ quasar targeting catalog}\label{sec:catalog}

Using the models of quasar and stellar fluxes described in the
previous \sectionname\ we create a quasar targeting catalog. For every
point source (\flag{objc\_type} = 6) in the \sdss\ Data Release 8
(DR8) with a reasonable detection\footnote{Defined to be those point
  sources for which the PSF magnitude in at least one of $(u, g, r, i,
  z)$ is brighter than (22.5, 22.5, 22.5, 22, 21.5) after correction
  for Galactic extinction following \citet{Schlegel98a}
  \citep{Aihara11a}.} in at least one band that is \flag{primary} and
has a dereddened \iband\ magnitude 17.75 $ \leq i < 22.45$ we
calculate the probability that the object is a quasar or star using
\eqnname~(\ref{eq:bayes}). Specifically, we use the models from the
previous \sectionname\ as follows. For an object with dereddened
\iband\ magnitude $i$ we first find the bin for which this magnitude
is contained within the central 0.1 mag of the bin (this is always
possible since neighboring bins overlap by 0.1 mag). We use this bin
to evaluate the relative-flux density for this object's dereddened
fluxes. We convolve the underlying \ngauss\ Gaussian mixture model
with the object's uncertainties as in
\eqnname~(\ref{eq:errconv}). This uncertainty convolution is simply
adding the object's uncertainty variance to the intrinsic model
variance for each component.

We further evaluate the number density as a function of apparent
magnitude for the object's \iband\ flux. We do this for each of the
classes (\starc, and the three \qsoc\ classes) and normalize the
probabilities as in \eqnname~(\ref{eq:normalize}). Code that can be
used to reproduce this information is made publicly available and is
described in Appendix~\ref{sec:code}.

For each object we also calculate the bitmask used for \boss\ quasar
target selection. This bitmask is calculated using version
\versiontag{bosstarget v2\_0\_10} of the \boss\ quasar target
selection code. For the convenience of the user, we summarize this
bitmask into a simple \flag{good} flag. When this flag is zero the
object passes all of the \boss\ flag cuts; when the flag is one, the
object fails one of the basic \boss\ targeting flag criteria
(\flag{interp\_problems}, \flag{deblend\_problems}, or
\flag{moved}). An object with a \flag{good} flag equal to two fails
one of the more esoteric \boss\ quasar selection cuts. No objects with
a \flag{good} flag $> 0$ would be targeted by the \boss. These flag
cuts are discussed in more detail in Appendix~\ref{sec:flags}.

In addition to the quasar and star probabilities and flag-logic
described above, we also report each of the factors in
\eqnname~(\ref{eq:exdmodel}) for all of the target-classes for each
object in the catalog. This allows the user of the catalog to
substitute different number-count or relative-flux density models for
any of the components.

A description of the catalog is given in
\tablename~\ref{table:catdescription}. The full catalog contains
\ncatentries\ objects. Summing the probabilities, we expect this
catalog to contain about 5,058,716 quasars, 2,551,146 of which are
low-redshift, 1,410,405 are mid-redshift, and 1,097,165 are
high-redshift. However, as we show below, this is probably a slight
overestimate, especially for the high-redshift class.

\begin{figure}
\includegraphics[width=0.5\textwidth,clip=]{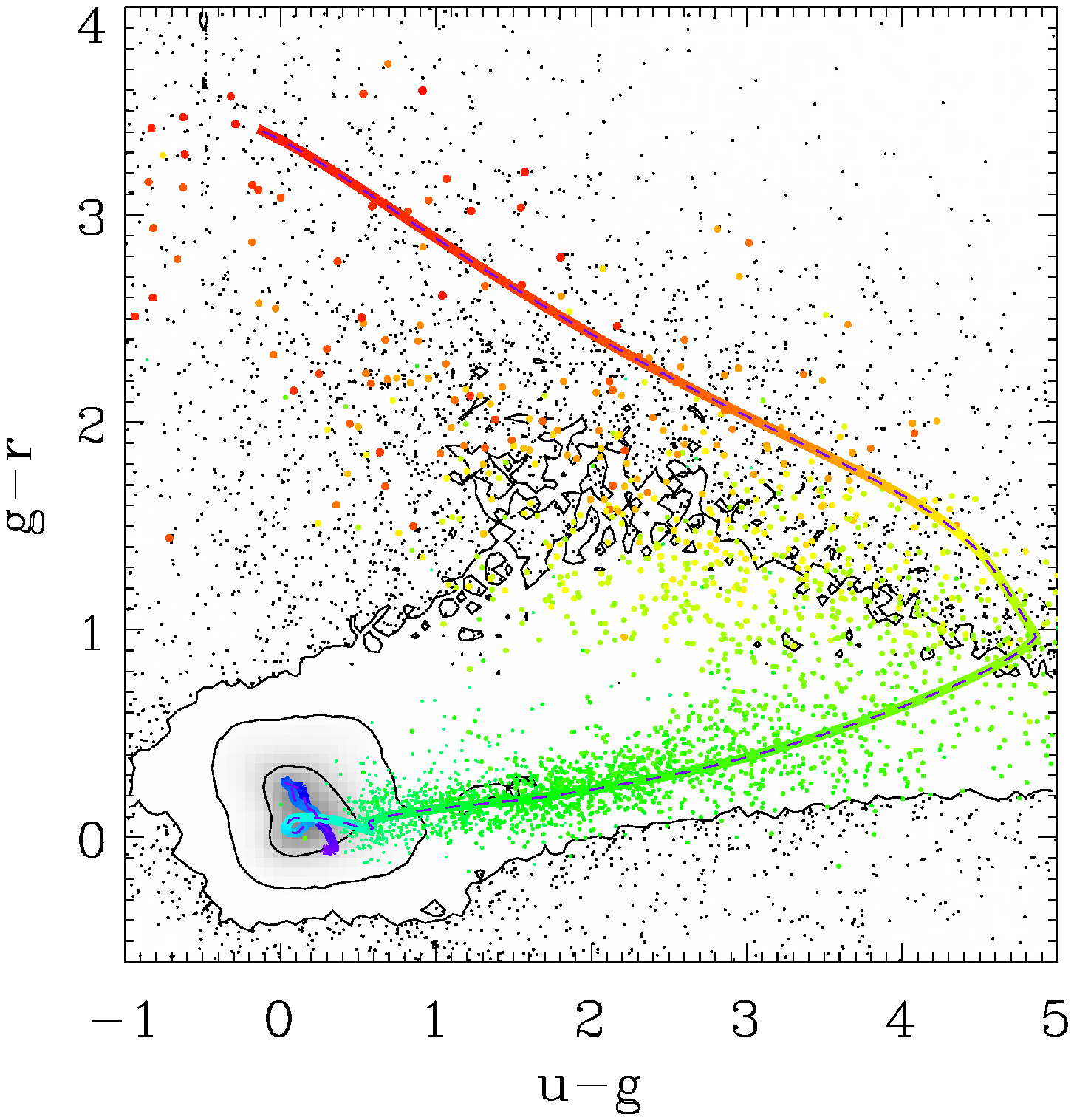}
\includegraphics[width=0.5\textwidth,clip=]{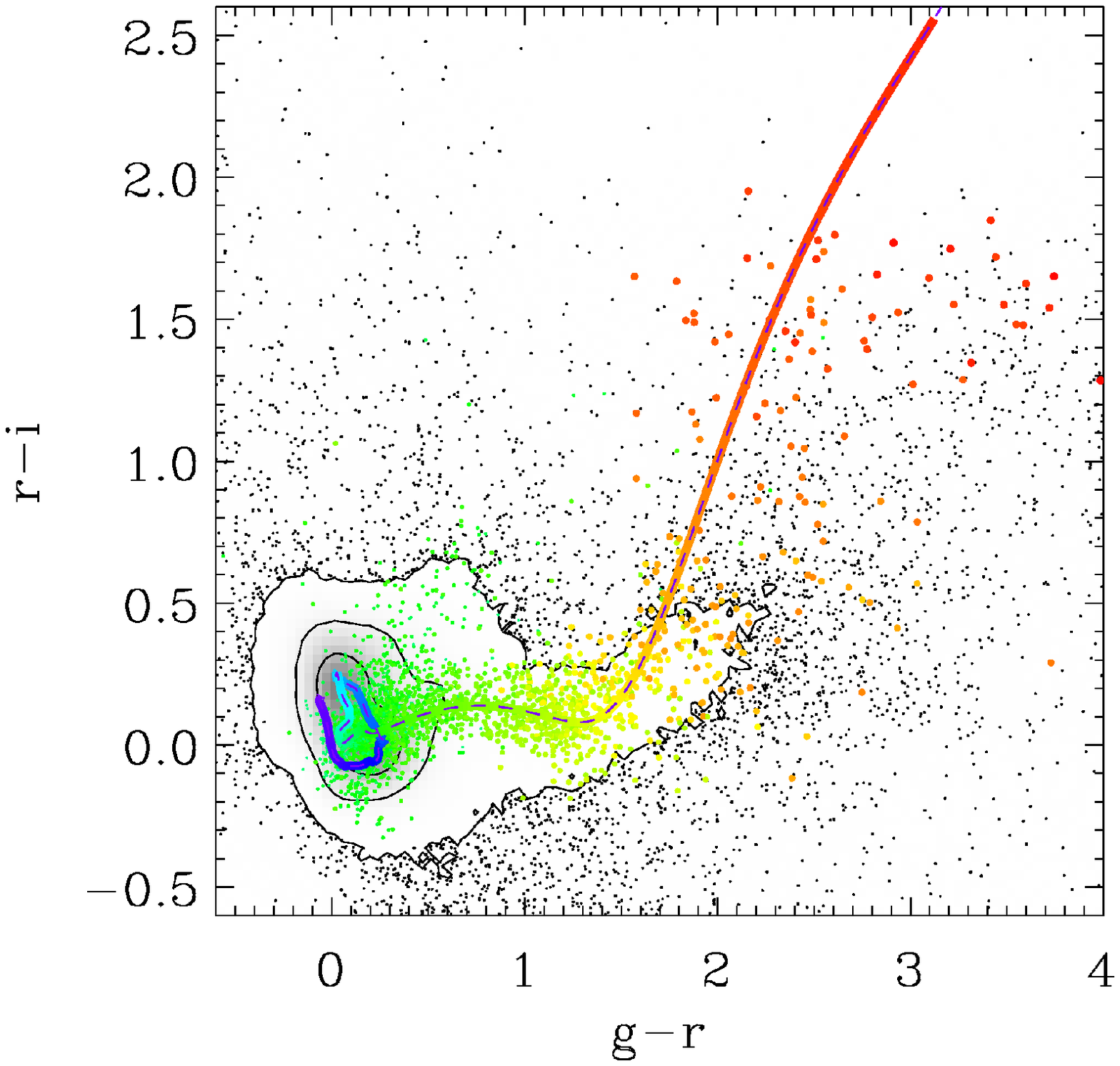}\\
\includegraphics[width=0.5\textwidth,clip=]{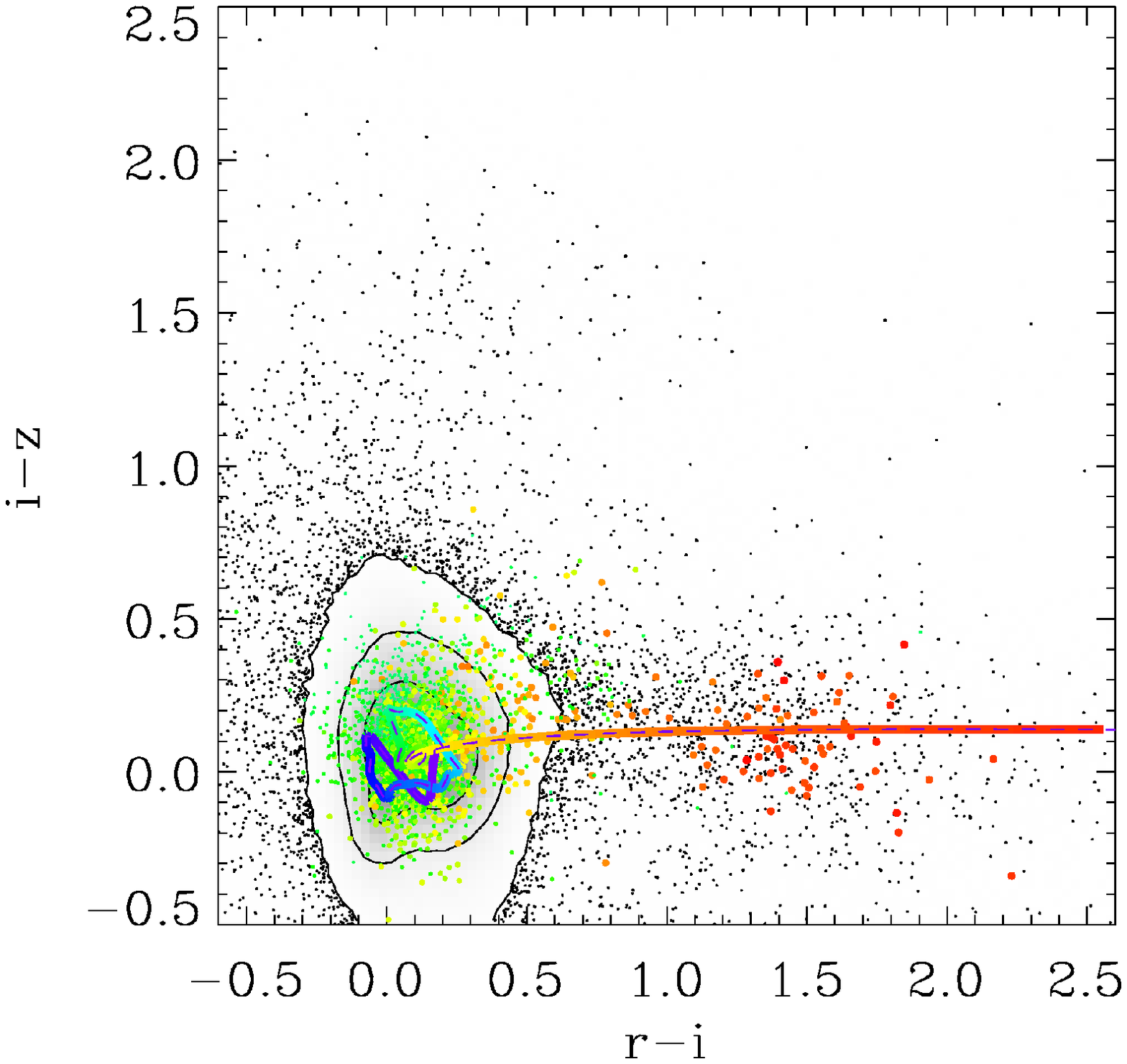}
\hspace{20pt}\raisebox{8pt}{\includegraphics[width=0.5\textwidth]{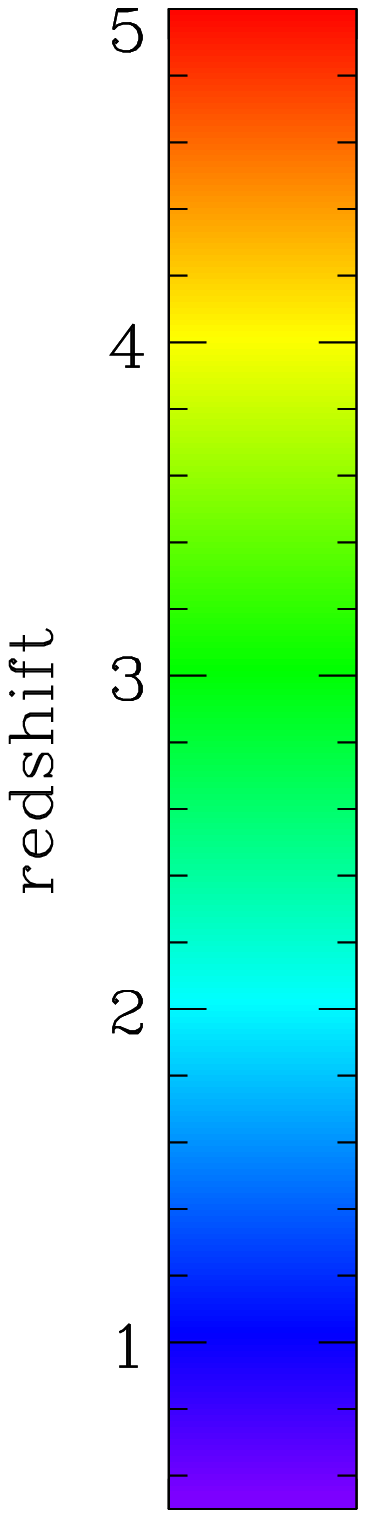}}
\caption{Color-color distributions of \flag{good} objects in the
  \sdssxdqso\ catalog with $P(\mathrm{quasar}) \geq 0.8$ and
  dereddened $i < 21$ mag. The grayscale is linear in the density and
  the contours contain 68, 95, and 99\,percent of the distribution. A
  sparse sampling of objects falling outside the outermost contour is
  shown as individual black points. A twenty-percent random sampling
  of objects with $z \ge 2.5$ in the \sdss\ DR7 quasar catalog
  \citep{Schneider10a} is plotted as redshift color-coded points
  according to the color-bar at the lower right (lower redshift
  quasars are omitted for clarity). Higher redshift objects are
  plotted as larger points. A fit to the quasar locus from
  \citet{Hennawi10a} is shown by the dashed black line, similarly
  color-coded to indicate redshift.}\label{fig:qsocatex}
\end{figure}

\begin{figure}
\includegraphics[width=0.5\textwidth,clip=]{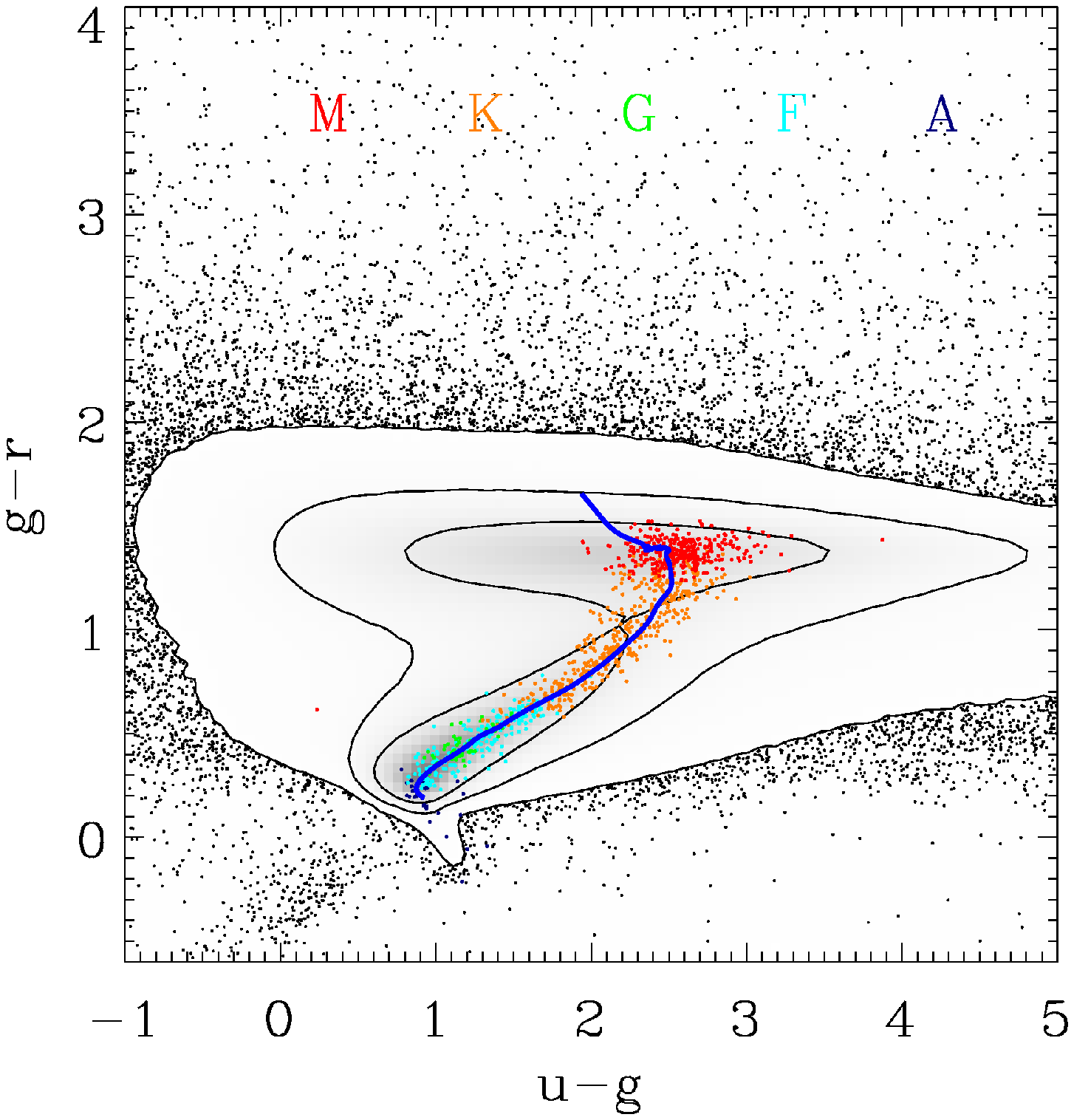}
\includegraphics[width=0.5\textwidth,clip=]{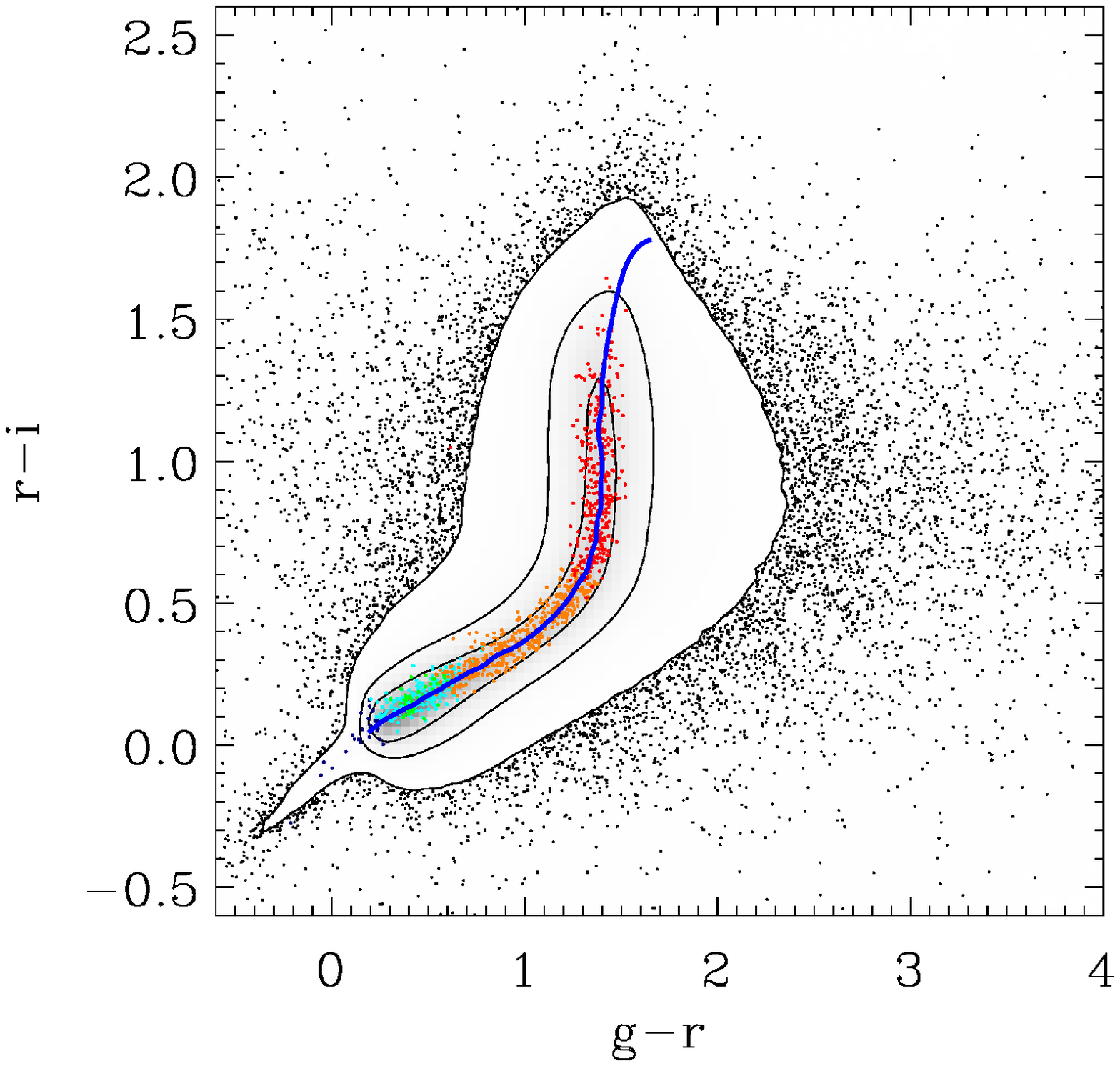}\\
\includegraphics[width=0.5\textwidth,clip=]{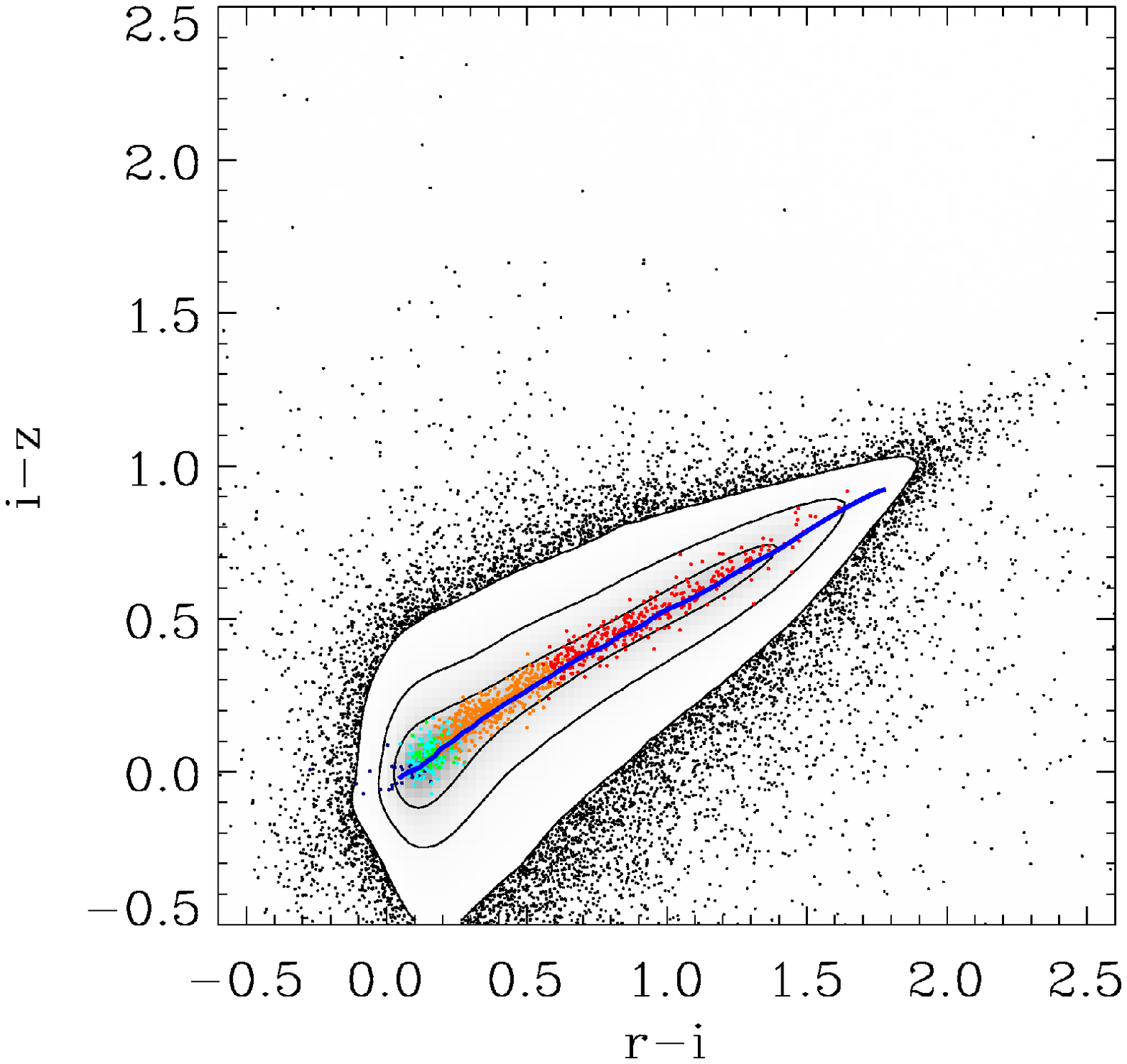}
\caption{Color-color distributions of \flag{good} objects in the
  \sdssxdqso\ catalog with $P(\mathrm{star}) \geq 0.95$ and dereddened
  $i < 21$ mag. The grayscale is linear in the density and the
  contours contain 68, 95, and 99\,percent of the distribution. A
  sparse sampling of objects falling outside the outermost contour is
  shown as individual black points. A fit to the stellar locus using
  spectroscopically confirmed stars from \citet{Hennawi10a} is shown
  in blue. Some representative classes of stars along the stellar
  locus from \sdss\ plates 323 and 324 \citep{dr4} are shown as colored
  points.}\label{fig:starcatex}
\end{figure}

We stress that this quasar targeting catalog is entirely probabilistic
and that the majority of objects are certainly stars. To create
high-confidence quasar samples it is necessary to select those objects
with the largest \pqso. As an example, we show in \figurename
s~\ref{fig:qsocatex} and \ref{fig:starcatex} color--color plots of
those objects in the \sdssxdqso\ catalog with $\pqso \ge 0.8$ and
$\pstar \ge 0.95$. For the photometric quasars in
\figurename~\ref{fig:qsocatex} a sparse sampling of $z \ge 2.5$
quasars from the \sdss\ DR7 quasar catalog \citep{Schneider10a} and a
model for the quasar locus from \citet{Hennawi10a} is shown for
comparison. For the photometric stars in
\figurename~\ref{fig:starcatex} a model for the stellar locus from
\citet{Hennawi10a} and some representative classes of stars along the
stellar locus are shown. While we cannot be sure for any of the
objects in either of these photometric samples whether they are truly
quasars or stars, the color distribution of the ensemble does resemble
that of a set of quasars and stars, respectively.

\subsection{Catalog availability}

The \sdssxdqso\ catalog is available through the \sdssiii\ DR8
\emph{Science Archive Server} at
\begin{quote}
\url{http://data.sdss3.org/sas/dr8/groups/boss/photoObj/xdqso/xdcore/}\ .
\end{quote}
The catalog consists of a single fits file for each imaging run in the
DR8 release \citep{Aihara11a}; filenames have the form
xdcore\_RUN6.fits, where RUN6 is the six-digit imaging run number. The
catalog entries are described in
\tablename~\ref{table:catdescription}, as well as in the
\sdssiii\ datamodel at
\begin{quote}
\url{http://data.sdss3.org/datamodel/files/BOSS\_PHOTOOBJ/xdqso/xdcore/xdcore\_RUN6.html}
\end{quote}

%BOVY: reference point

\section{Testing the catalog}\label{sec:properties}

\begin{figure}
\includegraphics[width=\textwidth]{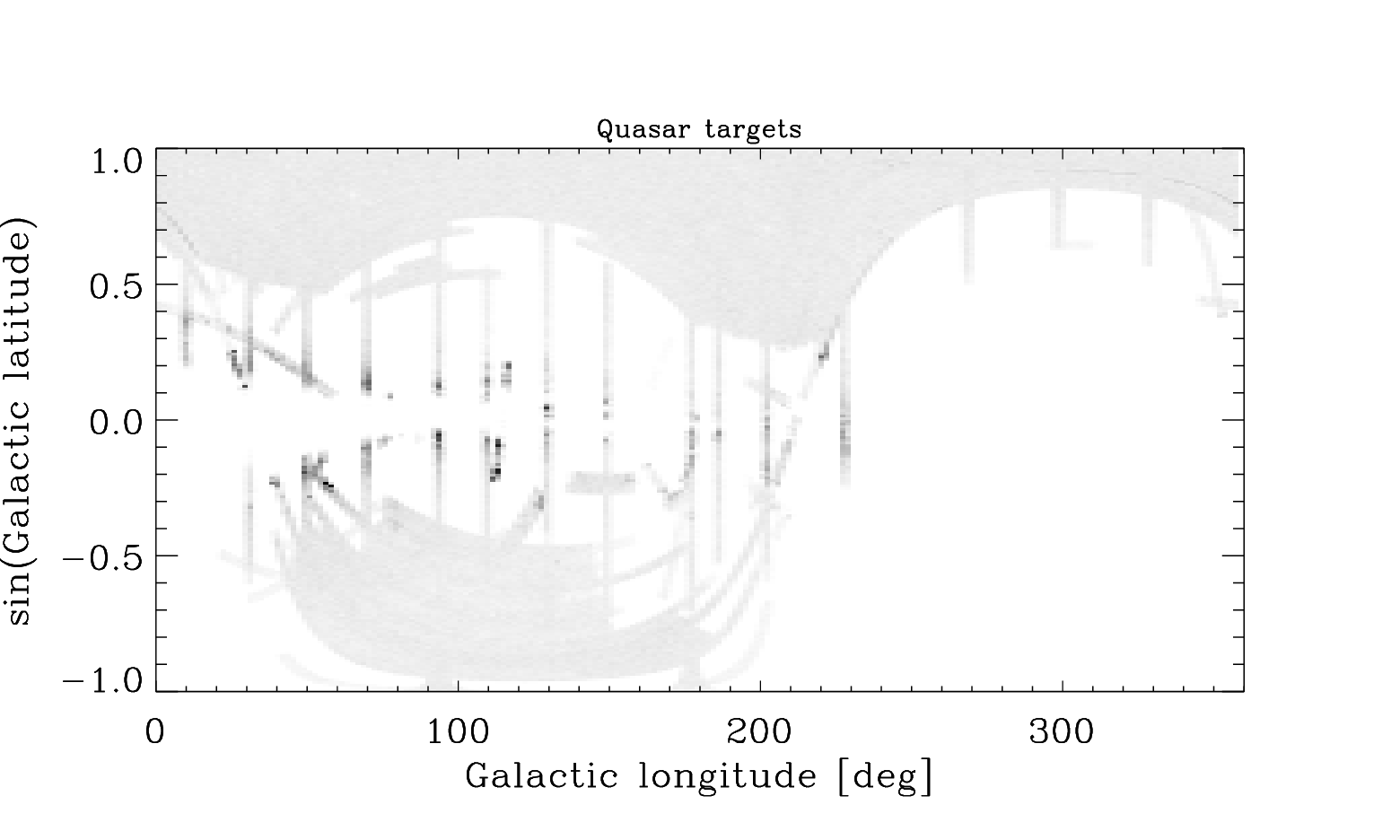}\\
\includegraphics[width=\textwidth]{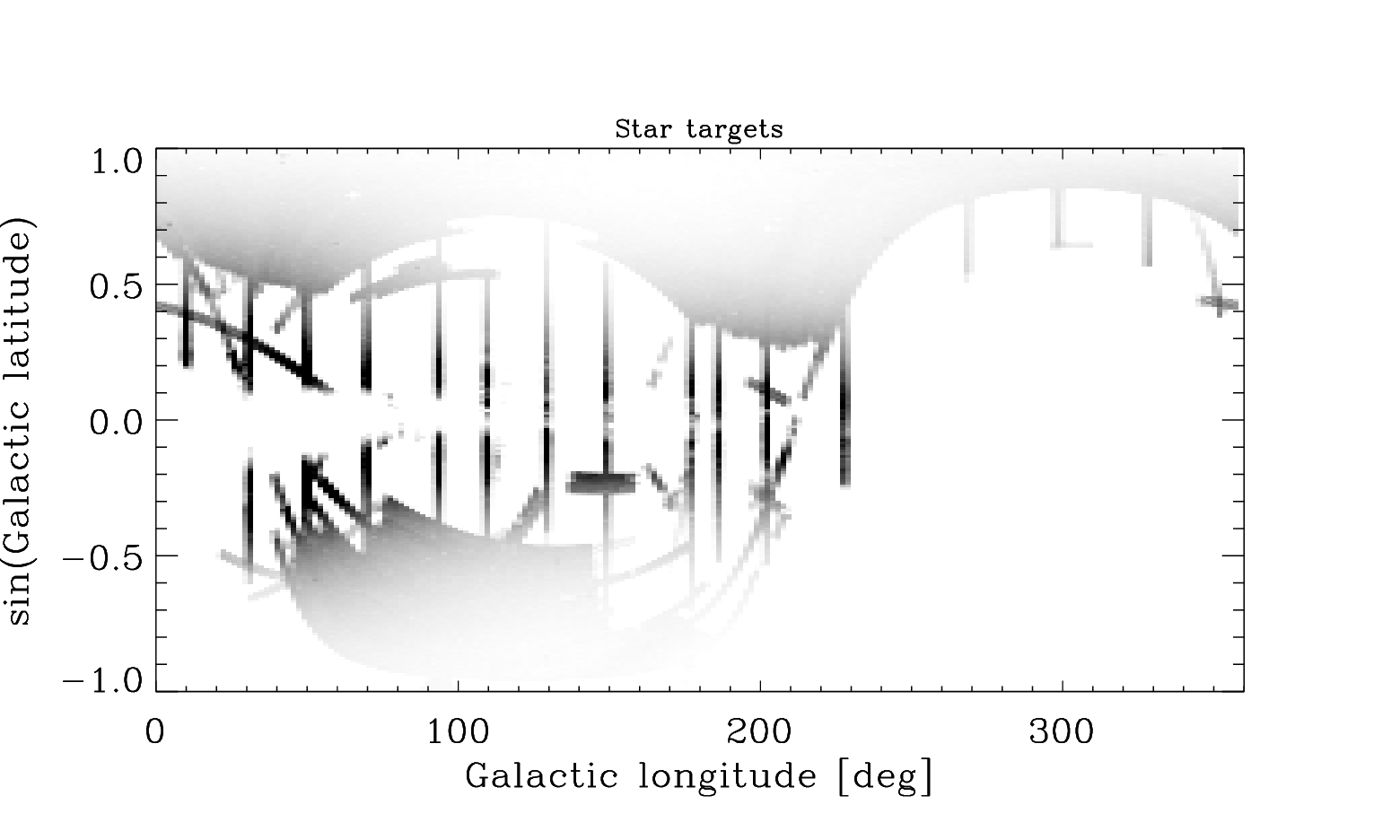}
\caption{Sky distribution of quasar ($P(\mathrm{quasar}) \geq 0.5$)
  and star ($P(\mathrm{star}) \geq 0.95$) targets. The contrast for
  the star targets is saturated near the Galactic
  plane.}\label{fig:radec}
\end{figure}

While there are many detailed considerations in constructing a
specific quasar sample from the \sdssxdqso\ catalog, for the purposes
of testing we use an arbitrarily chosen cut of $\pqso > 0.5$ to define
a \qsoc\ sample and $\pstar > 0.95$ to define a \starc\ sample. We
show the sky distribution of quasars and stars in
\figurename~\ref{fig:radec}. As expected for a quasar catalog, the
distribution is mostly flat, with only minor increases in density near
the Galactic plane where the density of stars and hence potential
contaminants is extremely high. This expected behavior is satisfying
since our model did not include a prior depending on Galactic
longitude and latitude. The stars selected by the star cut do show the
expected strong dependence on Galactic latitude. If we had included a
better prior depending on the Galactic latitude and longitude of
targets, there would have been less quasar targets in regions closer
to the Galactic plane, since objects would then get a higher
probability of being stars. Quasar are targeted predominantly at high
Galactic latitude, such that the inclusion of such a prior does not
lead to great improvements to the targeting efficiency; therefore, we
have not pursued this here.

\begin{figure}
\includegraphics[width=0.4\textheight]{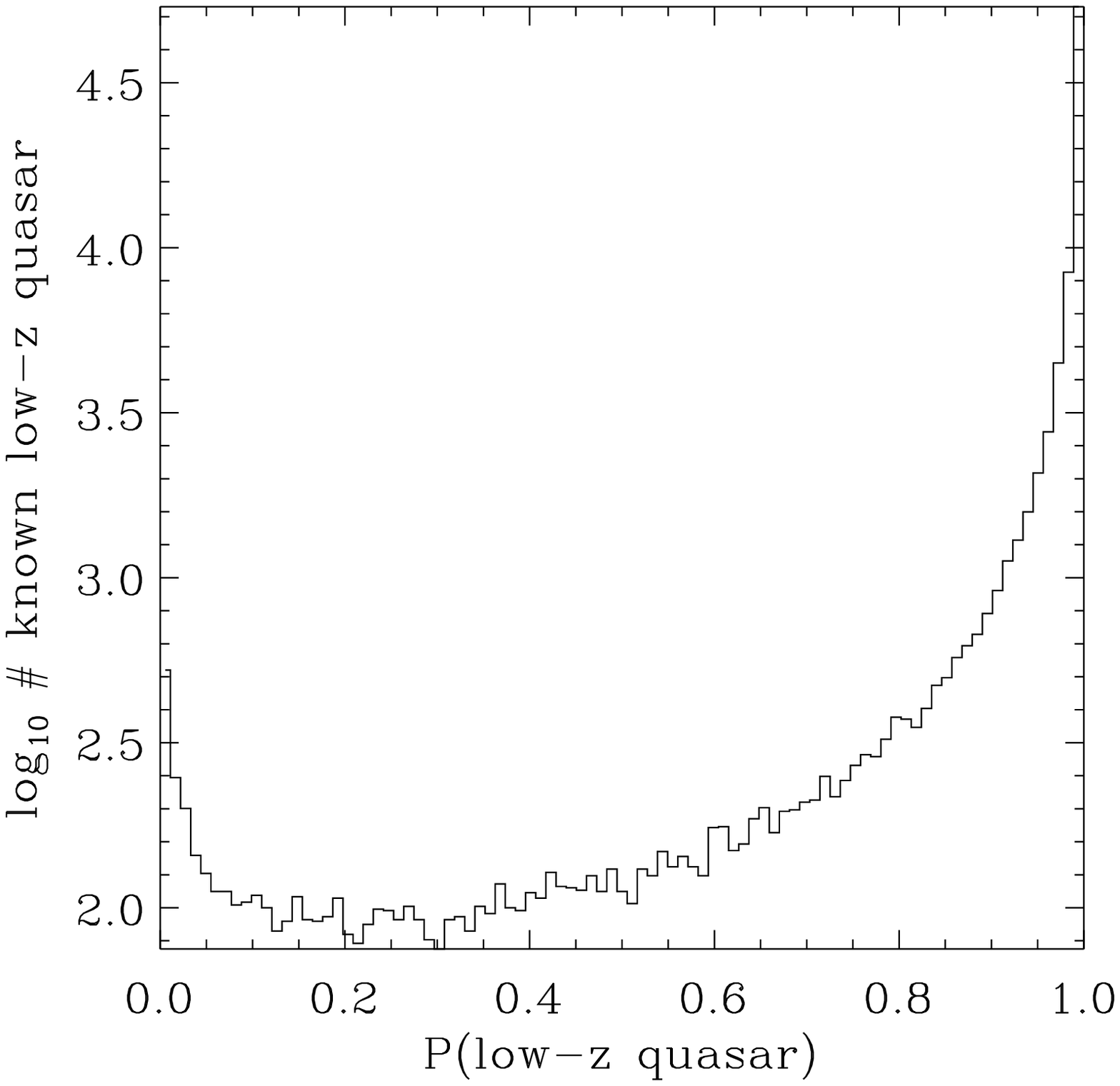}
\includegraphics[width=0.4\textheight]{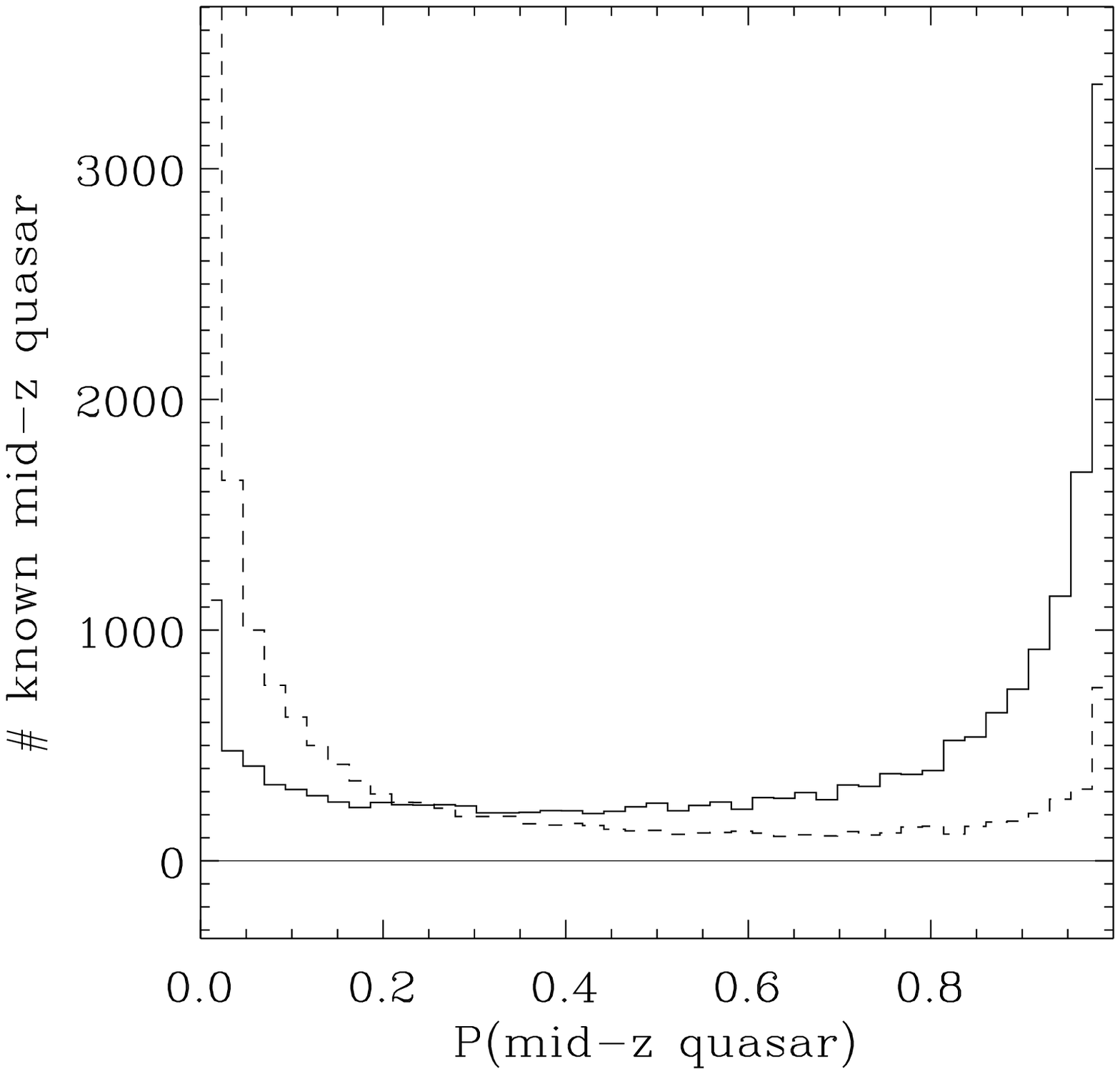}\\
\includegraphics[width=0.4\textheight]{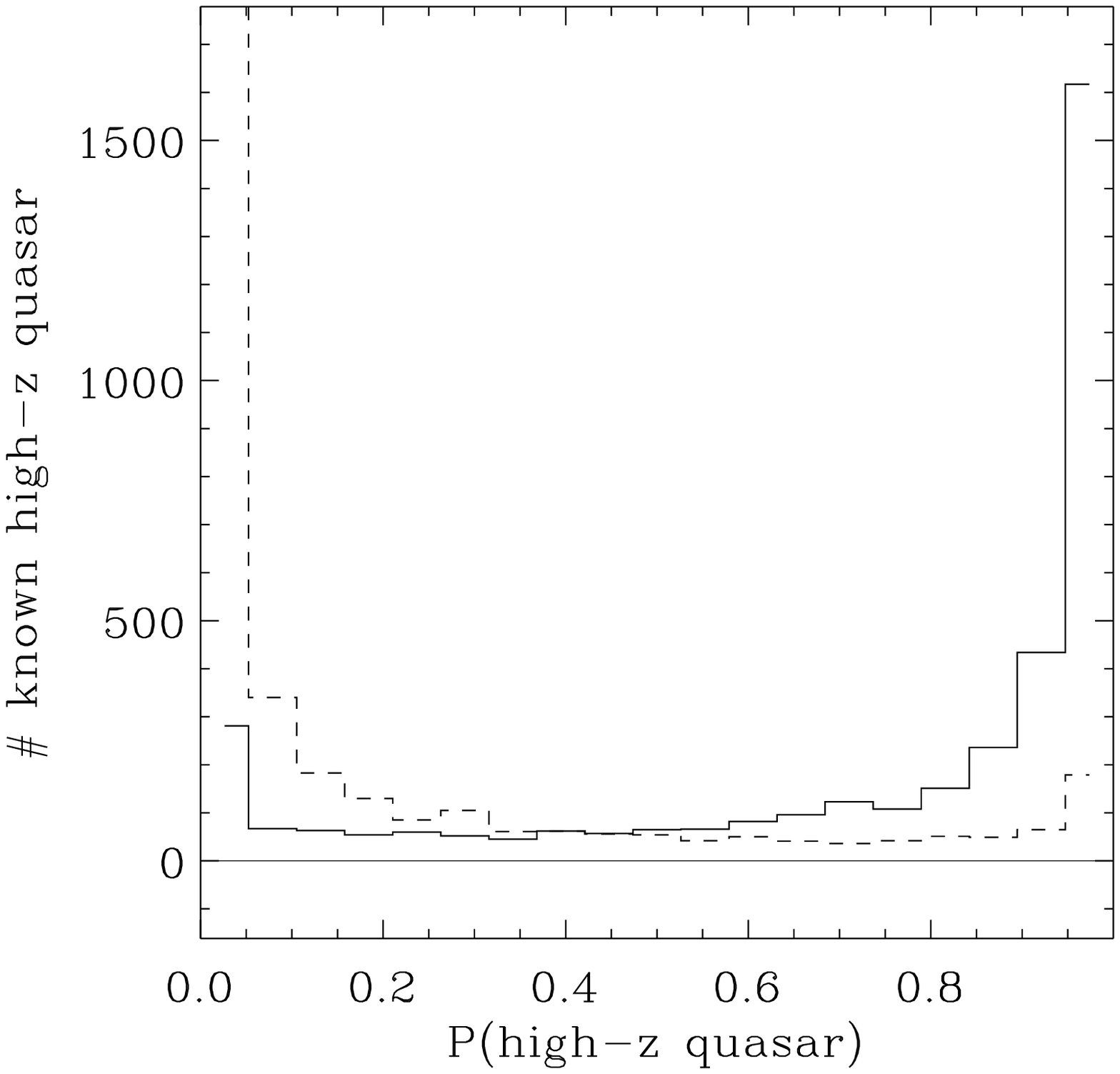}
\includegraphics[width=0.4\textheight]{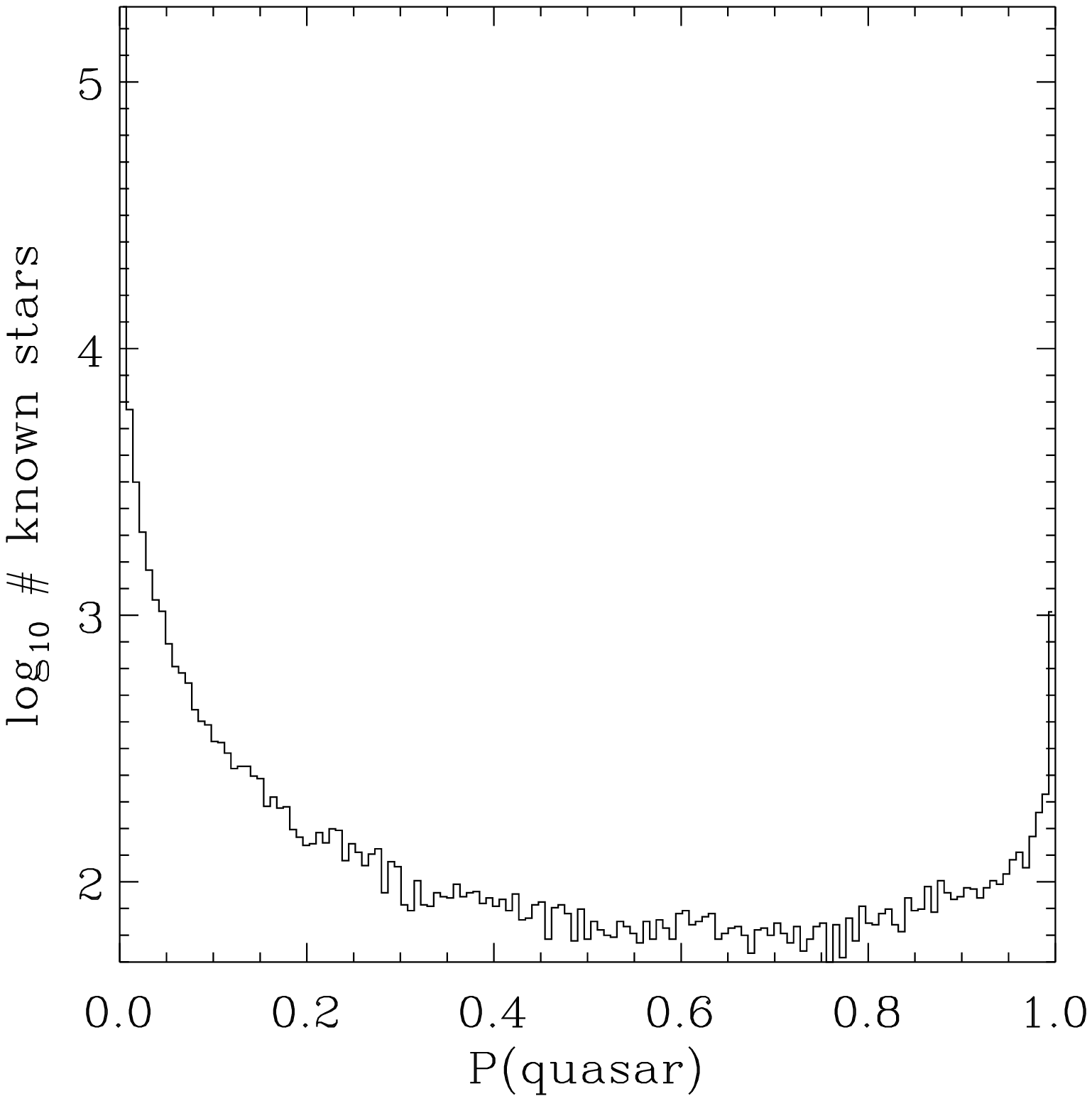}
\caption{\xdqso\ probabilities of known quasars and stars. The dashed
  line in certain panels is the distribution of $P(\mathrm{star})$ for
  the same objects. Note that the top left and bottom right panels
  show the \emph{logarithm} of the histogram.}\label{fig:known}
\end{figure}

In \figurename~\ref{fig:known} we show the \xdqso\ probabilities of
known, \ie, spectroscopically confirmed, quasars and stars. These come
from a compilation of known quasars and stars from various surveys,
including some early \boss\ data. We split the sample of known quasars
into our three redshift bins. We see that both known low-redshift
quasars and known stars are assigned very high probabilities for the
correct class, with no significant populations of low-probability
objects for a given class. The distribution of mid-redshift
probabilities for known mid-redshift quasars shows that most of these
quasars have high correct probabilities, but that there is a
population of objects around $P \approx 0$. Inspection of the
probabilities of being stars for these objects shows that not all of
them get high star probabilities, such that at least some of these are
`misclassified' into the wrong redshift range. Most of the known
high-redshift quasars are assigned high high-redshift probabilities,
with only a small population being `misclassified' as stars. The
color-color distributions of `misclassified' quasars show that many of
these are $z \approx 2.8$ quasars that overlap the stellar locus, such
that it is unsurprising that color selection fails. The sky
distribution of `misclassified' quasars shows that many of these
objects lie in \sdss\ Stripe-82, where deeper photometry for quasar
selection is available than the single-epoch fluxes used in this
comparison. These Stripe-82 quasars were probably targeted based on
higher signal-to-noise co-added photometry, whereby they could be
better distinguished from stars.

\begin{figure}
\includegraphics[width=0.5\textwidth]{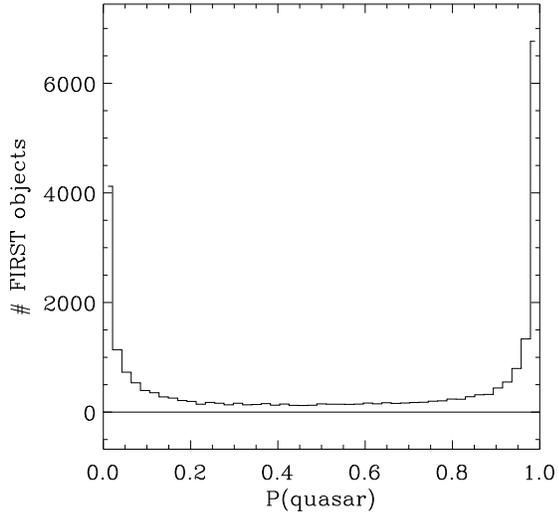}
\caption{\xdqso\ probabilities of \first\ sources.}\label{fig:first}
\end{figure}

We have cross-correlated our catalog with the \first\ point-source
catalog \citep{Becker95a} with a matching radius of 0$''$.5
\citep{Mcgreer09a}. We expect that most of these objects---optically
unresolved point sources with compact radio morphology---are quasars,
with only a small fraction being compact radio galaxies
\citep{Mcgreer09a}. In \figurename~\ref{fig:first}, we show the quasar
probabilities for the matching \first\ objects. The majority of these
sources have high quasar probabilities. Those that do not have high
quasar probabilities either overlap with the stellar locus or have
very unusual colors. An inspection of \first\ sources targeted as
quasars by the \boss\ shows that \first -selected quasars with small
\xdqso\ quasar probabilities are significantly redder than quasars
with high \xdqso\ probabilities; they are missed by the \xdqso\ method
because it is trained on optically-selected quasars which are
typically bluer than radio-selected quasars.

\begin{figure}
\includegraphics[width=0.5\textwidth]{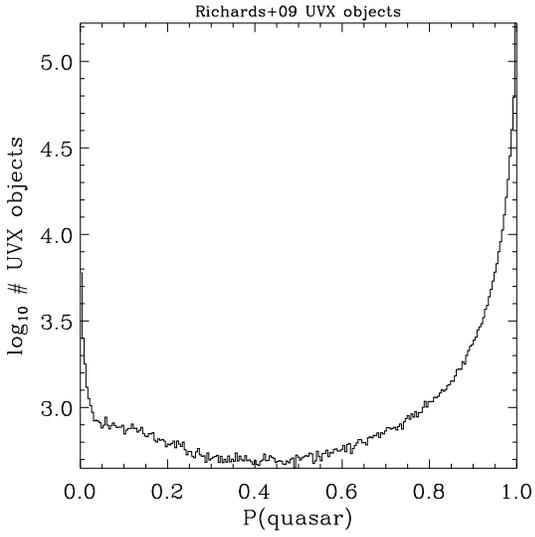}
\caption{\xdqso\ probabilities of \flag{UVX} sources in the \citet{Richards09a} photometric quasar catalog.}\label{fig:richards}
\end{figure}

Finally, we show in \figurename~\ref{fig:richards} the quasar
probabilities for the \flag{UVX} objects in the \citet{Richards09a}
photometric quasar catalog. We expect $\approx 95$ percent of these
objects to be $z < 2.4$ quasars. Only a small fraction of objects are
assigned low quasar probabilities, which is unsurprising given the 95
percent efficiency of the photometric catalog. We discuss the
\citet{Richards09a} technique further in \sectionname~\ref{sec:kde} in
the specific context of target selection.

\section{\boss\ quasar target selection}\label{sec:boss}

As discussed in the introduction, the \boss\ aims to obtain spectra of
$\approx 160,000$ quasars in the redshift range $2.2 \leq z \leq 3.5$
to observe the baryon acoustic feature in the Lyman-$\alpha$ forest
and obtain a percent-level measurement of the angular diameter
distance to redshift 2.5. This requires observing at least 15 quasars
deg$^{-2}$ over the \boss\ survey area. Since only about 40 targets
deg$^{-2}$ will be allocated for quasar targets, target selection must
reach 50\,percent efficiency to achieve this goal. Achieving this
efficiency is complicated by the substantial overlap between the
quasar and stellar loci in color-space in the \boss\ redshift range
and the need to target to $g \approx 22$ mag where photometric
uncertainties are significant.

While a uniform target selection of the \boss\ quasars is unnecessary
for the cosmological distance measurement because the quasars are only
used to reveal the intervening Lyman-$\alpha$ forest, the
\boss\ quasar target selection reserves 20 targets deg$^{-2}$ for a
``CORE'' sample of objects that is selected by a single method over
the course of the survey, to allow statistical studies of quasar
properties. Since only single-epoch \sdss\ \ugriz\ fluxes are
available for the bulk of the \boss\ survey footprint, this
\core\ sample must be based on these single-epoch fluxes alone.

The \sdssxdqso\ catalog presented in \sectionname~\ref{sec:catalog}
allows such a uniform sample to be defined while maintaining a high
target efficiency. In practice we use the medium-redshift quasar
probabilities \flag{pqsomidz} (see
\tablename~\ref{table:catdescription}) and target all \flag{good}
objects with dereddened ($g < 22$ mag or $r < 21.85$ mag) and $i >
17.8$ mag---the \boss\ flux limits---with \flag{pqsomidz} larger than
a threshold set to give 20 or 40 targets deg$^{-2}$ over a certain
region (\eg, a part of the survey such as Stripe-82 or the entire
survey footprint).

Even though the \xdqso\ targeting technique was not used during the
first year of \boss\ data acquisition, we can nevertheless evaluate
the \xdqso\ performance by determining how many of the \xdqso\ quasar
targets are spectroscopically confirmed \boss\ quasars. This is a
conservative test as it presumes that the \boss\ quasar sample is
highly complete; any incompleteness would lead us to underestimate our
efficiency since there would then be real quasars selected by \xdqso\,
but which did not receive a \boss\ fiber. In
\figurename~\ref{fig:comparelike_efficiency} we show the
\xdqso\ targeting efficiency in an area from the \boss. This area
consists of \boss\ observations of Stripe-82. Since over eight epochs
are available for most sources in Stripe-82, \boss\ quasar target
selection was based on deeper co-added data and included variability
selection (\citealt{palanque11}). Therefore, it is highly complete and it is
therefore the best test of the \xdqso\ technique. We only use the
$\approx 205$ deg$^2$ region of Stripe-82 where at least 15 2.2 $\leq
z \leq 4$ quasars deg$^{-2}$ were observed by the \boss; on average 27
$z > 2.15$ quasars deg$^{-2}$ are available. This is close to the
total number of quasars in this redshift range expected based on
various quasar luminosity functions
\citep{Richards06a,Jiang06a,Hopkins07a}. We see that at 20 targets
deg$^{-2}$ the efficiency is about 50\,percent; the efficiency drops
rapidly at higher target densities.

A further test of the \xdqso\ technique was performed during the Fall
of 2010. Approximately 20 medium-redshift quasar targets (catalog tag
\flag{pqsomidz}) per square degree were observed as part of the
\boss\ in a region of $\approx 200$ deg$^2$ just north of
\sdss\ Stripe-82. As of 2010 November 22, 4,593 objects were targeted
and 2,194 of these were classified as 2.2 $\leq z \leq 3.5$ quasars,
corresponding to an efficiency of 48\,percent. In
\sectionname~\ref{sec:compare} we compare the \xdqso\ selection to
other targeting algorithms which were tested during the first year of
\boss\ commissioning.

\begin{figure}
\includegraphics[width=0.5\textwidth]{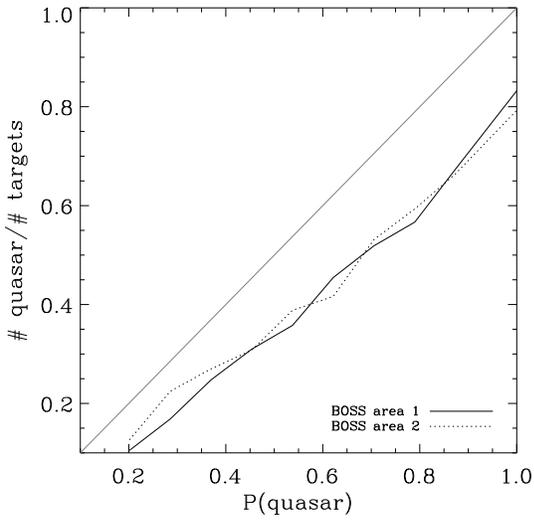}
\caption{Target efficiency of confirmed 2.2 $\leq z \leq 3.5$ quasars
  as a function of the P(2.2 $\leq z \leq 3.5$ quasar) of the targets
  for two areas from the \boss\ quasar survey: area 1 consists of
  \boss\ observations of Stripe-82 sources and area 2 is an $\approx
  150$ deg$^2$ area around $\alpha_{\mathrm{J2000}} = 120\degree,
  \delta_{\mathrm{J2000}} = 45\degree$.}\label{fig:consistency}
\end{figure}

We can also use the early \boss\ data to evaluate whether the
\xdqso\ probabilities given to objects are correct in the ensemble
sense. That is, we can bin objects in $P(2.2 \leq z \leq 3.5
\ \mathrm{quasar}| \setoffj)$ and evaluate the fraction of these
objects that are quasars. This test again assumes that the test sample
of quasars is highly complete. The results from this exercise are
shown in \figurename~\ref{fig:consistency}. We see that the quasar
probabilities appear to be overestimated. Several shortcomings of the
model for the quasar and stellar populations could be responsible for
this result. Since we used relatively small samples of stars in each
bin compared to the number of quasars---the proportion being around
one to one rather than reflecting the true number densities according
to which stars outnumber quasars by a factor of $\approx 100$ down to
$g \approx 22$ mag---we probe much lower density regions for the
quasars than we can for the stars. Thus, in low stellar-density
regions, the stellar density is likely to be underestimated with
respect to the quasar densities, leading to overestimated quasar
probabilities in these regions. This is unlikely to change the
\emph{ranking} of quasar targets significantly---most targets will be
affected in the same way. Thus, quasar targeting using the
\sdssxdqso\ catalog is robust with respect to this modeling
problem. However, this aspect of our analysis does merit caution when
taking the specific value of the quasar probabilities serious in an
analysis of quasar properties. Our model also does not include
variations in the stellar number density with Galactic latitude and
longitude, so the stellar density will be underestimated in regions
closer to the Galactic plane. Several of the extensions described in
\sectionname~\ref{sec:extensions} can be used to mitigate these
problems with the catalog.

\section{Discussion}\label{sec:discussion}

\subsection{Comparison with other target selection methods}\label{sec:compare}

Several sophisticated methods already exist to select quasars from
photometric \sdss\ data, especially for the mid-redshift range
targeted in the \boss. 

\subsubsection{Richards et al. (2009a) NBC-KDE}\label{sec:kde}

The NBC-KDE photometric quasar catalog of \citet{Richards09a} also
uses density estimation to calculate relative number densities of
stars and quasars to determine quasar probabilities. It differs from
the technique described in this paper in that it (a) uses colors
rather than relative fluxes in \eqnname~(\ref{eq:exdmodel}), (b)
assumes a flat probability distribution for $p(\fluxi | i \in
\mathrm{class})$, (c) uses a simple kernel density estimation
technique of the \emph{observed} distribution of colors, thus ignoring
the flux uncertainties of the training data, and (d) ignores the flux
uncertainties of the data at evaluation. Since the catalog is limited
to $i < 21.3$ mag, ignoring the flux uncertainties is mostly harmless,
although medium and high redshift quasars have very low $u$- and
$g$-band fluxes with large associated uncertainties. This assumption
also makes it harder to extend the catalog to fainter magnitudes. The
NBC-KDE catalog also does not compute exact kernel-density likelihoods
for all of the \sdss\ data because of computational limitations. The
\exd\ technique permits these likelihoods to be computed for all of
the data (see above).

We compare the \xdqso\ catalog with the NBC-KDE catalog by (a)
limiting the \xdqso\ catalog to the \sdss\ DR6 footprint, since the
NBC-KDE catalog only covers this area, (b) limiting the
\xdqso\ catalog to $i < 21.3$ mag, and (c) limiting the NBC-KDE
catalog to $i \geq 17.75$, the bright limit of the \xdqso\ catalog. We
select those targets in the NBC-KDE catalog with \flag{good} $\geq 0$
and then select equal numbers of targets from the \flag{good} entries
in the \xdqso\ catalog in the three shared target classes of the two
catalogs: \flag{lowz}, \flag{midz}, and \flag{highz}. Note that the
\flag{good} flags in the two catalogs are different. The results of
this comparison are given in \tablename~\ref{table:kde}. The confirmed
quasars in this \tablename\ come from a compilation of all
spectroscopically confirmed quasars. The bulk of these are
\sdss\ quasars that were selected to be brighter than $i = 20.2$ mag.

We stress that this comparison is significantly biased in favor of the
NBC-KDE catalog: the NBC-catalog was allowed to set the target
threshold and its \flag{good} flag includes cuts on Galactic latitude
that we did not apply to the \xdqso\ catalog. Nevertheless, we see
that the two catalogs perform similarly for low-redshift quasars, and
that the \xdqso\ technique performs better in the mid-redshift range
where the quasar and stellar loci overlap and photometric target
selection is difficult. 

The NBC-KDE catalog performs better at high-redshift than the
\xdqso\ catalog. This is not unexpected as we only used 150 deg$^2$ of
stellar data to train the \xdqso\ algorithm. The red-star
stellar-locus outliers that contaminate $z > 3.5$ quasar selection are
very rare on the sky: If we assume that they are ten times more
abundant than $z > 3$ quasars we expect our stellar training set to
only contain about 2,000 of these. This is not enough to model their
color distribution since they are spread over all magnitude bins. In
the following \sectionname\ we discuss improvements to the
\xdqso\ technique that can improve high-redshift quasar selection.

\subsubsection{Yeche et al. 2009's neural network}\label{sec:nn}

The quasar-selection technique of \citet{Yeche09a} uses a neural
network (NN) to select quasars in the \boss\ mid-redshift range. This
technique uses as the input variables the four colors $u-g$, $g-r$,
$r-i$, and $i-z$, the $g$-band magnitude, and the five magnitude
uncertainties. These 10 variables are propagated through a simple NN
that is trained on sets of known quasars and point-like objects from
\sdss\ DR7 to obtain an output parameter \flag{xnn} on which potential
targets can be ranked. A similar NN is used to find a photometric
redshift \flag{znn} for the quasar targets. To select targets in the
mid-redshift range they recommend the cuts \flag{znn} $> 2$, $u-g >
0.4$, and $g-i < 2$. While this technique uses the photometric
uncertainties, it does this in a black-box manner that is quite
different from the probabilistic way in which \exd\ treats the
uncertainties; the NN approach cannot work well when the test set is
noisier than the training set.

We use the version of the NN technique that is part of the official
\boss\ quasar selection framework to compare the \xdqso\ technique and
the NN technique at different target densities using early
\boss\ data. For the NN we first make the cuts listed above and then
rank the targets on \flag{xnn} until we reach the desired target
density. For \xdqso\ we rank the \flag{good} targets based on
\flag{pqsomidz} (see \tablename~\ref{table:catdescription}). The
results from this comparison for \boss\ year-one observations of
Stripe-82 are shown in
\figurename~\ref{fig:comparelike_efficiency}. We see that the neural
network performs significantly worse than \xdqso\ at all target
densities.

\subsubsection{Kirkpatrick et al.'s \Likelihood}\label{sec:like}

The \Likelihood\ technique of J.~A.~Kirkpatrick et al. (2011, in
preparation) uses an approach similar to the NBC-KDE catalog. Rather
than colors it uses fluxes, such that the apparent magnitude factor
also used by \xdqso\ is automatically taken into account, and rather
than a kernel with an optimized bandwidth, \Likelihood\ uses a delta
function---corresponding to a model of the underlying density
consisting of delta functions centered at the location of each object
in the respective training set. These delta functions are convolved
with the flux-uncertainties at evaluation such that a smooth density
is obtained nevertheless. The \Likelihood\ technique uses a similar
stellar training data as the \xdqso\ catalog. The quasar training set
is also modeled by re-sampling the quasar luminosity function, but
these are essentially the same set of quasars used to train the
\xdqso\ technique. Rather than simulating relative fluxes only, a full
quasar catalog with five-dimensional \ugriz\ fluxes in the relevant
magnitude range is simulated.

The \Likelihood\ technique uses the flux uncertainties to smooth the
discrete underlying delta-function distribution of its training
sets. However, since it does not use an optimized bandwidth, there is
the danger that the density might be undersmoothed in certain
regions. At the faint end, where the training set has a significant
contribution from the uncertainties, the \Likelihood\ method also
effectively convolves with the uncertainties twice. This follows
because the training set is a sample from the observed distribution of
fluxes, rather than the true underlying distribution; the former is
the latter convolved with the uncertainty distribution. Finally, as
with the NBC-KDE catalog, the calculation of the quasar and star
probabilities is extremely slow compared to \xdqso. In our comparison
below we use cached versions of the \Likelihood\ catalog created by
the \boss\ target selection team.

\begin{figure}
  \includegraphics[width=0.5\textwidth]{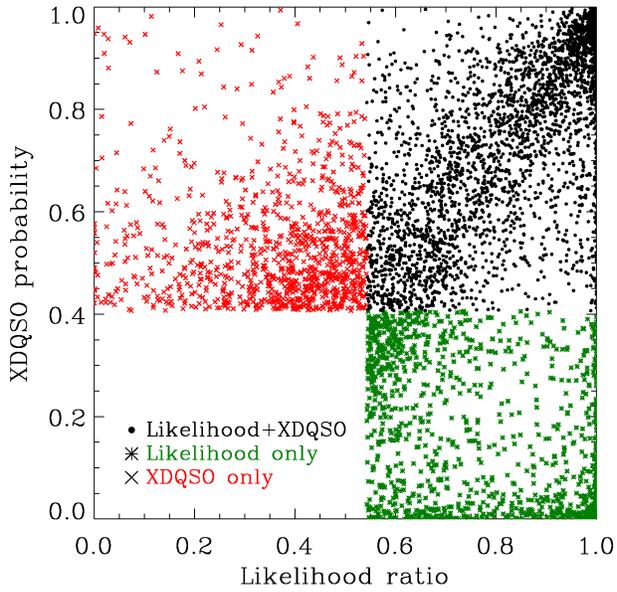}
  \caption{Comparison of the mid-redshift (2.2 $\leq z \leq 3.5$)
    quasar probability for the \xdqso\ and \Likelihood\ methods at 20
    targets deg$^{-2}$ for sources in \sdss\ Stripe-82. Targets
    selected by both methods are on the upper right, \Likelihood -only
    targets are on the lower right, and targets exclusive to
    \xdqso\ are on the upper left.}\label{fig:comparelike_ts}
\end{figure}

In \figurename~\ref{fig:comparelike_ts} we first compare the
\xdqso\ quasar probabilities in the \boss\ mid-redshift range with the
probabilities calculated using the \Likelihood\ method for sources in
Stripe-82. We select targets at 20 targets deg$^{-2}$ and show those
targets selected by both techniques or by only one of the
techniques. While many of the targets cluster around the one-to-one
line, there is a distinct population of targets that receive high
\Likelihood\ probabilities, yet low \xdqso\ probabilities. A similar
population of high \xdqso -only targets is absent, indicating that the
\Likelihood\ method indeed has problems with undersmoothing.

\begin{figure}
\includegraphics[width=0.5\textwidth]{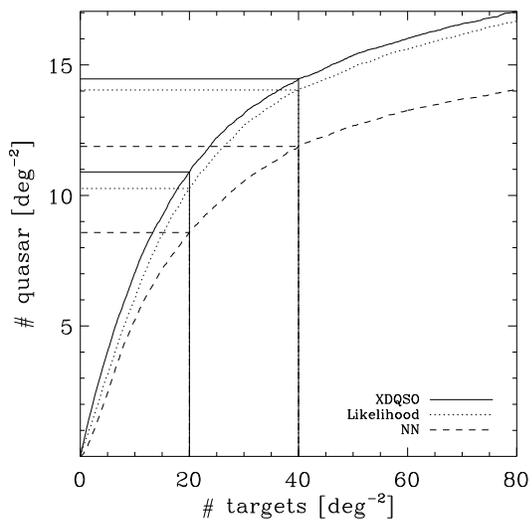}\\
\includegraphics[width=0.5\textwidth]{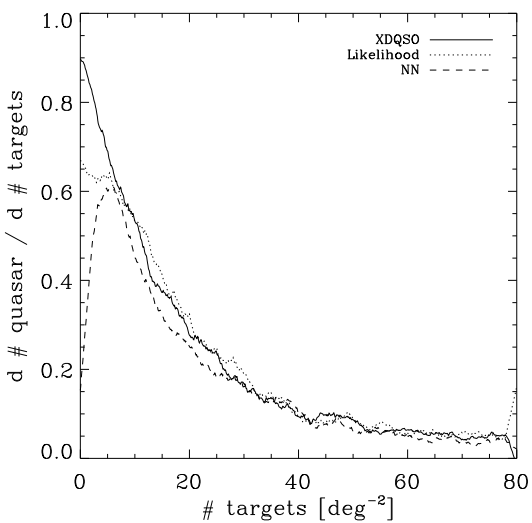}
\caption{Number of confirmed 2.2 $\leq z \leq 3.5$ quasars as a
  function of the target density for different target selection
  methods used in the \boss\ (\xdqso : this paper; \Likelihood:
  J.~A.~Kirkpatrick et al., 2011, in preparation; NN:
  \citealt{Yeche09a}). Input target densities relevant to the
  \boss\ target selection are highlighted. This uses
  \boss\ observations of sources in
  \sdss\ Stripe-82.}\label{fig:comparelike_efficiency}
\end{figure}

In \figurename~\ref{fig:comparelike_efficiency} we compare the
\xdqso\ catalog with the \Likelihood\ technique at different target
densities. The two methods perform similarly, with a slightly better
performance for the \xdqso\ catalog over the whole range.

A further comparison between the \xdqso\ and the \Likelihood\ methods
for medium-redshift quasar selection was performed during the Fall of
2010 using \boss\ observations of an $\approx 200$ deg$^2$ region just
north of Stripe-82 (see \sectionname~\ref{sec:boss}). Both methods
were given similar target densities to allow for a direct
comparison. As of 2010 November 22, \xdqso\ was given 4,593 targets,
while \Likelihood\ received 4,853 targets. Of the 4,593
\xdqso\ targets 2,194 were classified as 2.2 $\le z \le 3.5$ quasars,
while of the 4,853 \Likelihood\ targets only 2,056 turned out to be
medium-redshift quasars. From this test and that in
\figurename~\ref{fig:comparelike_efficiency} we conclude that the
\xdqso\ technique's performance is about 10\,percent better than that
of the \Likelihood\ method for the selection of medium-redshift
quasars.

\subsection{Extensions}\label{sec:extensions}

One of the main advantages of the general target selection technique
and in particular of the specific \xdqso\ implementation described in
this \documentname\ is that it can easily be extended in a variety of
ways. These extensions can be changes to the model---such as different
number count priors, the addition to the model of other data such as
\nir\ or \uv\ observations---or the combination of the flux-based
selection described here with target selection based on variability
information. All of these extensions are described briefly here.

Most of these extensions involve only some of the factors in \eqnname
s~(\ref{eq:bayes}) and (\ref{eq:exdmodel}). Since we provide all of
these factors separately in the \sdssxdqso\ catalog, extensions that
do not change all of the factors can use some of the information in
the catalog. For example, extensions that only change the number count
priors, \eg, $p(\fluxi | i \in \mathrm{\starc})$ or $P(i \in
\mathrm{\starc})$, will not need to re-calculate the relative-flux
likelihoods---the most expensive of the factors in
\eqnname~(\ref{eq:exdmodel})--but can instead re-use the catalog
values.

\subsubsection{Additional \nir\ or \uv\ data}

Quasar selection from broad-band fluxes can be improved by the
addition of \nir\ or \uv\ fluxes to the optical fluxes used to create
the \sdssxdqso\ catalog
\citep[\eg,][]{Warren00a,Maddox08a,Richards09b,Jiminez09a,Worseck10a}. For
example, the \emph{Galaxy Evolution Explorer} (\galex;
\citealt{Martin05a}) has completed a near full-sky survey in the
ultraviolet (UV) and the \emph{UKIRT Infrared Deep Sky Survey}
(\ukidss; \citealt{Lawrence07a}) is observing a large part of the
\sdss\ footprint in the \nir. However, this situation is complicated
by the fact that these surveys are generally shallower than the
optical fluxes available from \sdss, such that most of the objects in
the \sdss\ catalog are not detected at high significance in these
surveys.

Since these surveys have point-spread functions that are worse than
that of the \sdss, low signal-to-noise measurements of the \nir\ and
\uv\ fluxes of many of the objects in the \sdss\ catalog can be
obtained by forced photometry of the \galex\ or \ukidss\ images at the
\sdss\ positions, which can be regarded as truth because of the
difference in resolution. Because there are gaps in these surveys, it
will still be the case that some objects in both the training set and
the evaluation set will not have measured \nir\ or \uv\ fluxes.

To use these low signal-to-noise or non-existent fluxes, it is
necessary to employ a classifier than can handle the data
uncertainties correctly and that can handle missing data, both for
training and for evaluation of the quasar probabilities. The
\xdqso\ method described in this \documentname\ is the only technique
to date that can do this task naturally and it is therefore the only
method available that---in a straightforward way---can use all of the
available information for an object to classify it as a star or
quasar.

The procedure to use the \nir\ or \uv\ data is the following: We
combine the optical \ugriz\ fluxes in our training sets of quasars and
stars with the available \nir\ or \uv\ data. We then train the
\exd\ model for the relative fluxes on the combined relative optical
plus \nir\ or \uv\ fluxes, using missing relative fluxes---\eg, by
using a very large uncertainty variance for these---where \nir\ or
\uv\ fluxes do not exist. Then for those objects in the evaluation set
(\eg, the \sdss\ DR8 catalog) with measured \nir\ or \uv\ fluxes, we
replace the relative-flux likelihoods based on \ugriz\ fluxes in the
\sdssxdqso\ catalog with those likelihoods calculated using optical
plus \nir\ or \uv\ fluxes. The apparent \iband\ magnitude factors do
not change, as these are functions of $i$ alone.

This selection can then be used to select targets based on combined
\galex\ and \sdss\ data, or to use \galex\ fluxes instead of the
highly discriminating $u$-band fluxes for surveys such as
\panstarrs\ that do not have a $u$ band (\citealt{Jiminez09a}; J.~Bovy
\etal, 2011, in preparation). It can also be used to select targets
based on \sdss\ and \ukidss\ fluxes where the \sdss\ and
\ukidss\ overlap, or to use only some of the redder optical bands plus
\nir\ data to avoid potential biases due to the inclusion of bluer
optical bands \citep{Worseck10a}.

\subsubsection{Variability}

With the opening of the time-domain in the near-future with surveys
such as \panstarrs\ \citep{Kaiser02a,Morgan08a},
\lsst\ \citep{Ivezic08a,Abell09a}, \skymapper\ \citep{Keller07a}, and
\wfirst, the selection of quasars based on their variability has
recently received some attention
\citep{Kozlowski10a,Schmidt10a,Butler10a,Macleod10a}. Some of these
techniques currently amount to drawing the equivalent of
``color-boxes'' in variability space to select quasars
\citep{Schmidt10a,Macleod10a}, while others perform more sophisticated
model selection \citep{Butler10a}. However, it is clear that the
variability technique could be brought under the umbrella of
probabilistic target selection by doing density estimation in the
space of variability attributes (such as parameters of the structure
function) in a similar manner as we did in flux space in this
\documentname.

The variability data can be used to form a variability-likelihood
similar to the relative-flux likelihood used in
\eqnname~(\ref{eq:exdmodel}). If we assume that the variability of a
quasar is independent of its (relative) flux---not necessarily a good
assumption---we can combine the relative-flux and variability
likelihoods by simply multiplying the quantities. Alternatively, we
can perform density estimation in the combined space of relative
fluxes and variability parameters, and use the combined likelihood
instead of the relative-flux likelihood in
\eqnname~(\ref{eq:exdmodel})---this will capture any (relative) flux
dependence of the variability of quasars. Combining variability and
flux information is our best hope to create extremely efficient
quasars surveys in the future that are free from the biases associated
with color or variability selection alone.

\subsubsection{Other extensions}

All of the factors in \eqnname~(\ref{eq:exdmodel}) can be improved
upon using existing data and the \xdqso\ target selection
technique. For example, we used a star count model that is not a
function of Galactic coordinates, but we know that the number density
of stars is a strong function of Galactic latitude and Galactic
longitude. Both the total stellar number counts in
\tablename~\ref{table:numcounts} and the stellar number counts as a
function of apparent magnitude in \figurename~\ref{fig:numcounts}
could be re-calculated using models for the stars counts in different
directions. However, this will not lead to significant changes in the
calculated quasar and star probabilities, as the distribution on the
celestial sphere of photometrically selected quasars and stars already
follows the expected celestial distributions of quasars and stars
quite well (see above and \figurename~\ref{fig:radec}).

We also used relative-flux distributions for the stars that do not
depend on the celestial location of an object. However, the colors of
stars do change with proximity to the plane of the Galaxy. Therefore,
we could have used a model that reconstructs the relative flux
distribution of stars as a function of the position on the sky. Such a
model is hard to produce from the current data, if we do not want to
rely on theoretical models for this dependence, since we have no way
\emph{a priori} to separate quasars from stars all over the sky. What
we can do is model the relative-flux distribution of all point sources
as a function of position on the sky, from the single-epoch
\sdss\ fluxes available on the \sdss\ footprint. In order to use the
noisy single-epoch fluxes properly, it is again necessary to use a
technique such as \xdqso\ that uses the uncertainties correctly. Using
the model for the quasar fluxes that we have been using in this
\documentname\ we can calculate quasar probabilities by using the
quasar model in the numerator of \eqnname~(\ref{eq:normalize}) and the
model for all point sources in the denominator. However, it is then
possible for the probability to exceed one, since the probability is
not explicitly normalized.

We could also use a training set consisting of point sources over the
entire \sdss\ footprint rather than the 150 deg$^2$ area of Stripe-82
to increase our sampling of rare stellar-locus outliers. As mentioned
above, red stellar-locus outliers outnumber high-redshift quasars and
are a significant contaminant for high-redshift quasar selection. Our
current training set does not contain enough of these red stellar
outliers to model their relative flux distribution. By extending our
stellar training sample to the full $\approx 10^4$ deg$^2$
\sdss\ footprint we would have about 100 times more stellar outliers
and we could model their color distribution. This would significantly
improve high-redshift quasar target selection.

\subsection{Decision theory}

Inference (Bayesian or frequentist) only assigns relative
\emph{probabilities} on outcomes or models. It does not tell you what
to do; this also involves your capability, \eg, the number of fibers
you can use per square degree and their minimal spacing, and your
utility, \eg, signal-to-noise in the Lyman-$\alpha$ forest for the
\boss\ quasar baryon-acoustic-feature detection.  The decision to cut
metal in the spectroscopic plate (in the case of the \boss) is a hard
decision that cannot, in the end, be probabilistic.  The probabilistic
results permit calculations of expected utility, which can be used to
set decisions, such as whether to prefer quasar targets to
luminous-red-galaxy targets in the case of fiber collisions in the
\boss.  The \boss\ targeting decisions are unlikely to be made this
way, but in situations in which inference is accurate, and (this is
rarer) utility can be calculated, it should improve overall efficiency
and capability.

To be more specific, different quasars with different properties will
have different value for any individual scientific project.  For
example, for the Lyman-$\alpha$ baryon acoustic feature analysis the
\boss\ is expected to obtain more information from quasars that are
brighter, and quasars in certain redshift ranges.  At the same time,
different quasars with different properties will be easier and harder
to select, because of their photometric differences and similarities
with stellar sources or other contaminants.  It will usually be a
frequent occurrence that one will decide, when comparing two sources,
to target the object less likely to be a quasar because the lower
probability is compensated by greater value to the survey if the
object is indeed a quasar.

Additionally, some classes of contaminants can have higher utility
than others, and one might decide to preferentially target quasars
that are more likely to be confused with useful contaminants. For
example, some contaminants for the \boss\ quasar target selection are
BHB stars, which are very useful for studies of Galactic structure
\citep[\eg,][]{Xue08a}, and hypervelocity stars. Future surveys might
decide to combine different science goals with different classes of
targets, and probability-based utilities are then essential to most
efficiently achieve all of the science goals.

\section{Conclusion}\label{sec:conclusion}

The \xdqso\ method described in this \documentname\ is a new general
target selection technique that can be used to design highly efficient
surveys for faint objects that are hard to distinguish from
contaminants based on their measured attributes. This high efficiency
is achieved through the use of the extreme-deconvolution
density-estimation to classify objects based on the number densities
of desired objects and contaminants in attribute space. We used this
approach to create an input catalog for \boss\ quasar target
selection, which aims to target $\approx 160,000$ $2.2 \leq z \leq
3.5$ quasars over $\approx 10,000$ deg$^2$ at 50-percent efficiency
down to $g \approx 22$ mag. Quasar colors in this redshift range
overlap strongly with stellar colors and classification is made even
more difficult by the large photometric uncertainties at these faint
magnitudes. We demonstrated that at the 20 targets deg$^{-2}$ required
for the \boss\ `CORE' sample the \xdqso\ technique is indeed
approximately 50\,percent efficient and that it outperforms all other
medium-redshift quasar target selection techniques.

We release the full \ncatentries-object \sdssxdqso\ quasar targeting
catalog containing star and low-, medium- and high-redshift quasar
probabilities for all primary objects in the $17.75 \leq i < 22.45$
dereddened \iband\ magnitude range in \sdss\ DR8. We also release code
to reproduce the catalog from the \sdss\ point-source catalog.

The \xdqso\ target selection technique can be extended to include low
signal-to-noise data in \nir\ and \uv\ filters and to include other
information such as quasar variability. It is the low signal-to-noise
ratio regime at the faint edge of surveys that often contains the most
interesting objects. \xdqso\ is the only target selection technique
currently available that can calculate robust quasar probabilities
taking the data-uncertainties fully into account. Since the most
successful photometric quasar catalogs are based on calculating good
photometric quasar probabilities \citep{Richards09a}, the
\xdqso\ technique or similar will be essential to create the best and
largest photometric quasar catalogs in upcoming surveys such as
\panstarrs\ and \lsst.

%BOVY: reference point

\acknowledgements It is a pleasure to thank Gordon Richards, Michael
Strauss, and Christophe Yeche for helpful comments and
assistance. J.B. and D.W.H. were partially supported by NASA (grant
NNX08AJ48G) and the NSF (grant AST-0908357). D.W.H. is a research
fellow of the Alexander von Humboldt Foundation of Germany. J.F.H
acknowledges support provided by the Alexander von Humboldt Foundation
in the framework of the Sofja Kovalevskaja Award endowed by the German
Federal Ministry of Education and Research.

Funding for the SDSS and SDSS-II has been provided by the Alfred
P. Sloan Foundation, the Participating Institutions, the National
Science Foundation, the U.S. Department of Energy, the National
Aeronautics and Space Administration, the Japanese Monbukagakusho, the
Max Planck Society, and the Higher Education Funding Council for
England. The SDSS Web Site is http://www.sdss.org/.

The SDSS is managed by the Astrophysical Research Consortium for the
Participating Institutions. The Participating Institutions are the
American Museum of Natural History, Astrophysical Institute Potsdam,
University of Basel, University of Cambridge, Case Western Reserve
University, University of Chicago, Drexel University, Fermilab, the
Institute for Advanced Study, the Japan Participation Group, Johns
Hopkins University, the Joint Institute for Nuclear Astrophysics, the
Kavli Institute for Particle Astrophysics and Cosmology, the Korean
Scientist Group, the Chinese Academy of Sciences (LAMOST), Los Alamos
National Laboratory, the Max-Planck-Institute for Astronomy (MPIA),
the Max-Planck-Institute for Astrophysics (MPA), New Mexico State
University, Ohio State University, University of Pittsburgh,
University of Portsmouth, Princeton University, the United States
Naval Observatory, and the University of Washington.

SDSS-III is managed by the Astrophysical Research Consortium for the
Participating Institutions of the SDSS-III Collaboration including the
University of Arizona, the Brazilian Participation Group, Brookhaven
National Laboratory, University of Cambridge, University of Florida,
the French Participation Group, the German Participation Group, the
Instituto de Astrofisica de Canarias, the Michigan State/Notre
Dame/JINA Participation Group, Johns Hopkins University, Lawrence
Berkeley National Laboratory, Max Planck Institute for Astrophysics,
New Mexico State University, New York University, the Ohio State
University, the Penn State University, University of Portsmouth,
Princeton University, University of Tokyo, the University of Utah,
Vanderbilt University, University of Virginia, University of
Washington, and Yale University.

\appendix

\section{Flag Cuts}\label{sec:flags}
%ADM wrote v1.0 of this section October 8, 2010

As part of the information in our catalog, we include an entry
\flag{good} that facilitates two different sets of useful flag
cuts. One set is the \boss\ flag cuts, which are appropriate when
attempting to use the information in the catalog directly for
statistical studies. For objects that do {\em not} pass this set of
flag cuts, \flag{good} $> 0$. The second set of flag cuts is less
restrictive, and is more appropriate for follow-up observations where
high completeness is required (such as, for instance, a search for
high redshift quasars). For objects that do {\em not} pass this less
restrictive set of flag cuts, \flag{good} = 1. Objects that pass these
cuts but fail other \boss\ flag cuts have \flag{good} = 2.

We also provide a flag \flag{photometric}, which is not bitwise---it
is either True or False---and can be utilized at the user's
discretion. This is a standard \sdss\ flag: \flag{photometric} refers
to objects that were observed under good imaging
conditions. Restricting to \flag{photometric} affects the coverage of
the catalog and should only be considered when constructing a highly
restrictive sample.

\subsection{\flag{good} = 1}

Our less restrictive set of flag cuts (\flag{good} = 1) slightly
differs from the standard \boss\ quasar targeting flags. They use
broader magnitude limits of $17.75 \leq i < 22.45$ (extincted PSF
luptitudes) and only adopt a subset of the raw flag cuts. We adopt the
raw flag cut definitions detailed in \citet{Stoughton:2002ae}, except
for gerr, rerr and ierr, which denote errors on extincted PSF
luptitudes in the $g$, $r$ and $i$ bands, and for the column and row
movement information. First, two sets of flag cuts are defined from
the raw flag cuts, denoting whether the photometric pipeline had
difficulty deblending or interpolating a source:\\

\begin{quote}
    \flag{INTERP\_PROBLEMS} = (\flag{PSF\_FLUX\_INTERP} \&\& ($ {\rm gerr}>0.2 \parallel {\rm rerr} >0.2 \parallel {\rm ierr}>0.2$)) $\parallel$ \flag{BAD\_COUNTS\_ERROR} $\parallel$ (\flag{INTERP\_CENTER} \&\& \flag{CR}) \\
    \flag{DEBLEND\_PROBLEMS} = \flag{PEAKCENTER} $\parallel$ \flag{NOTCHECKED} $\parallel$ (\flag{DEBLEND\_NOPEAK} \&\& (${\rm gerr}>0.2 \parallel {\rm rerr}>0.2 \parallel {\rm ierr}>0.2$))\\
\end{quote}
where symbols (\&\&, $\parallel$, !) denote standard Boolean
logic. \flag{MOVED} is set if the raw flag
\flag{DEBLENDED\_AS\_MOVING} is set {\em and} if the value and error
on the row or column position in the SDSS imaging pipeline (rowv,
colv, rowverr, colverr) suggest a $> 3\sigma$ move across a row or a
column.  Then, if (\flag{INTERP\_PROBLEMS} $\parallel$
\flag{DEBLEND\_PROBLEMS} $\parallel$ \flag{MOVED}) we set \flag{good}
= 1.

\subsection{\flag{good} = $2$}

The flag cuts leading to \flag{good} = 1 are less restrictive than the
\boss\ quasar survey flags. Additional flag cuts to mimic the
\boss\ flags use magnitude limits of $(g \le 22 \parallel r \le
21.85)~\&\&~i \ge 17.8$ where magnitudes are extincted PSF
luptitudes. Then for objects passing the \flag{good} = 1 flag cuts we
have:

\begin{quote}
  IF ((\flag{!BINNED1} $\parallel$ \flag{BRIGHT} $\parallel$
  \flag{SATURATED} $\parallel$ \flag{EDGE} $\parallel$ \flag{BLENDED}
  $\parallel$ \flag{NODEBLEND} $\parallel$ \flag{NOPROFILE})
   \&\& !(\flag{INTERP\_PROBLEMS} $\parallel$
   \flag{DEBLEND\_PROBLEMS} $\parallel$ \flag{MOVED})) THEN
   (\flag{GOOD} = 2)
\end{quote}

All of our individual flag cuts are also recorded as a mask of bits
(called \flag{bitmask}). For most users, we recommend a cut at
\flag{good} != 1 for follow-up observations and \flag{good} = 0 for
direct statistical analyses or to mimic \boss\ targeting. For users
brave enough to use our catalog beyond this level of sophistication,
\flag{bitmask} can be used to back out individual flag
cuts. \flag{bitmask} conforms to the values listed in the
\boss\ Target Selection Framework paper (E.~S.~Sheldon et al., 2011,
in preparation).

\section{Target selection code}\label{sec:code}

\begin{figure}
\includegraphics[width=0.5\textwidth]{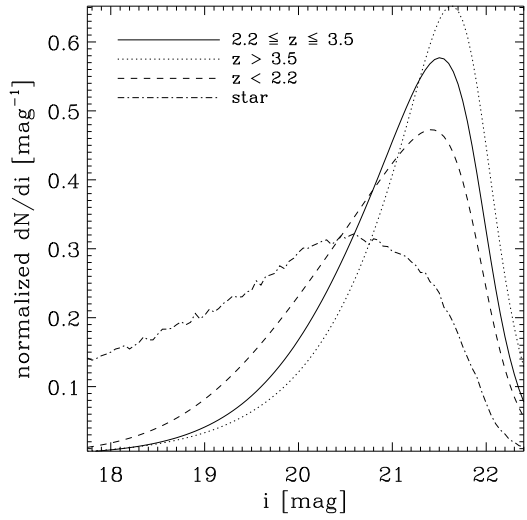}
\caption{Number counts $p(\fluxi | i \in \mathrm{class})$ for the
  various target classes. These have the expected property that higher
  redshift quasars are fainter since they are more
  distant.}\label{fig:numcounts}
\end{figure}

The \xdqso\ target-selection technique described in
\sectionname~\ref{sec:exdmodel} is made publicly available at
\begin{quote}
\url{http://data.sdss3.org/sas/dr8/groups/boss/photoObj/xdqso/xdqso-code.tar.gz}
\end{quote}
as a set of data files containing the parameters of the model and code
that calculates the \xdqso\ likelihoods and probabilities. The data
files need to be saved in a directory given by an environment variable
\texttt{\$XDQSODATA}. The data files consist of text files containing
the number count priors of \tablename~\ref{table:numcounts} and
\figurename~\ref{fig:numcounts} and FITS files containing the
\exd\ models for all of the bins with one file for each target
class. Each FITS file contains \nbins\ extensions, where extension $k$
contains a structure with the amplitudes (tag \flag{xamp}), means (tag
\flag{xmean}), and covariance matrices (tag \flag{xcovar}) for bin $k$
in \iband\ magnitude.

In the IDL implementation the procedure \flag{xdqso\_calculate\_prob}
takes as input a structure containing tags \flag{psfflux},
\flag{psfflux\_ivar}, and, if the option \flag{/dereddened} is not
set, \flag{extinction} containing the fluxes, inverse flux-variances,
and extinction values in the five \sdss\ bands. This is the structure
of the \sdss\ `sweeps' \flag{calibObj}\footnote{See
  \url{http://data.sdss3.org/datamodel/files/PHOTO\_SWEEP/RERUN/calibObj.html}~.} files. The
output is a structure that mirrors the input structure and contains
all of the \xdqso\ likelihoods, number count priors, and probabilities
as given in the catalog description in
\tablename~\ref{table:catdescription}. It is more efficient to run the
code on arrays of objects rather than on objects individually, as each
data file is read only once.

\clearpage
\begin{deluxetable}{cr@{.}lr@{.}lr@{.}lr@{.}l}
\tablecaption{Total Number Counts in 17.75 $\leq i < 22.45$.\label{table:numcounts}}
\tablecolumns{9}
\tablewidth{0pt}
\tablehead{\colhead{} & \multicolumn{2}{c}{$z < 2.2$} & \multicolumn{2}{c}{$ 2.2 \leq z \leq 3.5$} & \multicolumn{2}{c}{$z > 3.5$} & \multicolumn{2}{c}{Star}\\
\colhead{} & \multicolumn{2}{c}{Quasar} & \multicolumn{2}{c}{Quasar} & \multicolumn{2}{c}{Quasar} & \multicolumn{2}{c}{}}
\startdata
Number counts (deg$^{-2}$) & 140&72 & 50&70 & 6&13 & 5209&38\\
\enddata
\end{deluxetable}

\clearpage
\begin{deluxetable}{ll}
\tablecaption{\sdssxdqso\ catalog entries.\label{table:catdescription}}
\tablecolumns{2}
\tablewidth{0pt}
\tabletypesize{\scriptsize}
\tablehead{\multicolumn{1}{l}{Column} & \multicolumn{1}{l}{Description}}
\startdata
objId & Unique SDSS identifier\tablenotemark{\protect{\ref{objid}}}\\
run & \sdss\ imaging run number\\
rerun & \sdss\ processing rerun number\\
camcol & Camera column\\
field & Field number\\
id & The object id within a field\\
RA & Right ascension in decimal degrees (J2000.0)\\
Dec & Declination in decimal degrees (J2000.0)\\
photometric & Indicates whether this object was observed under good imaging conditions (True/False)\\
psfMag & PSF \"{u}bercalibrated \ugriz\ asinh magnitude (corrected for Galactic extinction)\\
psfMagErr & Error in PSF \ugriz\ asinh magnitude\\
extinction\_u & Extinction in $u$ band; $A_u/A_g/A_r/A_i/A_z = 5.155/3.793/2.751/2.086/1.479$\\
qsolowzlike & Relative-flux density factor $p(\setofrfj | \fluxi, z < 2.2 \ \mathrm{quasar})$\\
qsomidzlike & Same as previous, but for the $2.2 \leq z \leq 3.5$ quasar class\\
qsohizlike & Same as previous, but for the $z > 3.5$ quasar class\\
starlike & Same as previous, but for the \starc\ class\\
qsolowznumber & Number density as a function of apparent magnitude for this object for the\\
& $z < 2.2$ quasar class (in deg$^{-2}$ mag$^{-1}$; =$p(\fluxi | z < 2.2\ \mathrm{quasar})\,P(z < 2.2\ \mathrm{quasar})$)\\
qsomidznumber & Same as previous, but for the $2.2 \leq z \leq 3.5$ quasar class\\
qsohiznumber & Same as previous, but for the $z > 3.5$ quasar class\\
starnumber & Same as previous, but for the \starc\ class\\
pqsolowz & Probability that the object is a $z < 2.2$ quasar\\
pqsomidz & Probability that the object is a $2.2 \leq z \leq 3.5$ quasar\\
pqsohiz & Probability that the object is a $z > 3.5$ quasar\\
pqso & Probability that the object is a quasar (sum of previous three)\\
pstar & Probability that the object is a star\\
bitmask & \boss\ quasar target selection bitmask\\
good & \flag{good} flag: 0: passes all \boss\ cuts, 1: fails basic \boss\ cut (see text),\\
&  2: fails other \boss\ cuts (see Appendix~\ref{sec:flags})\\
\enddata
\setcounter{tabletwo}{1}
\makeatletter
\let\@currentlabel\oldlabel
\newcommand{\@currentlabel}{\thetabletwo}
\makeatother
\renewcommand{\thetabletwo}{\alph{tabletwo}}
\tablenotetext{\thetabletwo}{\label{objid}
Calculated using \versiontag{photoop v1\_9\_9}.\stepcounter{tabletwo}}
\end{deluxetable}

\clearpage
\begin{deluxetable}{crrr}
\tablecaption{Comparison of the NBC-KDE and \sdssxdqso\ catalogs.\label{table:kde}}
\tablecolumns{4}
\tablewidth{0pt}
\tablehead{\colhead{} & \colhead{confirmed $z < 2.2$} & \colhead{confirmed $ 2.2 \leq z \leq 3.5$} & \colhead{confirmed $z > 3.5$}\\
\colhead{} & \colhead{quasar} & \colhead{quasar} & \colhead{quasar}}
\startdata
NBC-KDE & 77144  & 11485  & 2802\\
\xdqso\ & 77517  & 12711  & 2338\\
\\
total \# targets & 569785 & 151860 & 9086
\enddata
\end{deluxetable}

\clearpage

\clearpage

\clearpage

\clearpage

\clearpage

\clearpage

\clearpage

\clearpage

\clearpage

\clearpage

\clearpage

\clearpage

\clearpage

\clearpage

\clearpage

\end{document}